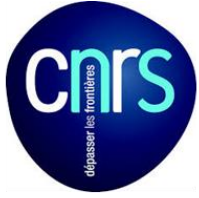
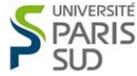
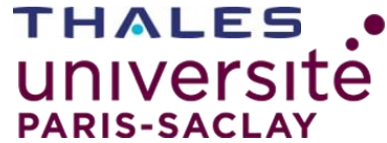

Unité Mixte de Physique CNRS/Thales
Univ. Paris Sud, Université Paris Saclay
91767 Palaiseau - France

# Habilitation à Diriger des Recherches
De l'Université Paris-Sud XI

# Ferroelectric tunnel junctions

Vincent Garcia

Unité Mixte de Physique CNRS/Thales
Palaiseau, France

# Table of Contents

# 1. Summary

My research is dedicated to the electronic properties of functional oxides. My activity specifically focuses on ferroelectric tunnel junctions in which an ultrathin layer of ferroelectric material is intercalated between two metallic electrodes. In these devices, polarization reversal induces large modifications of the tunnel resistance, leading to a non-destructive readout of the information. After a demonstration of the concept with scanning probe microscopy techniques, I have been exploring the properties of ferroelectric tunnel junctions with various ferroelectric materials ($BaTiO_3$, $BiFeO_3$, and PVDF). I showed that these devices possess attractive properties for applications as non-volatile binary memories. In addition, exploring the fact that polarization usually reverses by the nucleation and propagation of domains, I demonstrated a memristive behavior in the junctions associated to non-uniform configurations of ferroelectric domains. Such ferroelectric memristors can be used as artificial synapses in neuromorphic architectures. Coupled to magnetic electrodes, the resulting multiferroic tunnel junctions enable a non-volatile control of magnetism at the ferroelectric/electrode interface and of the spin-polarized current associated. Besides this main activity on tunnel devices, I explored the influence of ferroelectricity on magnetic orders and on the properties of functional oxides.

# 2. Résumé

Mon activité de recherche se focalise sur les propriétés électroniques d'oxydes fonctionnels. Ces travaux s'articulent notamment autour des jonctions tunnel ferroélectriques où une couche ultramince ferroélectrique est intercalée entre deux électrodes. Dans ces dispositifs, le renversement de la polarisation ferroélectrique induit de fortes modulations de la résistance tunnel, conduisant à une lecture non-destructive de l'information. Après une démonstration du concept par des techniques de microscopie de champ proche, j'ai exploré les propriétés de jonctions tunnel ferroélectriques à partir de différents systèmes ferroélectriques ($BaTiO_3$, $BiFeO_3$ et PVDF) et montré que ces dispositifs possèdent des propriétés attractives pour des applications en tant que mémoires non-volatiles binaires. De plus, en exploitant le fait que la polarisation ferroélectrique se renverse par la nucléation et la propagation de domaines, j'ai montré un comportement memristif associé à des configurations non uniformes de polarisations. Ces memristors ferroélectriques peuvent être utilisés comme synapses artificielles au sein d'architectures neuromorphiques. Couplées à des électrodes magnétiques, ces jonctions tunnel multiferroïques permettent un contrôle non-volatile du magnétisme d'interface et du courant tunnel polarisé en spin associé. Hormis cette activité centrale sur les jonctions tunnel, j'ai exploré l'influence de la ferroélectricité sur les ordres magnétiques et les propriétés d'oxydes fonctionnels.

# 3. SCIENTIFIC ACTIVITIES

## 3.1 Preface

Ferroelectric random access memories (FeRAMs), in which information is encoded through the ferroelectric polarization, are commercially available products with fast write speed, large read/write cycle endurance and low power consumption (*1*). In these ferroelectric capacitors, a ferroelectric thin film (typically 100 nm thick) is sandwiched between two electrodes and the remanent polarization is switched by applying an electric field between the electrodes. The widespread development of these memories is however limited due to the capacitive readout of the information (the ferroelectric polarization) that prevents the scalability of FeRAMs up to gigabit densities and necessitates the rewriting of information after readout (destructive readout) (*2*). In ferroelectric diodes, the current across a thick ferroelectric can be modulated by the polarization at the ferroelectric/electrode interface, giving rise to a non-destructive resistive readout of the information (*3*, *4*). But in these devices, the large thickness of the ferroelectric layer results in very low readout currents, limiting their miniaturization. As the ferroelectric layer thickness is decreased to a few nanometers, electronic conduction is greatly enhanced as quantum-mechanical tunneling through the ferroelectric becomes possible (*5*, *6*). These devices in which two electrodes sandwich a ferroelectric tunnel barrier (Figure 1) are called ferroelectric tunnel junctions (FTJs) (*7*, *8*). The tunnel transmission may be strongly modulated by the ferroelectric polarization producing giant tunnel electroresistance (TER) with OFF/ON resistance ratios reaching $10^4$ (*9–11*). Moreover, the readout current densities are much larger than in ferroelectric diodes, opening the path to possible applications as high-density ferroelectric memories with non-destructive readout.

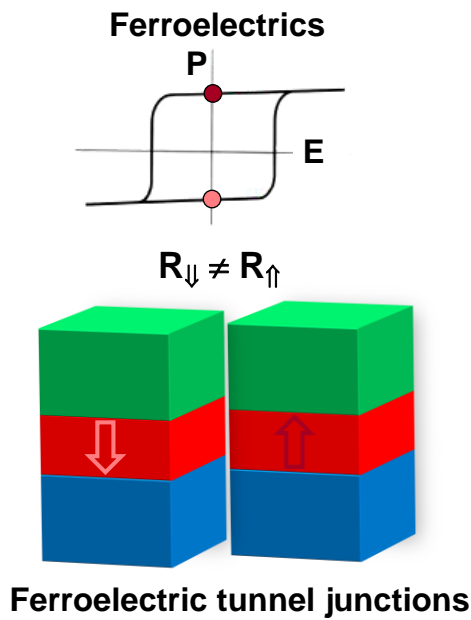

*Figure 1. Concept of ferroelectric tunnel junctions: an ultrathin layer of ferroelectric enables electrons to tunnel through. Depending on the polarization orientation the tunnel transmission varies, giving rise to a simple, non-destructive, resistive readout of the ferroelectric information.*

In this manuscript, I review our contribution to the nascent field of FTJs since 2009. First I briefly describe the mechanisms at the origin of tunnel electroresistance when voltage is applied across ultrathin ferroelectric films. Then I focus on our experimental reports that evidence a direct correlation between ferroelectric switching and resistive switching using scanning probe techniques, from experiments on ferroelectric surfaces to solid-state devices. Fundamental properties of these devices are discussed including microscopic investigations of polarization with electron microscopy, specific signatures of tunneling transport, and electrode/barrier interface engineering and its influence on tunnel electroresistance. With organic ferroelectric tunnel junctions, I show that simple electrostatic approaches enable a fully quantitative description of tunnel electroresistance owing to the weak metal-ferroelectric interfacial bonding. The key features of state-of-the-art ferroelectric tunnel junctions are then described in terms of potential applications as non-volatile memories, with a particular attention to the device reliability, endurance, scalability and retention. This careful crafting of the junction's structural and electronic details yields better performing devices, but also brings additional functionalities. One degree of freedom to engineer the response of these junctions is the domain structure in the ferroelectric barrier. We recently exploited it to achieve a new form of memristive behavior, expanding the scope of applications of ferroelectric junctions from digital information storage to brain-inspired computation. Through an in-depth investigation of ferroelectric domain dynamics combining scanning probe microscopy and tunnel transport, we demonstrate that the learning behavior of artificial synapses based on ferroelectric memristors can be modeled and anticipated by a nucleation-limited switching model. In addition, ferroelectric switching in the barrier can induce modifications of the magnetic properties when using ferromagnetic electrodes, or phase transitions in strongly correlated oxides. This functionality also opens new routes for the nanoscale control of their electronic and spintronic response. We show that spin-dependent tunneling is a powerful tool to probe such interfacial magnetoelectric coupling.

## 3.2 Interplay between electron transport and ferroelectric polarization

### 3.2.1 Tunnel electroresistance: the simple electrostatic picture

At the surface of a ferroelectric, polarization charges are usually present and, depending on their sign, will repel or attract electrons. This occurs over a short distance in the electrode beyond which the density of electrons resumes its normal value: the electrons near the interface screen the polarization charges. In the Thomas-Fermi theory, the screening length is a function of the electronic density of states at the Fermi level. For very good metals, the Thomas-Fermi screening length can be shorter than a tenth of nanometer. For semiconductors, it can reach tens of nanometers and screening is imperfect. However, as emphasized by Stengel *et al.*, the actual effective screening length differs from the one defined by Thomas-Fermi and strongly depends on the microscopic properties of the ferroelectric/electrode interface system (*12*).

This incomplete screening gives rise to an additional electrostatic potential at the ferroelectric/electrode interface (> 0 when P points to the interface and < 0 when P points away from the interface). We now consider an ultrathin ferroelectric layer sandwiched between two different electrodes with more efficient screening on the left side than on the right side (Figure 2). The larger the screening length over the dielectric constant of the electrode, the larger is the additional electrostatic potential at the electrode/ferroelectric interface. For simplicity, we suppose that the initial electronic potential barrier is rectangular (i.e. identical barrier heights for the left and right interfaces). Polarization charge effects induce an asymmetric modulation of the electronic potential profile (*7*, *8*). When P is reversed the asymmetry of the electronic potential profile is reversed. This results in the barrier height being in average higher when P points to the left than when P points to the right. Because the tunnel transmission depends exponentially on the square root of the barrier height (*13*), the junction resistance will depend on the direction of P. The asymmetry between the two ferroelectric/electrode interfaces is thus essential to modulate the current transmission of the ferroelectric barrier. The use of different electrode materials is however not mandatory and the asymmetry may come from the interfaces only, due to different interface terminations (*14*), an ultrathin dielectric layer at one of the interfaces (*15*), or pinned interface dipoles (*16*, *17*).

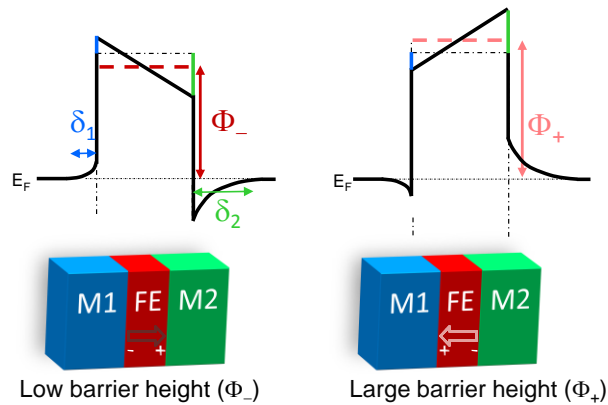

*Figure 2. Tunnel electroresistance induced by the modulation of the potential profile of the ferroelectric (FE) barrier when polarization is reversed in the case of asymmetric electrodes (M1 and M2).*

### 3.2.2 Mechanisms of electroresistance through ultrathin films

In this section, we review the influence of ferroelectric polarization (P) on the transport properties of an ultrathin ferroelectric layer sandwiched between two metallic electrodes. As described in the previous section, the imperfect screening of polarization charges at the interface results in a distorted potential profile whose asymmetry and average height can depend on the polarization direction. This is the main mechanism producing tunnel electroresistance in FTJs.

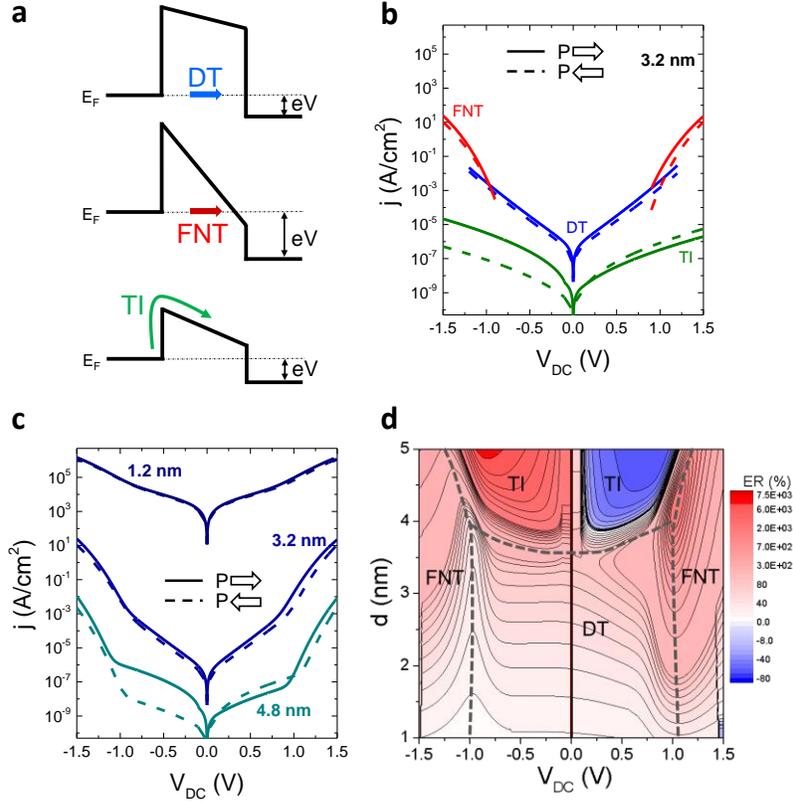

*Figure 3. Electron transport across ultrathin ferroelectrics. (a) Sketch of the three possible transport mechanisms through ultrathin ferroelectrics: direct tunneling (DT), Fowler-Nordheim tunneling (FNT) and thermionic emission (TI). (b) Current densities (J) vs. voltage ($V_{DC}$) of the contributions from DT, FNT, and TI through a 3.2-nm-thick ferroelectric. (c) Total current densities $J = J_{DT}+J_{FNT}+J_{TI}$ vs. $V_{DC}$ for a 1.2-, 3.2- and 4.8-nm-thick ferroelectric. In (b) and (c), solid and dashed lines stand for P right and P left, respectively. (d) Corresponding electroresistance (ER) as a function of voltage and d as the ferroelectric thickness. The parameters used in (b-d) are P = 3 µC/cm², $\varepsilon_{stat}$ = 60, $\varepsilon_{ifl}$ = 60, m* = 1, $A^{**}$ = $10^6$ $Am^{-2}K^{-2}$, $\lambda_1$ = 0.08 nm, $\lambda_2$ = 0.055 nm, $\varepsilon_1$ = 8, $\varepsilon_2$ = 2, $\Phi_{1,2}$ = 1 eV. Adapted from (18).*

Although this manuscript mainly deals with FTJs, direct tunneling is not the only possible electronic transport mechanism across ultrathin ferroelectric films. Pantel *et al.* (*18*) calculated the influence of polarization charge effects on the current flowing through ultrathin ferroelectrics (in the limit of fully depleted ferroelectrics) within three possible transport mechanisms: direct tunneling, Fowler-Nordheim tunneling, and thermionic emission (Figure 3a). Figure 3b illustrates the contribution of each transport mechanism for an intermediate barrier thickness: direct tunneling current is prominent at low voltage and Fowler-Nordheim tunneling dominates at large voltage. As the barrier thickness increases, the direct tunneling current decreases exponentially and the current due to thermionic emission dominates (Figure 3c). Interestingly, the electroresistance (ER, defined here as ($J_{P\rightarrow}-J_{P\leftarrow}/J_{P\leftarrow}$)) increases in the low bias regime ($V_{DC}$ = -0.1 V) as the barrier thickness increases (Figure 3d) (*7*, *18*). For large thicknesses when transport is dominated by thermionic injection, the ER reaches high values and its sign reverses at positive voltage. Nevertheless, the large increase of the ER in the thermionic injection regime is at the expense of a strong decrease of the read current (Figure 3c) (*18*).

Experimentally, the frontier between these different transport mechanisms is hard to draw as it depends on various specific parameters (dielectric constants, screening lengths, interface barrier heights, effective mass). Nevertheless, specific signatures of each transport mechanism can be identified through the voltage and temperature dependences of the electronic current. For example, thermionic emission is a thermally-activated process that should in principle lead to strong variations with temperature; this is not the case for direct tunneling.

The interplay between electronic transport and ferroelectricity is in reality much more complex than what can be inferred from classical electrostatic models based on polarization charge effects. As ferroelectrics are all piezoelectric, converse piezoelectric effects result in a modification of the ferroelectric thickness when a voltage is applied. The voltage dependence of this modulation is reversed when the ferroelectric polarization is switched. This influences the current across a ferroelectric when read at a large voltage (*8*). However, FTJs are usually read at low voltage to avoid destabilizing the ferroelectric state and the piezoelectric coefficient may be limited in clamped ultrathin films (*19*, *20*).

In depth investigations on model systems have been performed by ab initio calculations for direct tunneling transport. The tunnel current is dependent on the product of interface transmission functions and on the decay rate of the evanescent states within the barrier (related to its complex band structure) (*21*). In the case of Pt/$BaTiO_3$/Pt tunnel junctions, the modifications of the electronic structure related to ferroelectric displacements strongly influence the transmission of the tunnel barrier via a modulation of the complex band structure of the barrier (*22*). The interface transmission probability at the Pt/$BaTiO_3$ interface varies with the orientation of the polarization (*22*). For $SrRuO_3$/$BaTiO_3$/$SrRuO_3$ junctions, asymmetries in the interface terminations (BaO/$RuO_2$ and $TiO_2$/SrO) induce an asymmetry of the amplitude of the ferroelectric displacements when the polarization points toward one or the other interface. As a result, a change of the band gap of $BaTiO_3$ for the two orientations of the polarization modifies the tunneling decay rate in $BaTiO_3$, which leads to a change in the tunnel current (*14*).

**Related publication:**
*Ferroelectric tunnel junctions for information storage and processing*
V. Garcia and M. Bibes
**NATURE COMMUNICATIONS 5, 4289 (2014)**

## 3.3 Ferroelectric polarization control of resistive switching

### 3.3.1 Scanning probe microscopy on ferroelectric surfaces

The concept of FTJs was proposed by Esaki in 1971 (*23*). It was forgotten for about 30 years as ferroelectricity was believed to vanish at critical thicknesses well above the nanometer range (*24*). Due to tremendous improvements in the growth of high quality epitaxial films and in the ferroelectric characterization techniques, ultrathin films of ferroelectric materials were experimentally demonstrated in the early 2000s (*25–33*). First experiments performed in Jülich resulted in the demonstration of 400% resistive switching at room temperature in junctions based on ultrathin films of $PbZr_{0.52}Ti_{0.48}O_3$ sandwiched between $SrRuO_3$ and Pt (*34*). A few years later, the same group proposed a method to ascribe ferroelectric switching to resistive switching in ferroelectric capacitors and concluded that their initial results were probably not related to ferroelectricity (*35*). Combining local-probe techniques based on atomic force microscopy (AFM), three groups including ours unambiguously demonstrated in 2009 the correlation between ferroelectricity and resistive switching on bare surfaces of ferroelectrics (*36–38*). We showed large TER of up to 750 correlated with ferroelectricity in highly strained ultrathin films of $BaTiO_3$, by combining images of ferroelectric domains by piezoresponse force microscopy (PFM) and resistance maps of the same domains with conductive-AFM (C-AFM) (Figure 4).

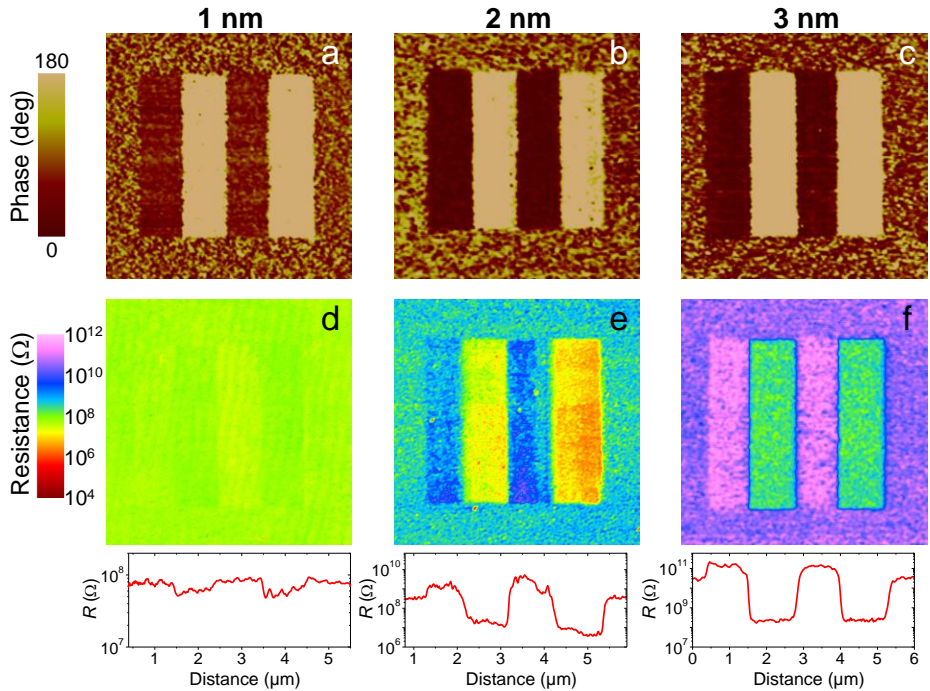

*Figure 4. Direct evidence for ferroelectricity-related TER with ultrathin $BaTiO_3$ films. Parallel (a-c) PFM phase and (d-f) C-AFM resistance maps of four ferroelectric stripes for $BaTiO_3$ films, 1, 2, and 3 nm thick. The corresponding resistance profiles of the poled area are displayed underneath (36).*

In addition, the resistance in the downward or upward polarized domains was found to increase exponentially with the thickness of $BaTiO_3$, for films between 1 and 3.5 nm (Figure 5a), but with different slopes. This suggests that electrons tunnel through the ultrathin films of $BaTiO_3$ and that polarization reversal results in a modification of the barrier transmission by electrostatic effects, resulting in an exponential increase of TER with the barrier thickness (Figure 5b) (*7*). Similar results were independently observed through ultrathin films of BaTiO$_3$ by Gruverman *et al.* with additional local current vs. voltage curves that were fitted by direct tunneling models (*38*). Maksymovych *et al.* connected a surface of PbZr$_{0.2}$Ti$_{0.8}$O$_3$ (30 nm) with a conductive AFM tip. They collected the local dc piezoresponse of the ferroelectric and measured simultaneously the current as a function of the voltage applied through the ferroelectric. A large change of current was directly correlated to ferroelectric switching. This 500-fold current contrast at large negative voltage was described within the Fowler-Nordheim tunneling regime (*37*). We note that for such a thick ferroelectric film, Fowler-Nordheim tunneling involves electron transport through the ferroelectric itself which may occur via hopping or other mechanisms, although the Schottky barrier resistance seems to be predominant. Within a collaboration with the group of Beatriz Noheda (Univ. Groningen), we demonstrated, combining scanning probe techniques on epitaxial films of PbTiO$_3$, large TER increasing with the barrier thickness up to maximal values of 500 for 3.6-nm-thick films (*39*). These and other demonstrations of ferroelectricity-induced electroresistance on bare surfaces of ferroelectrics using local probes (*36–41*) triggered the development of solid-state resistive devices based on ferroelectrics.

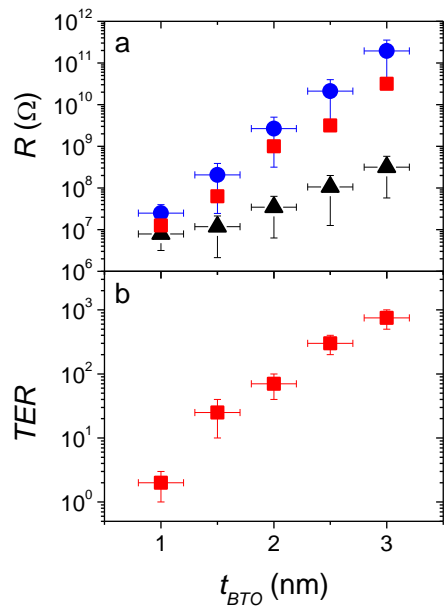

*Figure 5. (a) BaTiO$_3$-thickness dependence of the resistance of unpoled (red squares), positively (black triangles), and negatively (blue circles) poled regions. (b) Corresponding variation of the TER with thickness (36).*

**Related publications:**
*Giant tunnel electroresistance for non-destructive readout of ferroelectric states*
V. Garcia, S. Fusil, K. Bouzehouane, S. Enouz-Vedrenne, N. D. Mathur, A. Barthélémy, M. Bibes
**NATURE 460, 81-84 (2009)**
*Giant tunnel electroresistance with PbTiO$_3$ ferroelectric tunnel barriers*
A. Crassous, V. Garcia, K. Bouzehouane, S. Fusil, A. H. G. Vlooswijk, G. Rispens, B Noheda, M. Bibes, and A. Barthélémy
**APPLIED PHYSICS LETTERS 96, 042901 (2010)**

### 3.3.2 Probing ferroelectricity through top electrodes

In 2011, Pantel *et al.* fabricated submicron devices based on an ultrathin film of $PbZr_{0.2}Ti_{0.8}O_3$ of 9 nm sandwiched between a bottom electrode of $La_{0.67}Sr_{0.33}MnO_3$ and a top one of Cu (*42*). They observed the correspondence between ferroelectricity, via PFM vs. voltage loops acquired through the top Cu electrode, and the electroresistance measured by I vs. $V_{DC}$. An OFF/ON ratio of 1500 was found, with a relatively high current density of 10 A/cm$^2$ at low read voltage in the ON state. The transport properties were described by the thermionic emission mechanism (*42*).

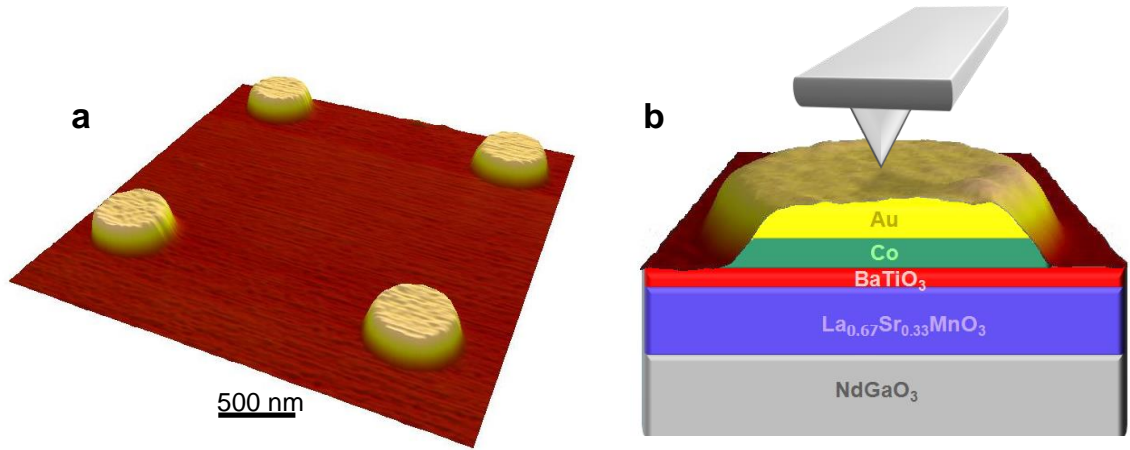

*Figure 6. Sketch of the BaTiO$_3$-based tunnel devices. (a) AFM image of four nanodevices defined by electron-beam lithography on a continuous film of BaTiO$_3$. (b) Sketch of the cross-section of one device contacted by the C-AFM tip.*

From 2010, we investigated tunnel junctions based on highly strained 2-nm-thick BaTiO$_3$ films grown on $La_{0.67}Sr_{0.33}MnO_3$ bottom electrodes with arrays of submicron Au/Co top electrodes defined by electron-beam lithography and lift-off (Figure 6) (*43*). Each solid-state junction is electrically connected by a conductive AFM tip which enables (i) to test a large number of devices without requiring a complex multiple-step lithography process and (ii) to probe the ferroelectric properties of the tunnel barrier through the top electrode (*44–46*). Indeed, in this geometry the whole junction area is electrically excited but the electromechanical response is locally detected under the AFM tip (*44*, *47*). As the piezoresponse is strongly attenuated by the top metallic electrode, we amplified the signal by tracking the contact resonance between the tip and the surface, in collaboration with Asylum Research (*43*). Local PFM phase and amplitude vs. voltage hysteresis are observed through the top metallic electrode suggesting that BaTiO$_3$ is ferroelectric inside the device (Figure 7a-b). The remanent resistance of the device is shown to change hysteretically with OFF/ON ratios of 100 after applying successive voltage pulses of 100 µs, with coercive fields matching ferroelectric switching events detected by PFM (Figure 7c) (*43*). These results suggest that switching the ferroelectric polarization induces large TER (*35*).

In some cases, probing the piezoresponse through the top electrode is too problematic. This is the case of poly(vinylidene fluoride) (PVDF)-based tunnel junctions, in which the ultrathin ferroelectric polymer is technically challenging to probe already without top electrodes. In such cases, we expect that the weak Van-der-Walls interactions at the interface between the polymer and the electrodes do not strongly affect the ferroelectric switching fields. Ferroelectric coercive fields probed on the free surface of the ferroelectric polymer (Figure 7d-e) can thus be compared to the resistive switching ones collected under dc voltage in the solid-state device (Figure 7f). Both switching events occur in the same voltage window, suggesting that the observed TER is correlated to the switching of polarization in the ultrathin PVDF (*48*).

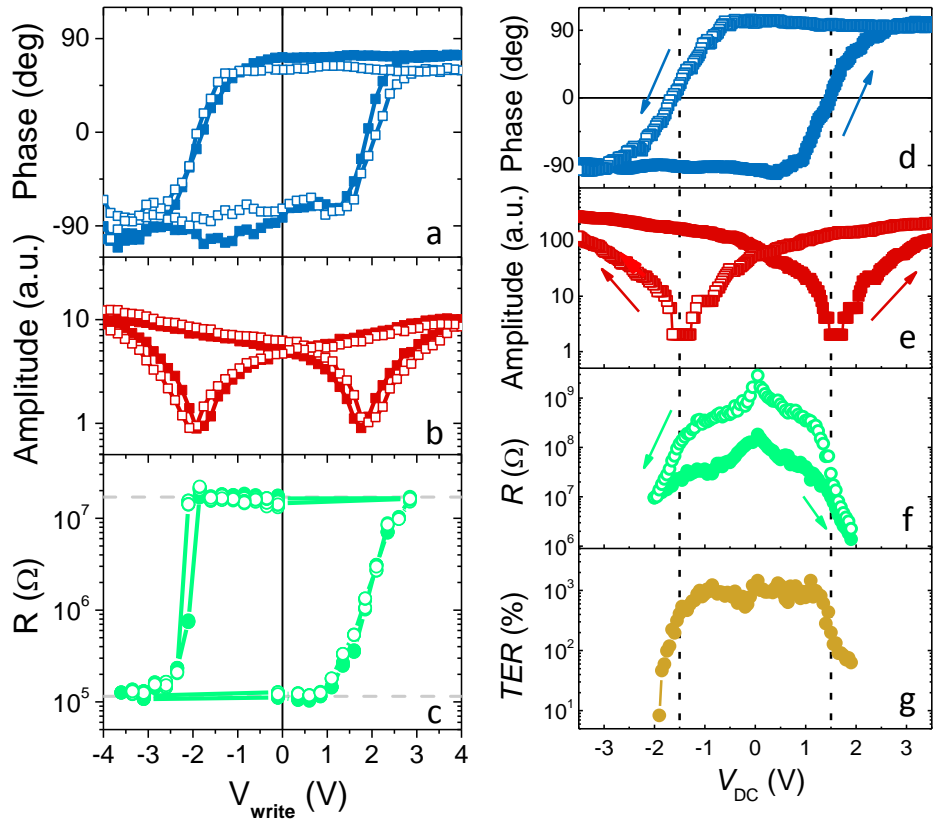

*Figure 7. Ferroelectric switching vs. resistive switching. Out-of-plane PFM (a) phase and (b) amplitude as a function of write voltage pulses across the top electrode of a typical Au/Co/BaTiO$_3$/La$_{0.67}$Sr$_{0.33}$MnO$_3$ junction. (c) Resistance at remanence ($V_{DC}$ = 100 mV) after applying successive voltage pulses of 100 µs. The open and filled circles represent two scans for reproducibility (43). (d-g) Polarization-induced resistance switching in PVDF tunnel junctions. Local PFM (d) phase and (e) amplitude vs. voltage on a 2 Ls PVDF film deposited on Au-coated Si substrates. Voltage dependence of (f) the resistance and (g) corresponding TER in an Au/PVDF (2 Ls)/W tunnel junction (48).*

**Related publications:**
*Solid-state memories based on ferroelectric tunnel junctions*
A. Chanthbouala, A. Crassous, V. Garcia, K. Bouzehouane, S. Fusil, X. Moya, J. Allibe, B. Dlubak, J. Grollier, S. Xavier, C. Deranlot, A. Moshar, R. Proksch, N.D. Mathur, M. Bibes, A. Barthélémy
**NATURE NANOTECHNOLOGY 7, 101-104 (2012)**
*Tunnel electroresistance through organic ferroelectrics*
B.B. Tian, J.L. Wang, S. Fusil, Y. Liu, X.L. Zhao, S. Sun, H. Shen, T. Lin, J.L. Sun, C.G. Duan, M. Bibes, A. Barthélémy, B. Dkhil, V. Garcia, X.J. Meng, J.H. Chu
**NATURE COMMUNICATIONS 7, 11502 (2016)**

## 3.4 Fundamental properties of ferroelectric tunnel junctions

### 3.4.1 Atomic-scale investigations

The realization of ferroelectric tunnel junctions necessitates the stabilization of ferroelectricity for films with thicknesses of few nanometers. Inspired the pioneering work of Choi et al. on the strain-induced improvement of ferroelectrics properties in thick BaTiO$_3$ films (*49*), I employed the strain strategy to stabilize ferroelectricity in ultrathin films of BaTiO$_3$, while I was a post-doc in the group of Neil Mathur (Univ. Cambridge, UK). The thin films are elaborated by pulsed laser deposition on NdGaO$_3$(001) substrates with well-matched buffer electrodes of La$_{0.67}$Sr$_{0.33}$MnO$_3$. These half-metallic electrodes can also act as spin-polarizers for spin-dependent transport. The lattice mismatch between BaTiO$_3$ and NdGaO$_3$ is large and the compressive strain (-3.2%) imposed by the substrate is quickly released by the formation of dislocations: 50-nm-thick BaTiO$_3$ thin films are fully relaxed and epitaxial with the c-axis (polarization axis) parallel to the growth axis (Figure 8a). As thickness decreases, X-ray diffraction characterizations indicate that compressive strain progressively increases (Figure 8a). High-resolution transmission electron microscopy (TEM) investigations reveal that 2-nm-thick BaTiO$_3$ films are fully strained by the substrate (Figure 8b); the large induced tetragonality (c/a = 1.05) is expected to favor ferroelectricity.

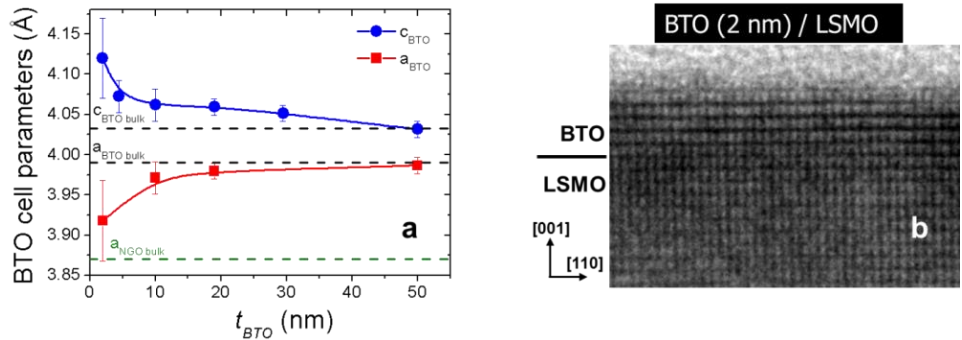

*Figure 8. (a) Evolution of the in-plane and out-of-plane cell parameters with thickness in epitaxial BaTiO$_3$ thin films grown on NdGaO$_3$ substrates, deduced from X-ray diffraction (5-50 nm) and high-resolution electron microscopy (2 nm). (b) Transmission electron microscopy cross-section image of a 2-nm BaTiO$_3$/30-nm La$_{0.67}$Sr$_{0.33}$MnO$_3$ bilayer grown on NdGaO$_3$ substrate. The crystallographic directions are given in pseudo-cubic notation (36).*

Since 2012, our research on FTJs has then been primarily dedicated to BiFeO$_3$ as a ferroelectric tunnel barrier. This material is fascinating given its multiferroic (antiferromagnetic and ferroelectric) character with ordering temperatures well above room temperature (*50*). In addition its large ferroelectric polarization of 100 µC/cm² (*51*) promises large tunnel electroresistance when used in the form of ultrathin films. Thin films of BiFeO$_3$ are elaborated by pulsed laser deposition on YAlO$_3$(001) substrates with buffer electrodes of Ca$_{0.96}$Ce$_{0.04}$MnO$_3$. The growth conditions of the heterostructures were optimized during the one-year visit of Hiroyuki Yamada (AIST, Japan) in the lab. The large epitaxial strain imposed by the substrate stabilizes a polymorphic phase of BiFeO$_3$, called the super-tetragonal phase (T-phase) because of its large c/a ratio (1.25) (*52*). Within the collaboration with the group of Frédéric

Pailloux (Univ. Poitiers), advanced TEM techniques suggested that the ultrathin films of $BiFeO_3$ are purely tetragonal (without any monoclinic distortion) (*53*). Thanks to high-angle annular dark field (HAADF) scanning TEM (STEM) techniques, in collaboration with the group of Alexandre Gloter (LPS), we evidenced the large c/a ratio of ultrathin films of $BiFeO_3$ (10 unit cells) as well as a large ferroelectric polarization (pointing toward the bottom electrode of $Ca_{0.96}Ce_{0.04}MnO_3$) from the relative displacement of Fe and Bi cations (Figure 9).

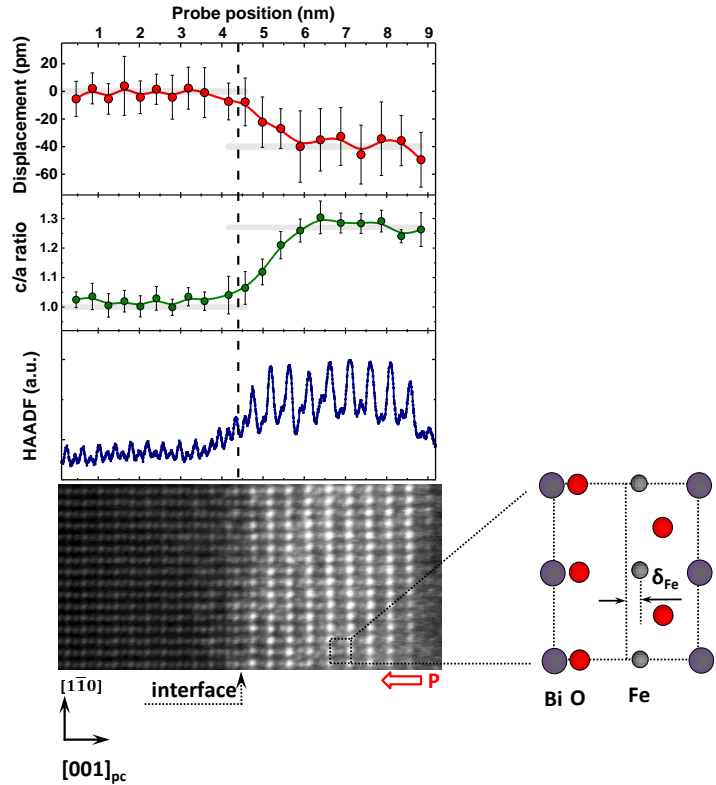

*Figure 9. HAADF-STEM image of a cross section of a $BiFeO_3$ (4.6 nm)/$Ca_{0.96}Ce_{0.04}MnO_3$ heterostructure along the pseudo cubic [110] zone axis. Evolution of the c/a ratio (green points) and of the B site displacement (Fe or Mn) relative to the A site (Bi or Ca) (red points) close to the interface obtained from the HAADF image (blue points). The relative displacement of Fe ($\delta Fe$) is schematized on the right panel (54).*

In addition, by correlating the energy shifts of the $L_{2,3}$ edge of Mn to the Ce doping in a $BiFeO_3$/ $Ca_{0.96}Ce_{0.04}MnO_3$/$Ca_{0.98}Ce_{0.02}MnO_3$/$CaMnO_3$ heterostructure using electron energy loss spectroscopy (EELS), we could directly visualize the excess of electrons at the $BiFeO_3$/$Ca_{0.96}Ce_{0.04}MnO_3$ interface that are screening the polarization charges of $BiFeO_3$ (Figure 10). Integrating this excess of electron charges suggests a ferroelectric polarization of the order of 75 µC/cm² for the ultrathin $BiFeO_3$, in agreement with the large displacement observed in the ferroelectric (40 pm in Figure 9).

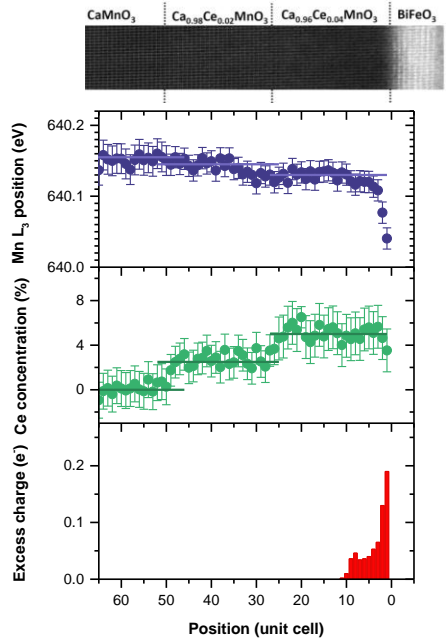

*Figure 10. HAADF image of the (Ca,Ce)MnO$_3$/BiFeO$_3$ heterostructure. The image is given as a guideline. (Top) Mn-L$_3$ peak position (blue points) extracted from EELS spectra unit cell by unit cell. (Middle) Ce concentration profile (green points) for the same heterostructure. (Bottom) Excess charge per unit cell at the interface between Ca$_{0.96}$Ce$_{0.04}$MnO$_3$ and BiFeO$_3$. The error bars correspond to 80% confidence level (54).*

**Related publications:**
*Depth Profiling Charge Accumulation from a Ferroelectric into a Doped Mott Insulator*
M. Marinova, J.E. Rault, A. Gloter, S. Nemsak, G.K. Palsson, J.-P. Rueff, C.S. Fadley, C. Carrétéro, H. Yamada, K. March, V. Garcia, S. Fusil, A. Barthélémy, O. Stéphan, C. Colliex, M. Bibes
**NANO LETTERS 15, 2533-2541 (2015)**
*Atomic structure and microstructures of supertetragonal multiferroic BiFeO$_3$ thin films*
F. Pailloux, M. Couillard, S. Fusil, F. Bruno, W. Saidi, V. Garcia, C. Carrétéro, E. Jacquet, M. Bibes, A. Barthélémy, G.A. Botton, J. Pacaud
**PHYSICAL REVIEW B 89, 104106 (2014)**
*Giant tunnel electroresistance for non-destructive readout of ferroelectric states*
V. Garcia, S. Fusil, K. Bouzehouane, S. Enouz-Vedrenne, N. D. Mathur, A. Barthélémy, M. Bibes
**NATURE 460, 81-84 (2009)**

### 3.4.2 Tunneling through ferroelectrics

In a previous section, we demonstrated the correlation between ferroelectric and resistive switching in FTJs based on BaTiO$_3$ (*43*) or PVDF organic ferroelectric (*48*) tunnel barriers. In order to get insights into the transport properties of such junctions and to rule out any other source of resistive switching such as the formation of conductive filaments (*55*), the evolution of the barrier transmission can be investigated as a function of the junction diameter, the ferroelectric barrier thickness, and temperature.

In the case of a ferroelectric tunnel barrier with homogeneous polarization state, we expect the tunnel transport to be independent of the junction size, unless discontinuities exist (either pinholes or conductive filaments). Previously-mentioned Co/BaTiO$_3$/La$_{0.67}$Sr$_{0.33}$MnO$_3$ tunnel junctions are always initially in a homogeneous upward polarization state (pointing toward the top electrode of Co). This corresponds to the junction resistance state being ON. Using C-AFM, we mapped FTJs with different diameters in order to get statistics on their transport properties (Figure 11a). The junction resistance varies with diameter and scales linearly with the inverse of the junction area (Figure 11b), indicating a constant resistance-area product in the ON state. Consequently, the formation of conductive filaments can be discarded as the cause of the TER observed in such BaTiO$_3$-based junctions.

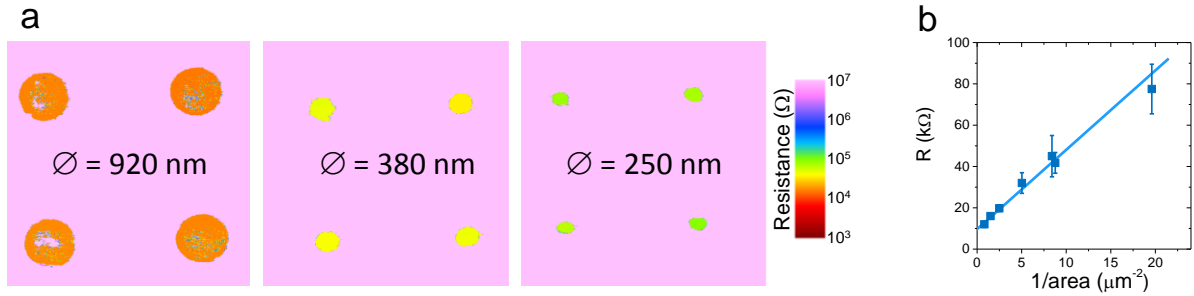

*Figure 11. (a) Resistance maps collected under a constant voltage (V$_{DC}$ = 300 mV) on matrices of Co/BaTiO$_3$/La$_{0.67}$Sr$_{0.33}$MnO$_3$ FTJs with different diameters (∅). (b) Linear dependence of the average resistance with the inverse of the junction area, showing a constant resistance-area product for the FTJs. All junctions are in the virgin ON state resistance (polarization pointing toward Co).*

The transport properties of the FTJs can be further investigated in both ON and OFF states by performing current vs. voltage measurements. The non-linearity of current with voltage in both states and its highly symmetric character are in agreement with tunneling transport (Figure 12a) and can be fitted by direct tunneling models through trapezoidal barriers (*56*). This suggests that mechanisms based on the modulation of the tunnel transmission by ferroelectricity are probably responsible for the TER effect.

In addition, low temperature AFM measurements show little variations of the transport characteristics from those obtained at room temperature (Figure 12b), with a very similar TER (inset of Figure 12a). The low temperature sensitivity is in line with direct tunneling rather than thermionic emission or other thermally-activated transport mechanisms. Moreover, complementary temperature-dependent measurements were performed on Fe/BaTiO$_3$/La$_{0.67}$Sr$_{0.33}$MnO$_3$ nanojunctions based on 1-nm-thick tunnel barriers (Supp. Info. of (*57*)). They show a non-monotonic variation, commonly observed for Co/SrTiO$_3$/La$_{0.67}$Sr$_{0.33}$MnO$_3$ magnetic tunnel junctions (*58*, *59*), with a maximum resistance around 250 K often attributed to a reduced ordering temperature for the La$_{0.67}$Sr$_{0.33}$MnO$_3$ at the interface with the tunnel barrier (*60*). Such peculiar signatures with temperature indicate that BaTiO$_3$ ultrathin films act as high quality tunnel barriers (*59*).

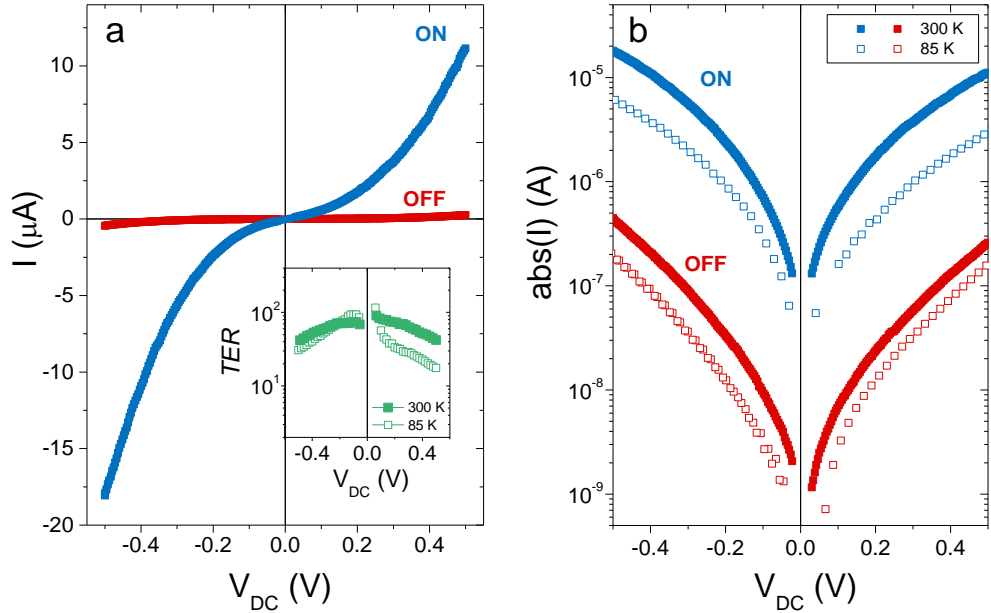

*Figure 12. (a) Current (I) vs. voltage ($V_{DC}$) of a typical Co/BaTiO$_3$/La$_{0.67}$Sr$_{0.33}$MnO$_3$ FTJ in the ON and OFF states at 300 K. (b) Same |I|($V_{DC}$) (solid symbols) plotted in semi log scale together with transport data collected at 85 K (open symbols) on a similar FTJ. The ON and OFF states were achieved with voltage pulses of 100 µs. The corresponding voltage dependence of the TER (calculated from I($V_{DC}$)) is displayed in the inset of (a) for both temperatures (43).*

**Related publication:**
*Solid-state memories based on ferroelectric tunnel junctions*
A. Chanthbouala, A. Crassous, V. Garcia, K. Bouzehouane, S. Fusil, X. Moya, J. Allibe, B. Dlubak, J. Grollier, S. Xavier, C. Deranlot, A. Moshar, R. Proksch, N.D. Mathur, M. Bibes, A. Barthélémy
NATURE NANOTECHNOLOGY 7, 101-104 (2012)

### 3.4.3 Electrode influence with BiFeO$_3$ tunnel barriers

During the PhD of Sören Boyn, we investigated the influence of the top electrode material on the properties of FTJs derived from BiFeO$_3$ (4.6 nm)/Ca$_{0.96}$Ce$_{0.04}$MnO$_3$/YAlO$_3$ heterostructures. Using W, Co, Ni, and Ir as top electrodes, we measured a systematic modification of the electron transport properties (Figure 13a). Adjusting the current-voltage characteristics with a direct tunneling model through a trapezoidal barrier enables to correlate the conductance changes to a variation of the interfacial barrier height ($\Phi_2$) between the ferroelectric and the top electrode. As the work function of the top electrode material increases, the tunnel barrier becomes more and more trapezoidal (Figure 13b). Consequently, the OFF state resistance, associated to polarization pointing toward the bottom electrode of Ca$_{0.96}$Ce$_{0.04}$MnO$_3$, is larger and larger, and the TER increases. However, for materials with large work function such as Ni or Ir, the large TER is associated to a poor reliability and a fast degradation of the junction. On the other hand, the TER is reliable and reversible for top electrode materials such as W or Co. This can be understood by the creation of a built-in field, caused by the mismatch between the two interfacial barrier heights, which favors one direction of polarization over the other (*61*) and requires large electric fields (associated with defect formation) to commute the resistance state. Therefore, there is a trade-off between large TER effects and reversible toggling between the ON and OFF states. So far, the best top electrode material is Co which gives TER greater than $10^4$ and a good reproducibility of ON/OFF commutations (*62*).

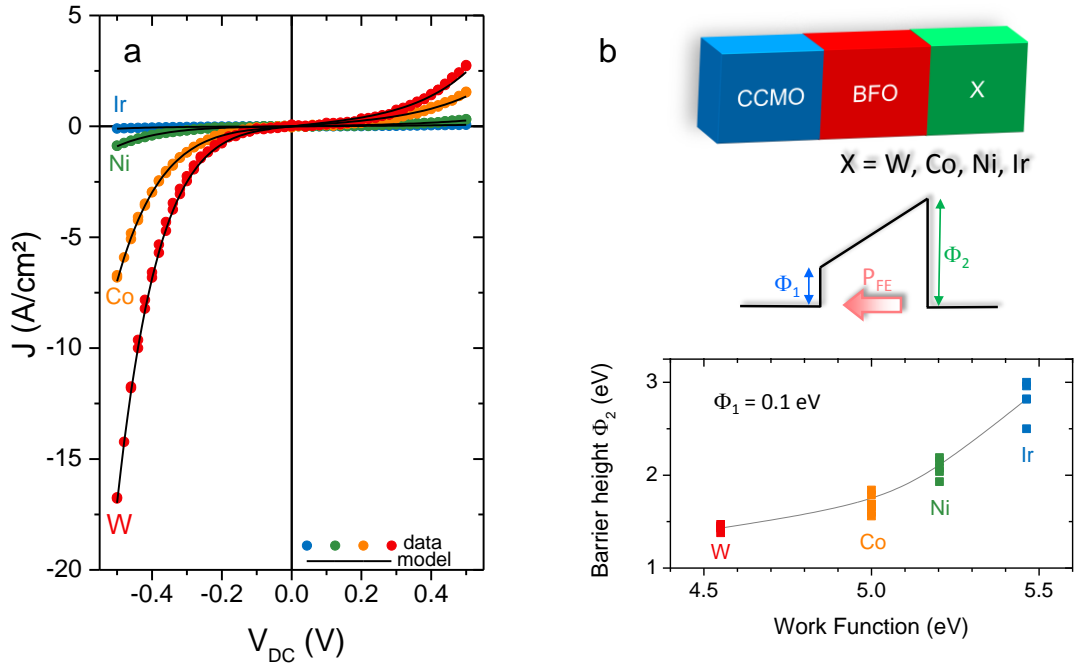

*Figure 13. (a) Current-voltage characteristics of the virgin (high-resistance with downward polarization) state for different top electrodes. Points correspond to measured values; lines represent the fitted model combining direct tunneling and Fowler-Nordheim tunneling. (b) Sketch of the device with the corresponding potential profile. Barrier heights at the top electrode interface as a function of the respective top-electrode work function. The line connects the mean values at each top electrode* (*63*).

After emphasizing the role of the top electrode material on the TER properties of FTJs, one may question the influence of the bottom electrode. As ultrathin films of super-tetragonal BiFeO$_3$ are epitaxially grown on perovskite substrates with small lattice parameters like YAlO$_3$ or LaAlO$_3$, the choice of buffer oxide electrode materials is limited. During the post-doc of Flavio Bruno, we investigated BiFeO$_3$-based FTJs combined with LaNiO$_3$ electrodes on LaAlO$_3$(001) substrates (*11*). The ultrathin films of BiFeO$_3$ are well-ordered with unit-cell steps and flat terraces (Figure 14a). The exponential increase of the resistance with the film thickness (Figure 14d) is similar to the observations with BaTiO$_3$ (Figure 5a); this indicates that the ultrathin film of BiFeO$_3$ acts as a tunnel barrier. Just as with Ca$_{0.96}$Ce$_{0.04}$MnO$_3$ bottom electrodes, the PFM signal indicates a preferential polarization state pointing toward the bottom electrode of LaNiO$_3$; it is switchable at low voltages with good retention properties (Figure 14b-c). Co/BiFeO$_3$/LaNiO$_3$ tunnel junctions were fabricated from these heterostructures. The homogeneous downward polarization is conserved in the devices and does not depend on the junction diameter. The associated resistance-area product is independent of the junction diameter (Figure 14e) suggesting that tunneling transport dominates. Furthermore, the low resistivity of the LaNiO$_3$ electrode leads to lower ON state resistances than for Ca$_{0.96}$Ce$_{0.04}$MnO$_3$-based FTJs. However, combined PFM and transport studies reveal a much larger sensitivity of the devices to voltage pulses, which leads to an irreversibility of the switching back to the OFF state (*11*). The low resistivity of LaNiO$_3$ generates large currents under voltage pulses and a fine control of the voltage pulse amplitude required to switch the polarization leads to a reduced irreversibility.

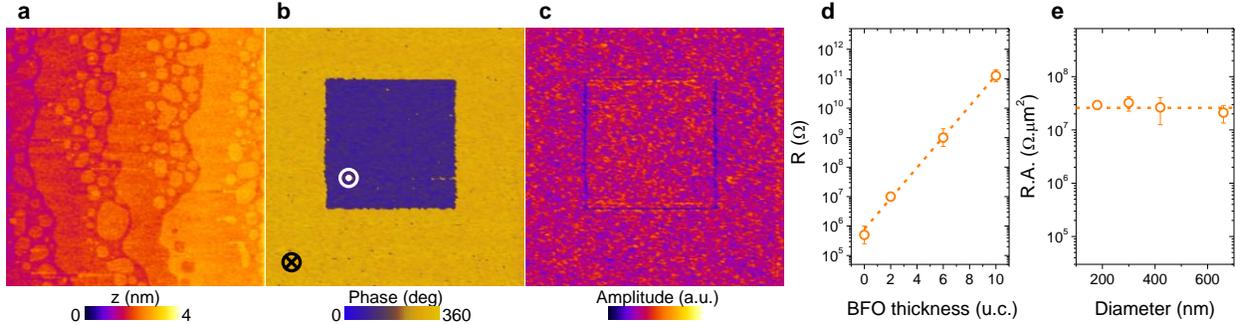

*Figure 14. (a) Topography, (b) out-of-plane PFM phase and (c) amplitude on a BiFeO$_3$ (10 u.c.)/LaNiO$_3$ (10 nm)//LaAlO$_3$ sample. (d) Average resistance deduced from C-AFM images performed on samples with various thicknesses of BiFeO$_3$. (e) Distribution of the resistance-area products of FTJs in the virgin state (polarization toward LaNiO$_3$) as a function of the device diameter ($V_{DC}$ = −200 mV) (11).*

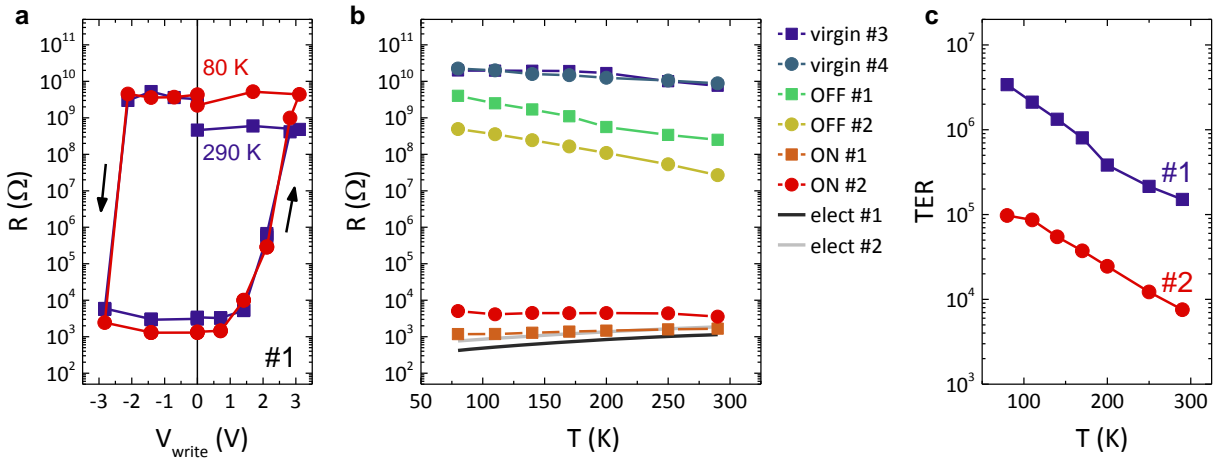

*Figure 15. Temperature investigations of Co/BiFeO$_3$/LaNiO$_3$ FTJs. (a) Hysteresis cycles of the resistance of junction #1 with write voltage pulses measured at 290 K and 80 K. (b) Temperature dependence of the virgin state of two junctions (#3 and #4), of the ON, OFF and bottom electrode states of junctions #1 and #2. (c) Temperature dependence of the TER for junctions #1 and #2, calculated from (b) (11).*

Using fully patterned FTJs, we investigated the temperature dependence of the virgin, ON and OFF resistance states (Figure 15). Although the weak temperature dependence of the virgin and ON states from 80 to 290 K is compatible with direct tunneling transport, the slightly larger one for the OFF state (Figure 15b) suggests the possibility of inelastic tunneling through localized states (*64*). Nevertheless, the low resistance of LaNiO$_3$ enables to obtain record TER of 10$^6$ at 80 K (Figure 15c). These results open the path to the ferroelectric control of metal-insulator transitions in FTJs combined with strongly-correlated nickelates (*65*).

### 3.4.4 Quantitative analysis of the tunnel electroresistance

Although experimental evidence suggests that TER is undoubtedly associated with ferroelectric polarization switching (*36–38*, *43*), there are still some fundamental issues to be addressed (*66*). Indeed, large TER in oxide-based tunnel junctions cannot be quantitatively interpreted by simple electrostatic models involving partial screening of polarization charges at the interfaces (*7*, *8*). Instead, it necessitates complex descriptions containing interfacial dielectric layers (*67*) or doped-semiconducting layers (*68*). For example, our experiments presented before with $BaTiO_3$ or $BiFeO_3$ tunnel barriers always show an OFF (ON) state for polarization pointing toward the oxide electrode (Co electrode) (*9*, *11*, *43*, *63*). This is counterintuitive as it indicates that the effective charge screening is better at the oxide interface than at the Co interface. It could suggest the presence of an interfacial dielectric layer at the Co/ferroelectric interface (*15*, *69*). Moreover, all the experimentally-reported FTJs are using ferroelectric barriers made of oxide perovskite thin films (*66*). To maintain ferroelectricity in nanometer-thick oxide films, sophisticated experimental approaches such as strain engineering (*49*) and careful control of epitaxial growth are generally required (*70*), which inevitably results in a complex fabrication process. En route toward more processable FTJs, organic ferroelectric materials (*71*) have also been considered as tunnel barriers (*72*, *73*). Organic FTJs may also exhibit different electronic transport properties from their inorganic counterparts because of the weak Van der Waals interfacial bonding with metal electrodes.

During the PhD of Bobo Tian in collaboration with Brahim Dkhil (Ecole Centrale-Supélec), we investigated the transport properties of FTJs based on ultrathin films of PVDF deposited layer-by-layer (one layer (L) is 2.2 nm thick) by the Langmuir-Blodgett technique. We fabricated nanoscale junctions by patterning the bottom electrode of W into 190-nm-wide pillars isolated by a $SiO_2$ matrix prior to the deposition of the PVDF barrier and then deposited and patterned Au top electrodes with evaporation and photolithography. From current-voltage (I-V) transport experiments with a dc voltage, we observed hysteresis with associated TER of 3 and 10 for films with thicknesses of 1 L and 2 Ls, respectively (Figure 7g). The weak temperature dependence of the I-V between 223 K and 290 K suggests that direct tunneling is the main transport mechanism.

In order to gain more insight into the interplay between polarization and tunneling, we performed I-V measurements in the low voltage range for both polarization orientations. The results are displayed for four typical junctions in Figure 16. The I-V are fitted using a direct tunneling model through a trapezoidal tunnel barrier (*38*, *56*). We fixed the barrier width as the thicknesses of PVDF films and the interfacial barrier heights at the PVDF/W and Au/PVDF interfaces are considered as free parameters. Transport measurements can be accurately reproduced by the model for both polarization orientations (solid lines in Figure 16b-e for downward polarization and in Figure 16f-i for upward polarization) considering a constant effective mass for tunneling electrons of m* = 0.12 $m_0$. The electron potential profiles resulting from the fits are displayed in Figure 16j-m (orange and brown for downward and upward polarization, respectively). Interestingly, the interfacial barrier heights vary similarly with polarization reversal in all four junctions: when the polarization switches from downward to upward, $\Phi_1$ increases while $\Phi_2$ decreases, and vice versa. The electron potentials decrease along the direction of the ferroelectric polarizations as expected within the framework of electrostatic models based on partial electrode screening of polarization charges (*7*, *8*). Moreover the polarization-induced variation of $\Phi_1$ is smaller than $\Phi_2$ in agreement with the lower screening length for W (~0.048 nm) (*74*) than for Au (~0.07 nm) (*75*). Such variations of potential profiles with polarization give an estimate of PVDF ferroelectric polarization of 8-18 µC/cm² in agreement with polarization measurements on thicker films (*76*). In conclusion, we demonstrated that the modulation of tunnel transmission through electrostatic variations of the potential barrier shape is responsible for TER in these PVDF-based FTJs.

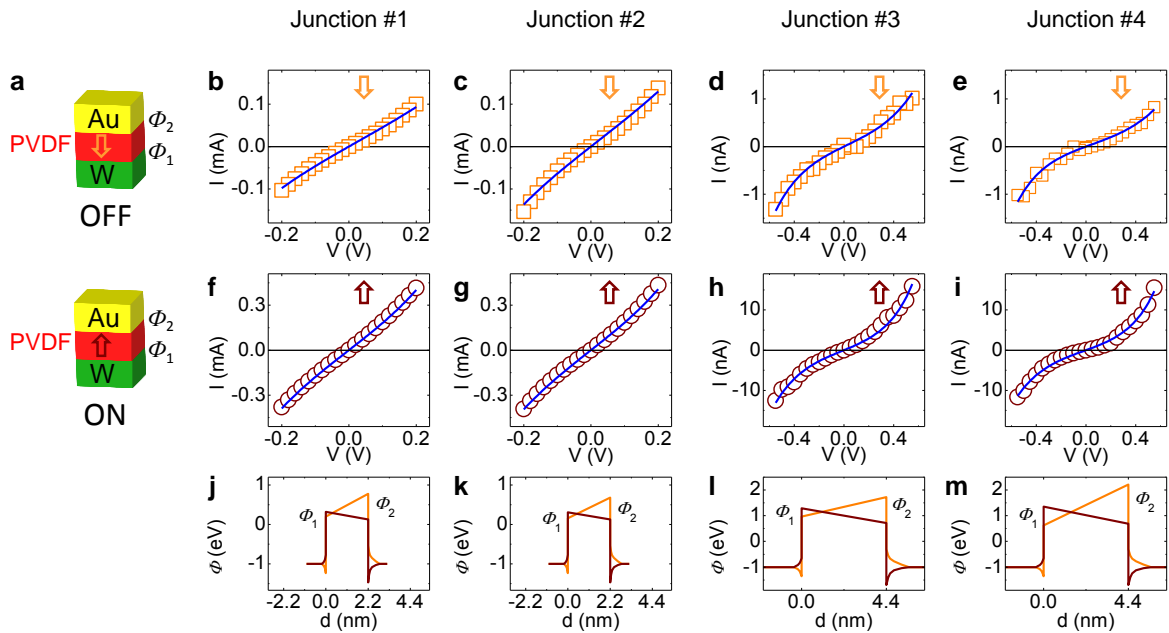

*Figure 16. Direct tunneling in Au/PVDF/W junctions. (a) Sketch of the two polarization configurations of the junctions with the corresponding ON and OFF states. (b-e) I-V curves in the downward polarization state for junctions #1, #2, #3 and #4. (f-i) I-V curves in the upward polarization state for the same junctions. Solid lines in b-i are fits from the direct tunneling model. (j-m) Calculated electron potential profiles in both polarization configurations (orange and brown lines for downward or upward polarization, respectively) for junctions #1, #2, #3 and #4. Junctions #1 and #2 are based on 1 L-thick PVDF and junctions #3 and #4 are based on 2 Ls-thick PVDF (48).*

## 3.5 Digital memories with ferroelectric tunnel junctions

Over recent years, solid-state FTJs with various ferroelectrics such as $BaTiO_3$ (*10*, *43*, *68*, *69*, *77–81*), $PbZr_{0.2}Ti_{0.8}O_3$ (*82*) and $BiFeO_3$ (*9*, *11*, *62*, *63*, *83*) have been demonstrated by several groups. Performance metrics such as OFF/ON ratios, write energies, data retention, scalability, endurance are critical to consider potential applications of FTJs as non-volatile memories.

### 3.5.1 Switching speed and reproducibility

Previously mentioned $Co/BaTiO_3/La_{0.67}Sr_{0.33}MnO_3$ tunnel junctions show large OFF/ON ratios (Figure 7c). A single voltage pulse is necessary to commute the resistance state of the FTJ between ON and OFF states and it requires low write current densities ($10^4$ A/cm$^2$) since the driving force for resistive switching is the electric field (*43*). Switching experiments on several tens of junctions show a low dispersion of the ON and OFF states (Figure 17a) suggesting a high yield and good uniformity of the sample. The ON and OFF states are stable over read/write cycles and do not exhibit resistance variations related to fatigue. Fatigue tests were performed over 900 read/write cycles (Figure 17b), after which electrical contact was lost due to the AFM tip drift. Complementary experiments were performed to show resistive switching in FTJs with OFF/ON ratios over 100 after applying write voltage pulses of 10 ns (Figure 17c), suggesting high speed with low write energy (100 fJ/bit estimated for these devices).

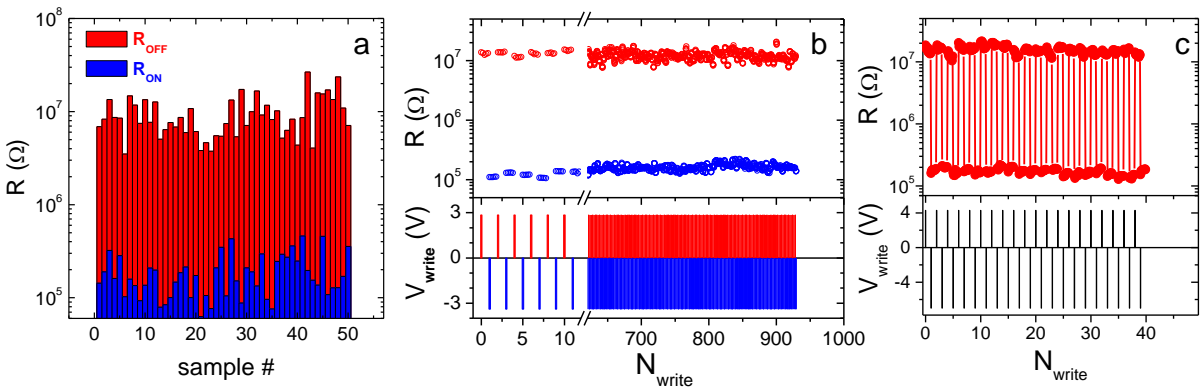

*Figure 17. (a) ON and OFF resistance states measured at 100 mV after applying 100 µs voltage pulses to 50 different $Co/BaTiO_3/La_{0.67}Sr_{0.33}MnO_3$ junctions. Average OFF/ON is 64. (b) Reversible resistance switching between ON and OFF resistance states of a typical junction for more than 900 cycles (voltage pulses of 100 µs). (c) OFF/ON resistance switching of a typical device after applying 10 ns voltage pulses (43).*

High-density integration of FTJs, as well as data endurance and retention are critical for memory applications. Up to now, submicron FTJs were patterned as matrices of top electrodes on a continuous ferroelectric tunnel barrier/bottom electrode bilayer, and were electrically connected using the conductive tip of an AFM (*43*, *78*, *82*). During the post-doc of Stéphanie Girod, we developed a chip of fully integrated $BiFeO_3$-based FTJs using a complex five-step lithography process (Figure 18), where each FTJ is connected with standard rf probes (*62*).

Among the 50 FTJs on the sample, 10% are either short-circuited or unswitchable. We obtain a high yield of working devices where 86% show OFF/ON ratios greater than 100 and 64% greater than 1000. The virgin state is a high resistance state as previously observed for FTJs contacted by conductive AFM tips. The ON state resistance is reached by successive 100 ns voltage pulses down to −2 V (Figure 19a) and it averages at $1 \times 10^5$ Ω with a standard deviation of $3 \times 10^4$ Ω (Figure 19b). Upon successive application of positive voltage pulses, the resistance increases to finally recover the virgin OFF state resistance with an average of $2 \times 10^8$ Ω and a standard deviation of $1.2 \times 10^8$ Ω. These statistics emphasize the reliability of ferroelectric switching and the good control of the fabrication process.

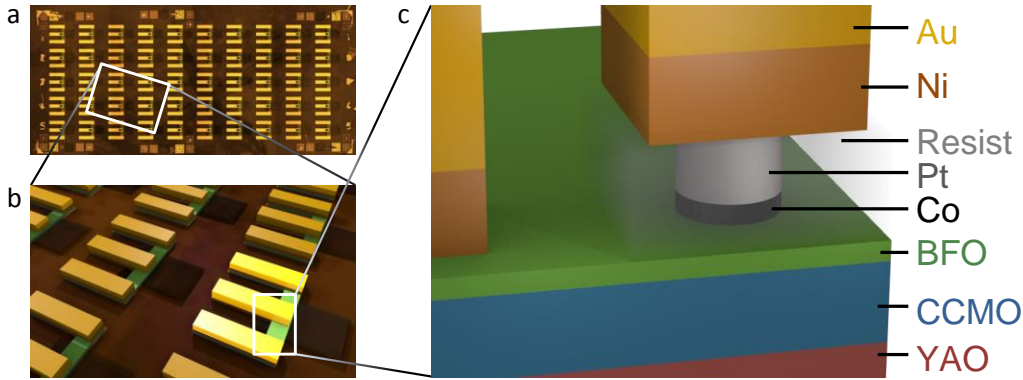

Figure 18. (a) Optical microscope image of the chip after patterning, showing 5 × 10 FTJs. (b) 3D representation of a zoomed area containing a few FTJs. The three parallel bars are the ground-signal-ground contact pads. (c) 3D sketch of one FTJ (62).

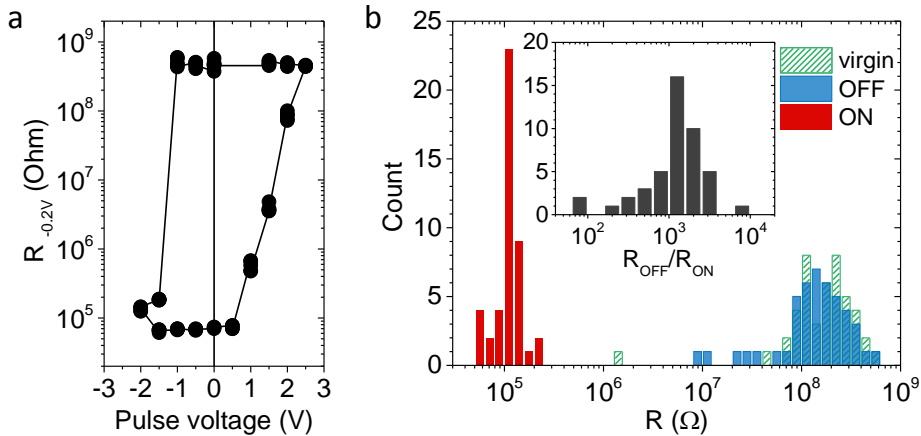

Figure 19. (a) Typical resistance hysteresis of a fully patterned Co/BiFeO$_3$/Ca$_{0.96}$Ce$_{0.04}$MnO$_3$ FTJ as a function of pulse voltage. The junction is initially in the high resistance (virgin) state and switches to the low resistance (ON) state upon application of negative voltage pulses with increasing amplitude and progressively back to the OFF state under positive voltage pulses. (b) Distribution of the virgin, ON and OFF states for the 45 working FTJs. The inset shows the distribution of the OFF/ON ratio (62).

### 3.5.2 Endurance, retention

Endurance in FTJs has only been demonstrated up to a few $10^3$ cycles (*9, 10, 43*), in contrast to the cycling performance of capacitive ferroelectric memories (>$10^{14}$ cycles) (*84, 85*). In the fully patterned FTJs described previously, the endurance was tested by applying voltage pulses of 100 ns. Each cycle consists of the application of 1.5 V to set the high resistance state and −1.3 V to set the low resistance state. Ideally the resistance has to be read after each pulse, but as the picoammeter we have used can take up to 2 s to perform a measurement, this procedure implies unreasonably long measurement time (>40 days for $10^6$ cycles). We therefore measured the high and low resistance states in increasing intervals of 1, 2, 4, 8, etc., cycle trains. We noticed that the FTJ is sometimes blocked in the high resistance state where the application of the negative voltage pulse does not lead to resistance switching. In that case, we discarded the preceding train of cycles and automatically applied negative pulses of increasing amplitude (down to −3 V) until the resistance dropped below a defined limit of $10^6$ Ω for the ON state. After this, we proceeded by applying the above mentioned procedure of cycle trains restarting at 1 cycle. A typical example of the endurance measurements is presented in Figure 20a. We reached up to 4 × $10^6$ cycles with a resistance contrast of about two orders of magnitude. This narrowed resistance contrast is due to the reduced pulse voltages which only partly switch the polarization of $BiFeO_3$. During this experiment, the resistance was blocked and unblocked 1039 times, i.e., in average every 4000 cycles.

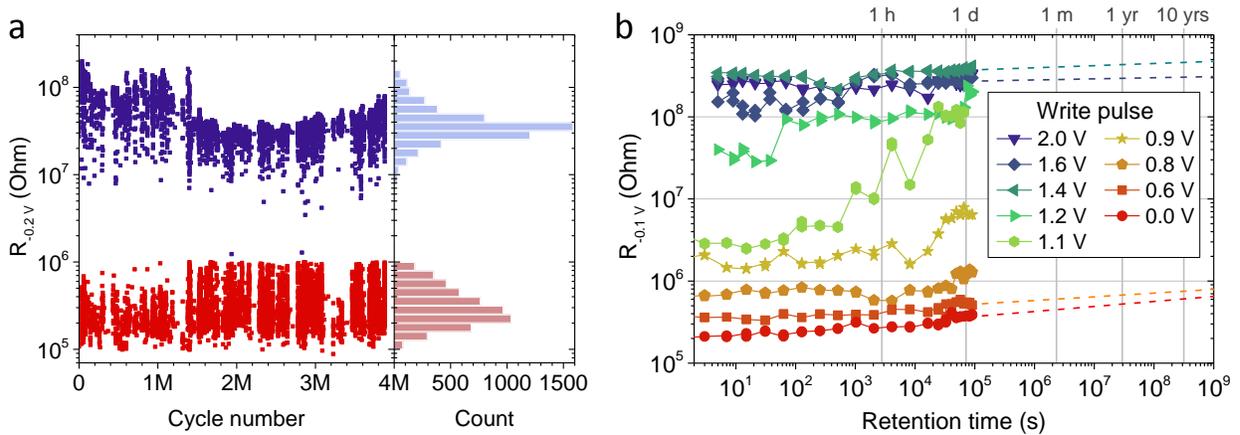

*Figure 20. (a) Resistance as a function of cycle number and distribution of the corresponding ON and OFF resistances. (b) Resistance as a function of time in different initial resistance states reached by application of write pulses of different amplitudes. The dashed lines show the extrapolation to $10^9$ s (62).*

After cycling, the FTJs can still be switched to the initial OFF and ON states. This indicates that the switchable polarization of $BiFeO_3$ is not affected (in contrast to Refs. (*86*) and (*87*)). The switching becomes, however, less deterministic, indicating a strong pinning of domain walls, probably due to creation or rearrangement of defects, such as oxygen vacancies. Additionally, the low resistance state requires higher voltage pulse amplitudes than before cycling as previously observed (*87*). This could be related to an increased internal field by the movement of oxygen vacancies (*85, 86, 88, 89*) which destabilizes the ON state. A possible means of slowing down the migration of oxygen vacancies and therefore enhancing this result is doping BiFeO$_3$ by La or Nb to reduce their mobility (*85*).

When used as a digital memory, good retention properties of FTJs are crucial. Prior to the measurements on fully patterned FTJs, we performed retention experiments on an array of four 180-nm-wide FTJs contacted by the AFM tip. All the devices were switched to different (ON, intermediates and OFF) resistance states and imaged with PFM (Figure 21a). Three days later (68 h), the same devices were imaged by PFM (Figure 21b) and measured electrically (Figure 21c). Here in the ON state (pad #1), small downward domains have nucleated in the bottom part of the junction, accompanied by a reduction of the PFM amplitude. The intermediate and OFF states (pad #2, #3, and #4) are very similar in both PFM phase and amplitude after three days. In addition, the resistance shows minor variations over that period of time for the four FTJs (Figure 21c). Overall, the BiFeO$_3$-based FTJs show good retention properties even in the intermediate states.

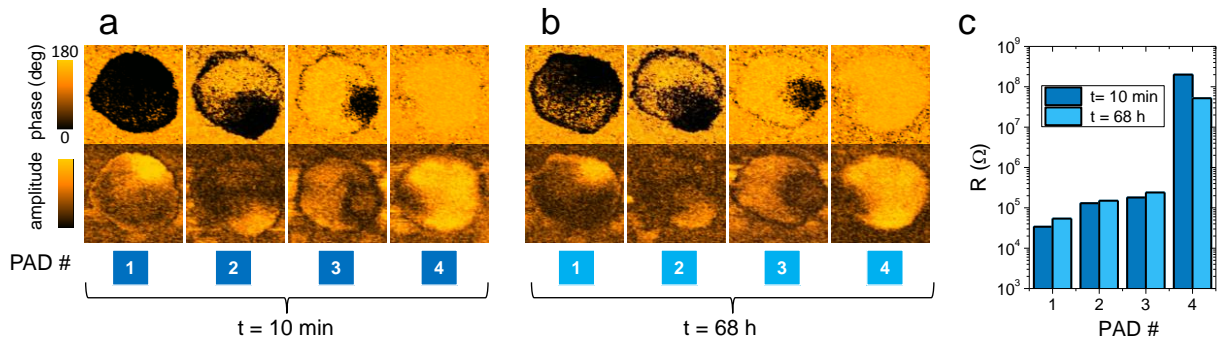

*Figure 21. Retention of Co/BiFeO$_3$/Ca$_{0.96}$Ce$_{0.04}$MnO$_3$ tunnel junctions. (a) Four 180-nm-wide FTJs are imaged with PFM (phase and amplitude, top and bottom, respectively) in four different resistance states. (b) The same devices are imaged after 68 hours. (c) Resistance of the FTJs initially and after 68 hours (9).*

In order to further evaluate those retention properties, we used the following measurement scheme for a representative junction. The FTJ is initialized in the ON resistance state and a 100 ns voltage pulse of variable amplitude is applied. After each write pulse, the resistance is recorded in increasing time intervals up to 25 h of total measurement time (Figure 20b). The ON and OFF states, i.e., when the polarization of BiFeO$_3$ is nearly saturated, show very good retention. By extrapolation, we estimate that the OFF/ON resistance ratio after ten years would still be more than two orders of magnitude (~375) on these 300 to 500 nm wide FTJs. These good retention properties have so far only been observed on 30 µm wide BaTiO$_3$-based FTJs (*10*) and our devices represent a 100 fold improvement for scalability.

The intermediate resistance states are less stable and tend to relax toward higher resistances (Figure 20b). The domain configuration in ferroelectric thin films is governed by two energies: the contribution from the depolarizing field and the domain wall energy. On the one hand, as the thickness of the ferroelectric film decreases, the depolarizing field increases, favoring a multi-domain polarization. On the other hand, if energy is needed for the creation of domain walls, the system will tend to lower their length, which is minimized in a mono-domain state. As the nearly saturated ON and OFF states are more stable than intermediate, multi-domain states, it seems that the domain wall energy is significant in our system. In addition, the voltage threshold to destabilize the ON state is lower than the voltage threshold of the OFF state (Figure 19a). This indicates a more energetically favorable downward polarization state which could result from the interfacial asymmetry of the FTJs (*90*). The relaxation of intermediate states toward the OFF state is in agreement with this scenario. The apparent discrepancy between the fully patterned junctions and the junctions connected by the C-AFM tip could originate from the smaller diameter of the latter.

> **Related publications:**
> *High-performance ferroelectric memory based on fully patterned tunnel junctions*
> S. Boyn, S. Girod, V. Garcia, S. Fusil, S. Xavier, C. Deranlot, H. Yamada, C. Carrétéro, E. Jacquet, M. Bibes, A. Barthélémy, J. Grollier
> APPLIED PHYSICS LETTERS 104, 052909 (2014)
> *Giant Electroresistance of Super-Tetragonal BiFeO$_3$-Based Ferroelectric Tunnel Junctions*
> H. Yamada, V. Garcia, S. Fusil, S. Boyn, M. Marinova, A. Gloter, S. Xavier, J. Grollier, E. Jacquet, C. Carrétéro, C. Deranlot, M. Bibes, and A. Barthélémy
> ACS NANO 7, 5385-5390 (2013)
> *Solid-state memories based on ferroelectric tunnel junctions*
> A. Chanthbouala, A. Crassous, V. Garcia, K. Bouzehouane, S. Fusil, X. Moya, J. Allibe, B. Dlubak, J. Grollier, S. Xavier, C. Deranlot, A. Moshar, R. Proksch, N.D. Mathur, M. Bibes, A. Barthélémy
> NATURE NANOTECHNOLOGY 7, 101-104 (2012)

### 3.5.3 Scalability, size effects

FTJs are promising to be used as memristors in neuromorphic architectures and as non-volatile memory elements. For both applications device scalability is essential, which requires a clear understanding of the relationship between polarization reversal and resistance change as junction size shrinks. In this context, we initially demonstrated switching down to diameters of 70 nm in ultrathin films of BaTiO$_3$ (*36*). Pantel et al. reported large TER (OFF/ON = 300) with Co/PbZr$_{0.2}$Ti$_{0.8}$O$_3$/ La$_{0.67}$Sr$_{0.33}$MnO$_3$ triangular-shaped tunnel junctions with an effective diameter of 220 nm (*82*). OFF/ON ratios of 10000 were shown in Co/BiFeO$_3$/Ca$_{0.96}$Ce$_{0.04}$MnO$_3$ tunnel junctions with diameters of 180 nm (*9*). Gao *et al.* showed indications of TER with I vs. V$_{DC}$ (OFF/ON > 100 at large voltage) on 20 nm wide Ag/BaTiO$_3$/SrRuO$_3$ junctions (*78*) but further investigations are needed to ensure switchable polarization for junctions with diameters smaller than 100 nm.

During the PhD of Sören Boyn, we investigated the properties of Co/BiFeO$_3$/Ca$_{0.96}$Ce$_{0.04}$MnO$_3$ FTJs with diameters ranging from 180 nm to 1200 nm. Figure 22 shows resistance hysteresis cycles as a function of the amplitude of write voltage pulses for five different junction diameters. All junctions show large TER with two well-defined ON and OFF resistance states and are initially in the OFF state. Strikingly, the TER decreases as the junction diameter increases. It reaches about $5 \times 10^4$ for 180-nm-wide junctions and only $6 \times 10^2$ for the 1200-nm ones.

In general, one would not expect the TER to depend on the lateral size of the junctions. Indeed, the resistance-area product of a junction in a homogeneous state of polarization should be constant if tunnel transport is involved. Figure 23 displays the ON and OFF resistance-area products as a function of the device diameter and the corresponding ferroelectric domain populations. The resistance-area product of the OFF resistance state does not vary with the junction diameter. This indicates a good structural and electrical homogeneity of the BiFeO$_3$ tunnel barrier under the assumption of a homogeneous polarization state up to the micron scale. This is corroborated by the bright PFM phase and constant PFM amplitude for all OFF resistance states (Figure 23b) which indicates a homogeneous polarization pointing downward (toward the Ca$_{0.96}$Ce$_{0.04}$MnO$_3$ electrode) for all device sizes.

In the ON resistance state however, the resistance-area product increases as the junction diameter increases (Figure 23a). This suggests that the switching behavior may change with the junction size. Indeed, the PFM phase images of junctions in the ON resistance state are drastically different (Figure 23b): for 180-nm junctions, the ON-state PFM phase is almost completely dark, suggesting that a majority of ferroelectric domains have switched from downward to upward (toward the Co electrode); for 290-nm junctions, the mixed PFM phase indicates that about half of the area is switched, while for 1200-nm junctions, a majority of domains remain in the downward state. In Figure 23c, the ON-state fraction of reversed domains, estimated from PFM phase images, is plotted as a function of the device diameter. We conclude that the decrease of TER with the junction diameter can be explained by a reduced fraction of switched ferroelectric domains in the ON resistance state.

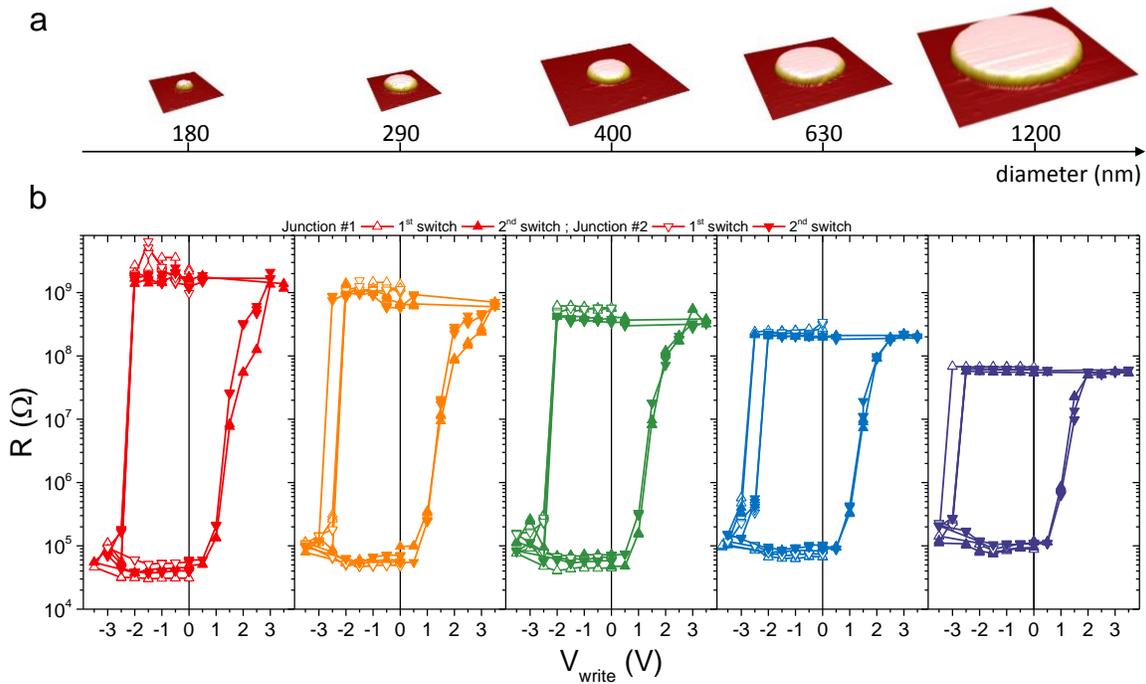

*Figure 22. (a) 3D topographic AFM images of BiFeO$_3$ tunnel junction top electrodes with increasing diameters. (b) Resistance hysteresis with write voltage pulses for each junction diameter. The first two hysteresis cycles of two typical junctions are displayed for each diameter (91).*

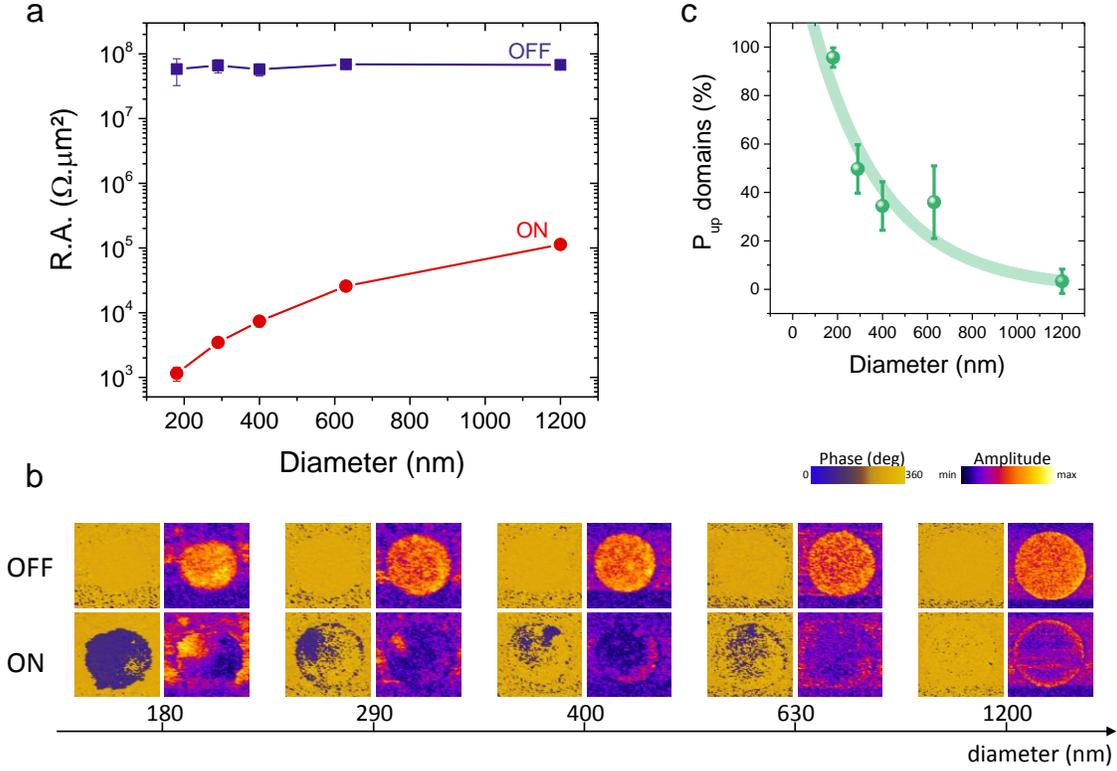

*Figure 23. (a) Resistance-area product in the OFF and ON states as a function of the junction diameter. (b) PFM images (out-of-plane phase (left) and amplitude (right)) in the OFF and ON states of junctions with different diameters. (c) Fraction of domains with polarization pointing up in the ON state as a function of the junction diameter estimated from PFM phases images of (b). The line is a guide to the eye (91).*

In order to get more insights into the size-dependent polarization reversal of these BiFeO$_3$ FTJs, we performed finite-size-element electric-field distribution simulations in collaboration with the group of Marty Gregg (Queen's University Belfast). In these 2D simulations, a voltage of −3.5 V is applied to the top of the Co electrode while each end of 10-µm-long bottom electrode of Ca$_{0.96}$Ce$_{0.04}$MnO$_3$ is connected to the ground (Figure 24a). We considered the actual values of the resistance-area products of BiFeO$_3$ junctions for the up (9 × 10$^2$ Ω.µm²) and down (2 × 10$^4$ Ω.µm²) domains in this voltage range using real-time transmission experiments and dc measurements as well as the resistivity of the Ca$_{0.96}$Ce$_{0.04}$MnO$_3$ (5 mΩ.cm) measured in patterned junctions (*62*). Owing to the non-linearity of tunnel transport, the resistance-area products are significantly lower during voltage pulses (−3.5 V) than in the low dc voltage range we use for the hysteresis (Figure 22b).

In the initial state of 180-nm-wide junctions (OFF resistance, downward polarization, Figure 24a), the voltage mainly drops across the BiFeO$_3$ tunnel barrier (Figure 24b), giving rise to a homogeneous electric field in the center of the junction (Figure 24c). This electric field strongly increases at the edge of the top electrode (Figure 24c) which should favor the nucleation of ferroelectric domains with up polarization, as observed experimentally with PFM (Figure 23b). The introduction of domains with up polarization at the edges of the junction (Figure 24d) during the OFF-to-ON switching induces a change of the voltage profile (Figure 24e) and a strong reduction of the electric field within the BiFeO$_3$ tunnel barrier (Figure 24f). This can be understood by considering the fact that as up domains switch, the equivalent resistance of the tunnel barrier comes close to that of the $Ca_{0.96}Ce_{0.04}MnO_3$ electrode. Consequently, a significant part of the voltage drops within the electrode (Figure 24g).

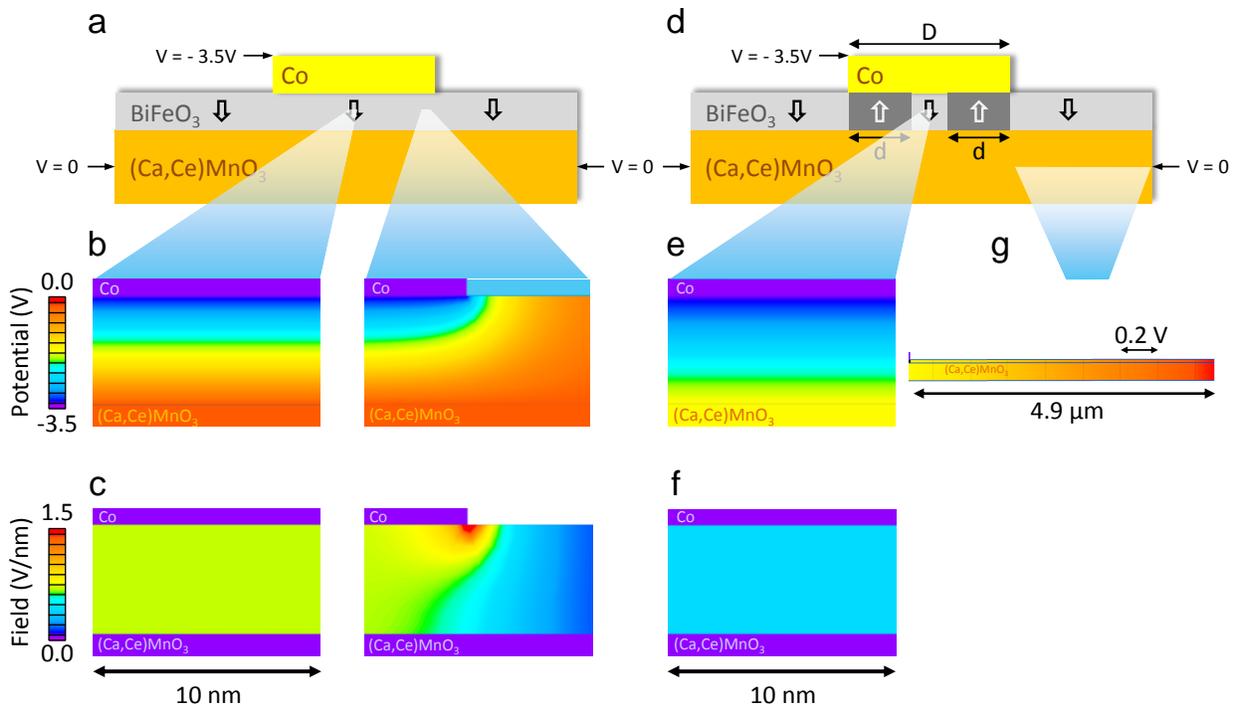

Figure 24. (a) Schematic of the structure (not to scale) used for the model in the case of a uniform polarization pointing down in $BiFeO_3$, considering a 180-nm-wide top electrode and 10-µm-long bottom electrode of $Ca_{0.96}Ce_{0.04}MnO_3$. The following parts (b-c) of the figure are zoomed into 10 nm × 6.5 nm regions. (b) The potential drop and (c) uniform electric field found within the center of the junction can be compared with the edge of the top electrode region, whereby the potential distribution of the fringes creates an enhanced electric field, facilitating the nucleation of domains from this point. (d) Schematic of the same structure considering that up domains (with width d = 80 nm) switched from the edges of the junction. (e) The potential and (f) electric-field distributions within the center of the junction show significant changes compared to the homogeneous state (b-c). Indeed, a significant part of the potential drops along the length of the bottom electrode of $Ca_{0.96}Ce_{0.04}MnO_3$ in the switched state as illustrated in (g) (91).

The electric-field simulations are repeated for various fractions of domains with up polarization and for 180-nm- and 1200-nm-wide top electrodes. The resulting evolution of the electric field in the center of the barrier as a function of domain population is plotted in Figure 25. In the case of a homogeneous polarization pointing downward, the electric field in $BiFeO_3$ is smaller for 1200-nm-wide junctions than for 180-nm ones. As the fraction of up domains increases, this difference becomes larger. Considering a critical field value below which no down domains can be switched (horizontal line in Figure 25), these simple simulations can qualitatively explain the experiments with $BiFeO_3$ FTJs: for a critical field of 0.48 V/nm, about 80% of up domains can be formed in 180-nm-wide junctions while less than 10% reverse in 1200-nm-wide junctions. These estimates fall within the range of our PFM observations (Figure 23b-c).

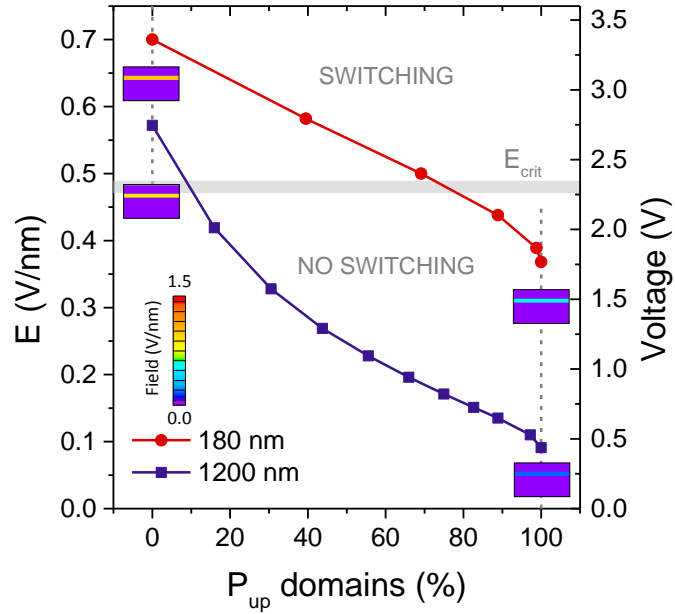

*Figure 25. Evolution of the effective electric field as up domains are introduced in the barrier, for 180-nm- and 1200-nm-wide junctions. The geometry used for the simulations is the same as in Figure 24d from which the fraction of upward domains is determined as $4d(D-d)/D^2$. The inset shows examples of the electric-field distributions in 10 nm × 6.5 nm regions of the center of junctions for homogeneous up of down polarization states. The horizontal line is an estimate of the critical field below which up domains cannot nucleate. For the 100-ns pulses of −3.5 V studied here, this corresponds to a fraction of up domains of 9% and 76% for 1200-nm and 180-nm junctions, respectively (91).*

The electric-field simulations thus demonstrate the influence of the electrode resistance on the efficiency of polarization switching in FTJs; this series resistance may impede the full reversal of polarization, limiting the TER. Owing to the stronger electric field at the edges, resorting to junction geometries with high perimeter-to-area ratios can improve the switching. We demonstrated that junctions with donut shapes show more efficient switching for similar areas. In contrast to thick insulating ferroelectric capacitors, the actual electric field across the ferroelectric tunnel barrier evolves dynamically with the domain population, resulting in a complex dynamic behavior. Our results indicate that FTJs are scalable and suggest that specific junction shapes facilitate complete polarization reversal.

**Related publication:**
*Tunnel electroresistance in BiFeO$_3$ junctions: size does matter*
S. Boyn, A.M. Douglas, C. Blouzon, P. Turner, A. Barthélémy, M. Bibes, S. Fusil, J.M. Gregg, V. Garcia
**APPLIED PHYSICS LETTERS 109, 232902 (2016)**

## 3.6 A ferroelectric memristor for neuromorphic computing

### 3.6.1 Memristor using ferroelectric domains

FTJs have potential applications as binary non-volatile memories, taking advantage of the non-destructive readout of polarization and simpler device architecture than conventional FeRAMs. Another degree of freedom is the ferroelectric domain structure of the tunnel barrier that can be exploited for analog devices such as memristors (*92–94*). Indeed, considering that in ferroelectrics, polarization reversal usually occurs by the nucleation and propagation of domains (*95*) and that the domain size scales with the square root of the film thickness (*96, 97*), non-uniform configurations of ferroelectric domains should in principle be achievable in FTJs. Using $BaTiO_3$ tunnel junctions first and more recently $BiFeO_3$-based FTJs, we investigated the interplay between the analog response of the junctions and the ferroelectric domain configurations.

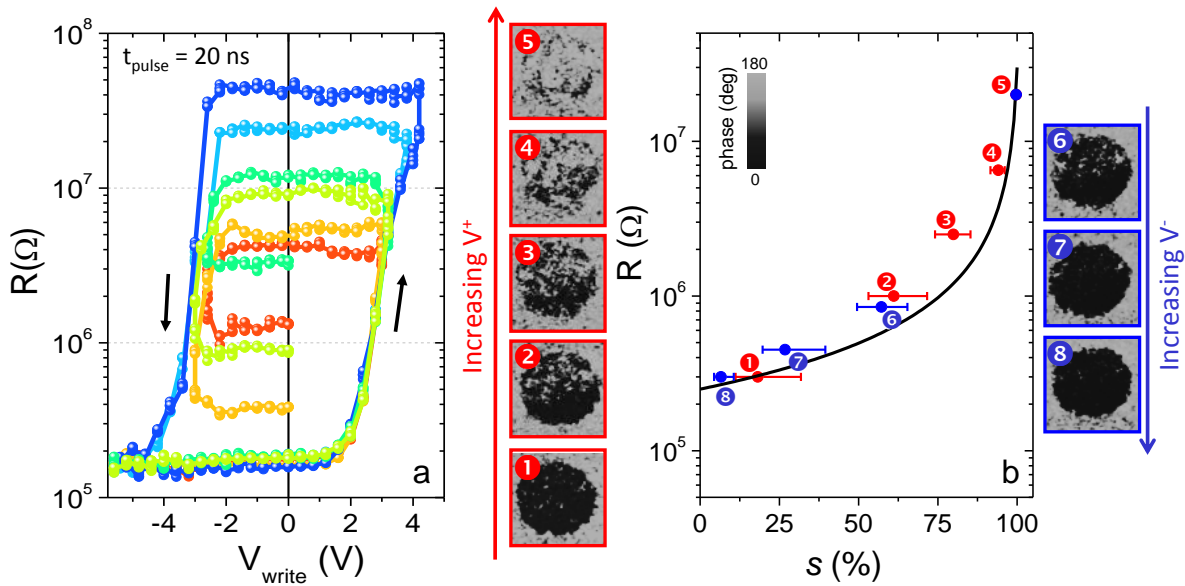

*Figure 26. Tuning resistance and ferroelectric domains with voltage amplitude. (a) Dependence of the junction resistance measured at $V_{DC}$ = 100 mV after the application of 20 ns voltage pulses ($V_{write}$) of different amplitudes. The different curves show different measurements with varying maximum $V_{write}$. (b) Variation of a similar capacitor resistance (red and blue symbols) with the relative fraction of down domains (s) extracted from PFM phase images. Red (blue) framed images show states achieved after the application of positive (negative) voltage pulses starting from the ON (OFF) state. The black curve is a simulation in a parallel resistance model (77).*

Hysteresis variations of the remanent resistance vs. write voltage pulses of 300-nm-wide $Co/BaTiO_3/La_{0.67}Sr_{0.33}MnO_3$ junctions show that intermediate resistance states can be stabilized between the ON and OFF states depending on the amplitude history of write voltage pulses (Figure 26a) (*77*). Using PFM imaging through the top electrode (*98–100*) of these junctions, we mapped the evolution of ferroelectric domains for various intermediate resistance states from ON to OFF and back to ON. The progressive resistance evolution between ON and OFF in $Co/BaTiO_3/La_{0.67}Sr_{0.33}MnO_3$ junctions is correlated to the switching of the ferroelectric polarization from up (toward Co, dark phase contrast) to down (toward $La_{0.67}Sr_{0.33}MnO_3$, bright phase contrast) through the progressive nucleation of domains in several zones with limited propagation (Figure 26b). A simple model of parallel conduction through up and down ferroelectric domains (Figure 27c) allows the correlation between resistance ($R$) and ferroelectric domain population as:

$$\frac{1}{R} = G = (1-S) \times \frac{1}{R_{ON}} + S \times \frac{1}{R_{OFF}} \quad (1)$$

where S represents the normalized fraction of domains with downward polarization ($S = 1$, for downward polarization and $R = R_{OFF}$; $S = 0$ for upward polarization and $R = R_{ON}$).

Besides, in Co/BaTiO$_3$/La$_{0.67}$Sr$_{0.33}$MnO$_3$ junctions the resistance can be progressively increased (or decreased) by applying cumulative 20 ns voltage pulses of the same intermediate positive (or negative) amplitude (*77*). Hence, this memristive behavior emulates the typical plasticity of synapses and could be implemented in neuromorphic architectures (*101*). Kim et al. reported alternative memristive behaviors in micron-size Co/BaTiO$_3$/La$_{0.67}$Sr$_{0.33}$MnO$_3$ junctions which they interpreted by the field-induced redistribution of oxygen vacancies at the CoO$_x$/BaTiO$_3$ interface (*81*). Considering the longer time scales involved (up to sec for >6 V amplitudes) and the opposite sign of the electroresistance they observed we discarded such effects from our interpretation.

Later on, our experiments with 180-nm-wide Co/BiFeO$_3$/Ca$_{0.96}$Ce0.$_{04}$MnO$_3$ tunnel junctions showed that the resistance variation of the junctions was correlated to the switching of polarization through images of ferroelectric domains (Figure 27a). In the virgin state, the polarization is pointing down (toward Ca$_{0.96}$Ce$_{0.04}$MnO$_3$) and the resistance is high, while, in the ON state, the polarization is pointing up (toward Co). Intermediate resistance states consist of a mixed population of up and down ferroelectric domains; the evolution of ferroelectric domains with the resistance is reversible. Once again, parallel domain conduction (Eq. (1)) was validated by the comparison of ferroelectric domain images and resistance measurements of multiple devices (Figure 27b).

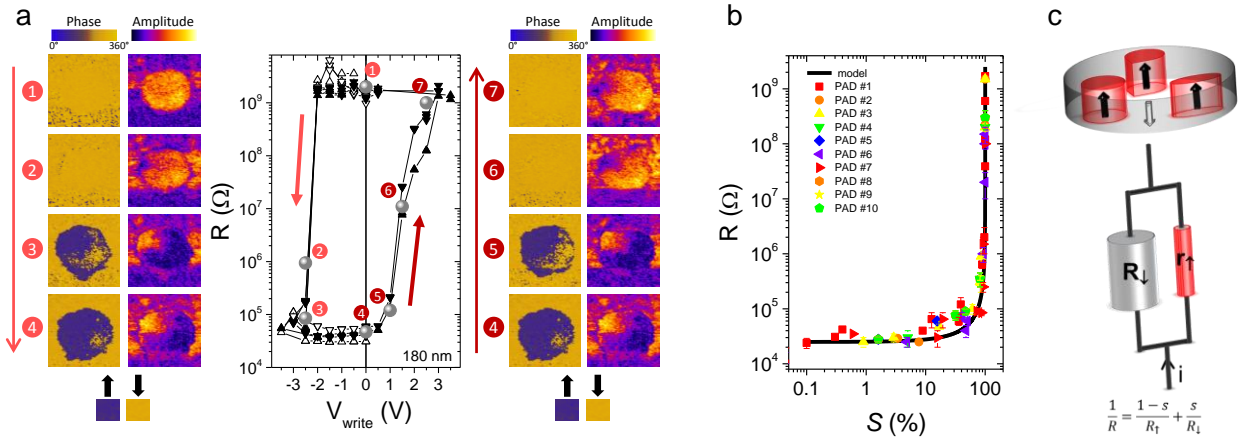

*Figure 27. (a) Switching of resistance and polarization in a 180-nm-wide Co/BiFeO$_3$/Ca$_{0.96}$Ce$_{0.04}$MnO$_3$ junction. PFM images show a progressive switching from downward to upward (1 to 4) as resistance switches from OFF to ON under negative voltage pulses of 100 ns. Reversibly the ON to OFF switching is accompanied by polarization switching from upward to downward (4 to 7) with positive voltage pulses. (b) Device resistance as a function of the fraction of down domains (S) estimated from PFM images for 10 devices in various resistance states. The black line is calculated using a parallel conduction model as in Eq. (1) sketched in (c) (9).*

The correlation between the resistance of BaTiO$_3$ (*77*) and BiFeO$_3$ (*9*) tunnel junctions in various intermediate states and the parallel resistance model suggests a negligible (if any) contribution from the ferroelectric domain walls to the tunnel resistance (*102*, *103*). In addition, the agreement between ferroelectric domain configurations and resistance indicates that the dynamics of ferroelectric domains in such nanostructures can be efficiently probed by highly-sensitive tunnel resistance measurements following voltage pulse excitations.

> **Related publications:**
> *A ferroelectric memristor*
> A. Chanthbouala, V. Garcia, R.O. Cherifi, K. Bouzehouane, S. Fusil, X. Moya, S. Xavier, H. Yamada, C. Deranlot, N.D. Mathur, M. Bibes, A. Barthélémy, J. Grollier
> **NATURE MATERIALS 11, 860-864 (2012)**
> *Giant Electroresistance of Super-Tetragonal BiFeO$_3$-Based Ferroelectric Tunnel Junctions*
> H. Yamada, V. Garcia, S. Fusil, S. Boyn, M. Marinova, A. Gloter, S. Xavier, J. Grollier, E. Jacquet, C. Carrétéro, C. Deranlot, M. Bibes, and A. Barthélémy
> **ACS NANO 7, 5385-5390 (2013)**

### 3.6.2  Ferroelectric domain dynamics

Modeling the ferroelectric memristor response to arbitrary voltage waveforms in artificial neural networks requires understanding the physical process underlying the time-dependent evolution of the FTJ resistance with voltage pulses. For this purpose, we image the pseudo real-time evolution of ferroelectric domains in Co/BiFeO$_3$/Ca$_{0.96}$Ce$_{0.04}$MnO$_3$ FTJs by stroboscopic PFM (*100*) while simultaneously measuring the electrical properties of the devices. Figure 28a shows the PFM phase and amplitude signals after cumulative 100-ns pulses of constant amplitude (1 V) from the ON to the OFF state. The evolution of the phase images reveals the gradual reversal of the polarization from up (dark domains) to down (bright domains) and is reminiscent of the polarization reversal with increasing voltages previously observed by PFM (*9*). The weak amplitude signal during reversal (e.g. stages 2 and 3) indicates that the ferroelectric system splits into many small ($\leq$ 10 nm) domains and that polarization switching is governed by the inhomogeneous nucleation of new domains rather than by the expansion of existing ones.

In collaboration with the group of Laurent Bellaiche (Univ. Arkansas), we complemented these experiments by effective-Hamiltonian-based atomistic molecular dynamics (MD) simulations on defect-free supertetragonal films of BiFeO$_3$ (*104*). These computations also yield an inhomogeneous character for polarization switching, therefore implying that this process is intrinsic in nature: one does not require the presence of defects to obtain an inhomogeneous switching in supertetragonal BiFeO$_3$. Such findings contrast with the case of bulk-like BiFeO$_3$ (rhombohedral phase) for which a homogeneous switching was recently predicted (*105*). In the rhombohedral phase, the large antiferrodistortive (AFD) tilts of oxygen octahedra provide a specific homogeneous path for polarization reversal. The computation results reveal that, while the AFD tilts are initially close to zero in the supertetragonal phase of BiFeO$_3$, regions with small polarization during polarization switching favor the emergence of significant AFD tilts. The supertetragonal phase can therefore develop an inhomogeneous AFD pattern during the switching, unlike the rhombohedral phase, explaining the possibility for inhomogeneous switching of polarization.

We extract the normalized reversed area *S* from the phase images and plot it as a function of the cumulated pulse duration in Figure 28b (black squares). Owing to the direct link between the junction resistance and this normalized reversed area S, well described by the previously-mentioned model of parallel resistances in Eq. (1) (*77*), one can also extract *S* from measurements of the junction resistance after consecutive pulses of 1 V (green squares in Figure 28b). Figure 28c shows transport data sets at different pulse amplitudes. They follow a systematic trend where, for a given cumulated pulse duration, the switched area is larger under higher voltages.

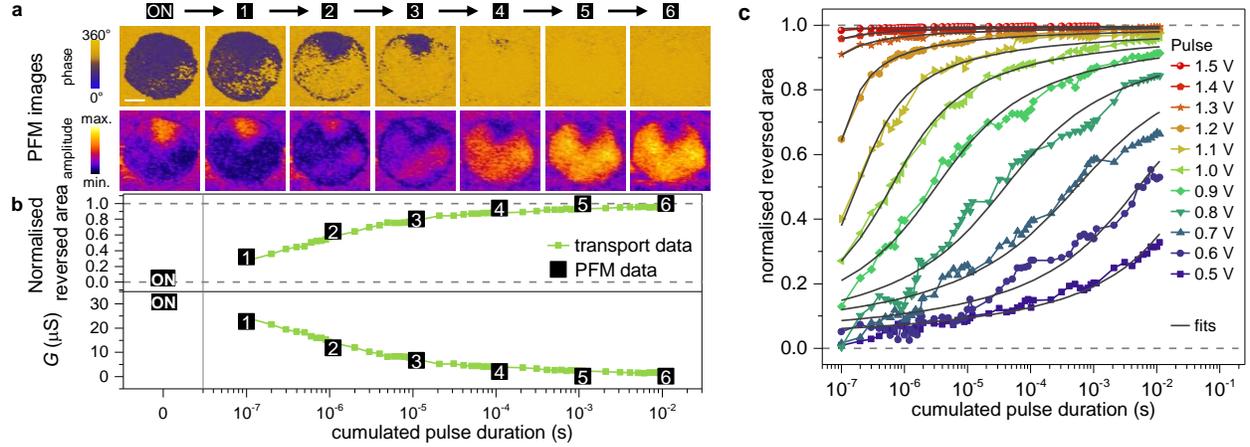

*Figure 28. (a) Evolution of the PFM phase and amplitude of a FTJ under cumulated pulses (1 V, 100 ns). Cumulative pulses induce a progressive switching with multiple nucleation areas and limited propagation of ferroelectric domains from up (dark phase) to down (bright phase) polarization. The scale bar is 50 nm. (b) (Top panel) normalized switched area as a function of cumulated pulse time calculated from transport measurements of a FTJ. The black squares indicate the normalized switched area obtained from PFM measurements in a. (Bottom panel) Corresponding conductance evolution measured as a function of cumulated pulse time. (c) Normalized switched area as a function of cumulated pulse time calculated from time-dependent transport measurements of a FTJ at different pulse amplitudes. The black lines are fit results from the nucleation-limited switching model (106).*

In ferroelectrics, inhomogeneous polarization switching can be described by a nucleation-limited model, which considers that the ferroelectric film is composed of different zones with independent switching kinetics (*107*, *108*). Assuming for each voltage *V* a broad Lorentzian distribution of the logarithm of nucleation times – with width $\Gamma(V)$ and centered at $\log(t_{\text{mean}}(V))$ – the normalized reversed area *S* can be approximated as function of time t and applied voltage *V* (*107*):

$$S_{\pm}(t,V) = \frac{1}{2} \mp \frac{1}{\pi}\arctan(\frac{\log(t_{\text{mean}}(V)) - \log(t)}{\Gamma(V)}) \quad (2),$$

where the index relates to the sign of the applied voltage. Fits obtained with this expression accurately reproduce the experimental data of ferroelectric domain dynamics probed by tunneling (black lines in Figure 28c). Figure 29a displays several representative distributions of switching times and illustrate the main trends: for larger voltages, switching occurs earlier and in a narrower time window (decrease of $t_{\text{mean}}$ and $\Gamma$), in agreement with previous results obtained on thick ferroelectric capacitors (*108*). As shown in the inset of Figure 29a, MD simulations confirm the relevance of Eq. (2) to characterize the switching process as well as the measured trend of $t_{\text{mean}}$ and $\Gamma$ with the magnitude of the electric field.

Figure 29b shows that the evolution of the switching time as a function of the inverse electric field, extracted from transport measurements (Figure 28c), follows the characteristic Merz's law (*109*) for ferroelectric switching, $t_{mean} \propto \exp\left(\frac{\alpha}{E}\right)$, where the activation field $\alpha$ is of the order of 3.0 V nm$^{-1}$. Due to the idealized nature of BiFeO$_3$ during MD simulations (no interface, no defects, no tunnel current), the timescale for polarization switching is much shorter than in experiments while the electric field is larger. Nevertheless, Merz's law can be applied to the MD simulation results (dotted line in Figure 29b) and indicates an activation field of 2.4 V nm$^{-1}$, i.e. in the same range as for experimental results. These simulation results strongly suggest that the experimentally observed inhomogeneous polarization switching in ultrathin films of BiFeO$_3$ has an intrinsic origin.

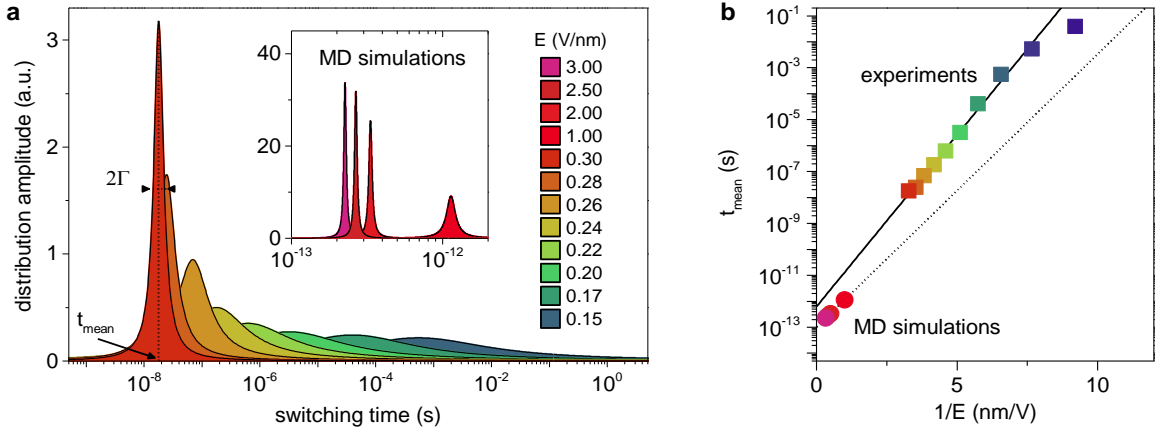

*Figure 29. (a) Examples of Lorentzian distributions of switching times extracted from the fits in Figure 28c at different pulse amplitudes and (inset) from the molecular dynamics simulations. (b) Evolution of the switching time ($t_{mean}$) as a function of the inverse of the electric field (1/E) obtained from the fits of the transport data in Figure 28c and molecular dynamics simulations. The solid and dotted lines are linear fits according to the Merz's law (106).*

### 3.6.3 Spike-timing-dependent plasticity with artificial synapses

Cortical information flows from neuron to neuron through synapses of variable connection strength. The overall distribution of the synaptic strengths provides the neural network with memory. Learning is achieved through the ability of synapses to reconfigure the strength by which they connect neurons (synaptic plasticity) (*110*). Several mechanisms regulating the evolution of the synaptic strengths have been proposed (*111*). A particularly promising one is spike-timing-dependent plasticity (STDP) (*112*) through which the synaptic strengths evolve depending on the timing and causality of electrical signals from neighboring neurons (*113*) (Figure 30a). As observed in biological systems (*114*), STDP enables learning without any external control on the synaptic strengths or any previous knowledge on the information to be processed. This makes STDP the basis for autonomous, unsupervised learning (*115*). The implementation of STDP in artificial neural networks thus emerges as a crucial milestone toward the realization of self-adaptive electronic architectures.

An electronic equivalent of the synapse for artificial neural networks is the memristor (*94*), a two-terminal nanoscale device whose resistance depends on the history of electric signals it was previously subjected to (*92*, *116*). Memristors thus exhibit plasticity and their conductance can emulate synaptic strength. Most memristors operate through the motion of ions or atoms in binary oxides ($TiO_2$ (*93*, *117*), $Ta_2O_5$ (*118*), etc.), Ag-Si/Ag-S nanocomposites (*119*) or phase change materials (*120*). Recently, STDP was demonstrated in individual devices based on such memristor technologies, confirming the potential of memristors for autonomous learning (*101*, *120–126*). However the connection between the STDP learning process and the physical mechanisms underlying resistance changes in these memristors is unclear (*116*). Future neuromorphic architectures will comprise billions of such synapses (*127*), which requires a clear understanding of the physical mechanisms responsible for plasticity. In addition, continual variations of the memristor conductance are required for learning new features under an incessant information flow. In the following, we show that resorting to purely electronic memristors with a high endurance, operating on well-established physical principles is a decisive asset for the future implementation of unsupervised learning (*128*) in high-density memristive crossbar arrays (*129*).

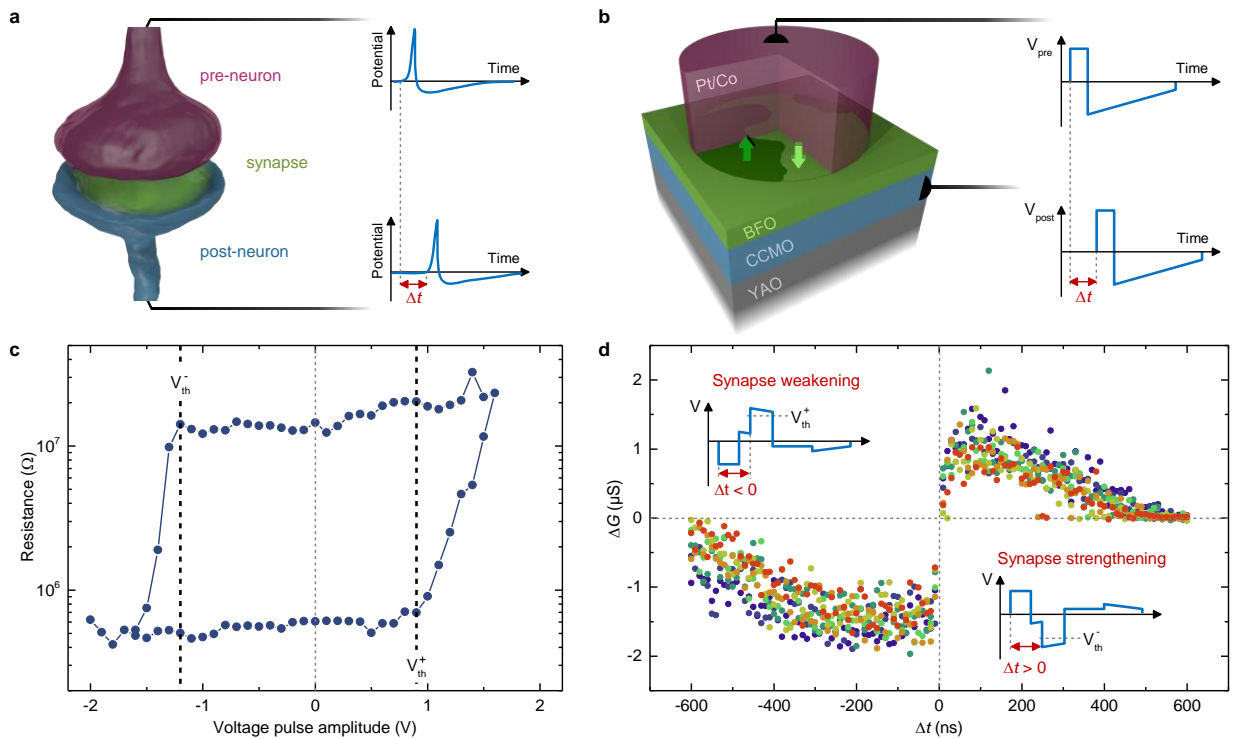

*Figure 30. Artificial synapses based on FTJs. (a) Sketch of pre- and post-neurons connected by a synapse. The synaptic transmission is modulated by the causality (Δt) of neuron spikes. (b) Sketch of the ferroelectric memristor where a ferroelectric tunnel barrier of $BiFeO_3$ is sandwiched between $Ca_{0.96}Ce_{0.04}MnO_3$ and Pt/Co electrodes. (c) Single pulse hysteresis loop of the ferroelectric memristor displaying clear voltage thresholds. (d) Measurements of spike-timing-dependent plasticity in the ferroelectric memristor. Modulation of the device conductance as a function of the time difference between pre- and post-synaptic spikes. Seven data sets were collected on the same device showing the reproducibility of the effect. The total length of each pre- and post-synaptic spike is 600 ns (106).*

As discussed previously, in the purely electronic memristors based on BiFeO$_3$ FTJs (sketched in Figure 30b), the junction resistance sensitively depends on the relative fraction of ferroelectric domains with polarization pointing toward one or the other electrode (*77*). Applying voltage pulses to the junction modifies the domain population, thereby inducing resistance changes. Figure 30c shows the dependence of the junction resistance with the amplitude of 100-ns voltage pulses in a typical Co/BiFeO$_3$/Ca$_{0.96}$Ce$_{0.04}$MnO$_3$ FTJ. From this hysteresis cycle, one clearly notes the existence of voltage thresholds $V_{th}^+$ ($V_{th}^-$) beyond which switching between low and high (high and low) resistance states occurs. The existence of such well-defined voltage thresholds (associated with the intrinsic coercivity of the ferroelectric) makes it possible to implement STDP in these FTJs (*130*). According to STDP, if the pre-neuron spikes just before the post-neuron – indicating a causal relationship – the synapse should be strengthened whereas if the pre-neuron spikes just after the post-neuron – indicating a non-causal relationship – the synapse should be weakened. We emulate the spikes from pre- and post-neurons (sketched in Figure 30a) by the waveforms shown in Figure 30b: rectangular voltage pulses followed by smooth slopes of opposite polarity. Importantly, the voltage never exceeds $V_{th}$ so that a single spike cannot lead to a change in resistance.

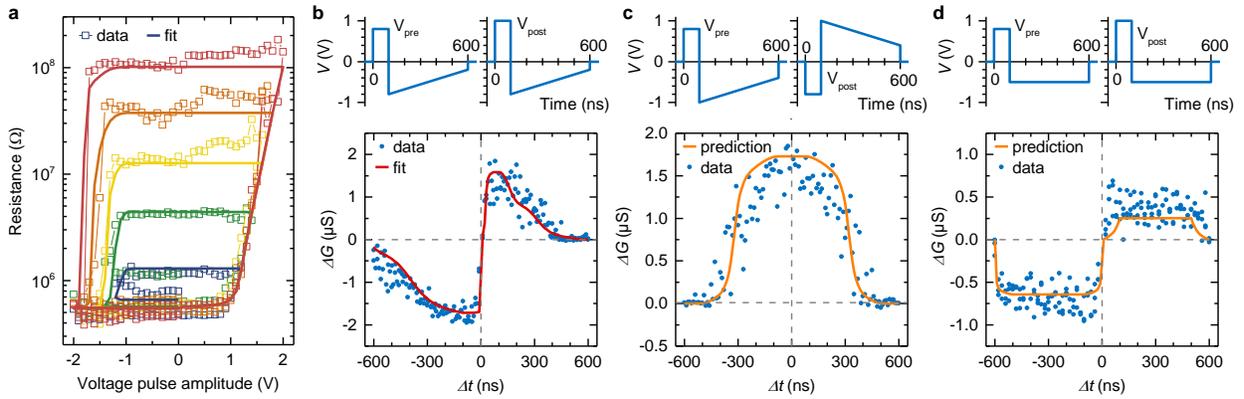

*Figure 31. Predicting spike-timing-dependent plasticity learning with ferroelectric synapses. (a) Multiple hysteresis loops of a ferroelectric memristor showing a clear dependence of resistance switching with the maximum pulse amplitudes. (b)-(d) Examples of spike-dependent-plasticity learning curves of different shapes. The pre- and post-synaptic spikes and conductance variations are in the top and bottom panels, respectively. For each device, simultaneous fits of data in (a) and (b) (solid lines) using Eq. (1) and Eq. (2) allow the prediction of new learning curves in (c) and (d) (orange lines) (106).*

When both pre- and post-neuron spikes reach the memristor with a delay $\Delta t$, their superposition produces the waveforms ($V_{pre} - V_{post}$) displayed in the inset of Figure 30d. The resulting combined waveform transitorily exceeds the threshold voltage, leading to an increase ($\Delta G > 0$, synapse strengthening) or a decrease ($\Delta G < 0$, synapse weakening) of the FTJ conductance, depending on the sign of $\Delta t$. As can be seen from the experimental STDP curve in Figure 30d, only closely timed spikes produce a conductance change whereas long delays leave the device unchanged.

Because of the direct relationship between S and R in these devices, the accurate description of ferroelectric switching by the nucleation-limited model (previous section) can be further extended to the modeling of resistance changes as a function of voltage amplitude. Figure 31a shows nested hysteresis loops of resistance as a function of voltage amplitude, characteristic of memristors. The STDP curve of the same device is displayed in Figure 31b. Combining Eq. (1) and Eq. (2), we simultaneously fit the resistance versus voltage cycles and the STDP curve. Both resistance changes can be accurately replicated (solid lines in Figure 31a-b) using the same set of parameters $t_{\mathrm{mean}}(V)$ and $\Gamma(V)$.

The full description for each memristor device makes it possible to predict the conductance changes driving STDP learning in ferroelectric synapses. In Figure 31b-d, we apply different voltage waveforms to our memristors to emulate various types of pre- and post-neuron activities. This procedure allows the generation of biologically realistic, though accelerated (Figure 31b-c), or artificially designed (Figure 31d) STDP learning curves. Using our model with the parameters extracted previously, we can now predict the conductance changes for these novel types of STDPs. Figure 31c-d shows the excellent agreement between these predictions and the measured conductance variations associated with different STDP curves.

### 3.6.4  Unsupervised learning simulations in artificial neural networks

We now use our understanding of ferroelectric domain dynamics in FTJs, and the subsequent modeling of STDP in the devices, to simulate unsupervised learning in spiking neural networks with ferroelectric synapses. These simulations serve a test bench to investigate the influence of the STDP curve shape on the ability of the network to recognize patterns in images (e.g., the horizontal, diagonal, and vertical bars labeled A, B, and C, in the inset of Figure 32a).

The simulated network is built around a crossbar of 9 × 5 ferroelectric memristors (Figure 32a). Each of the 9 spiking input neurons codes for one pixel of noisy images that contain one of the three patterns to recognize. The output neurons integrate the input signals flowing through the memristors along each row and fire when a threshold is reached. Lateral inhibition then resets all output neurons. Figure 32b shows that the recognition rate of the network increases with the number of image presentations, reaching 100% for low noise levels and almost 80% for high (of the order of the input amplitude) noise levels. The evolution of the conductance of the 9 memristors in each row is shown in the inset of Figure 32b. As the network is learning, the conductance images in row 1, 2, and 4 each converge toward one the three input patterns whereas the memristor conductances in rows 3 and 5 remain random.

After learning (corresponding to 200 input image presentations), three of the output neurons have specialized to each of the three input patterns and systematically fire when the corresponding input is presented (Figure 32c). In addition, our simulations reveal that successful unsupervised learning is highly dependent on the exact shape of the pre- and post-neuron spike waveforms. Figure 32d shows that, for the case presented before, the recognition rate reaches 100% for well-chosen post-neuron spike amplitudes in the range of $0.82 - 0.98$ V. However, when the amplitude of the post-neuron spike waveform is only slightly lower or higher, the recognition rate rapidly drops. Our simulations therefore emphasize the importance of a precise knowledge of the memristor dynamics and its accurate description on the basis of a physical model.

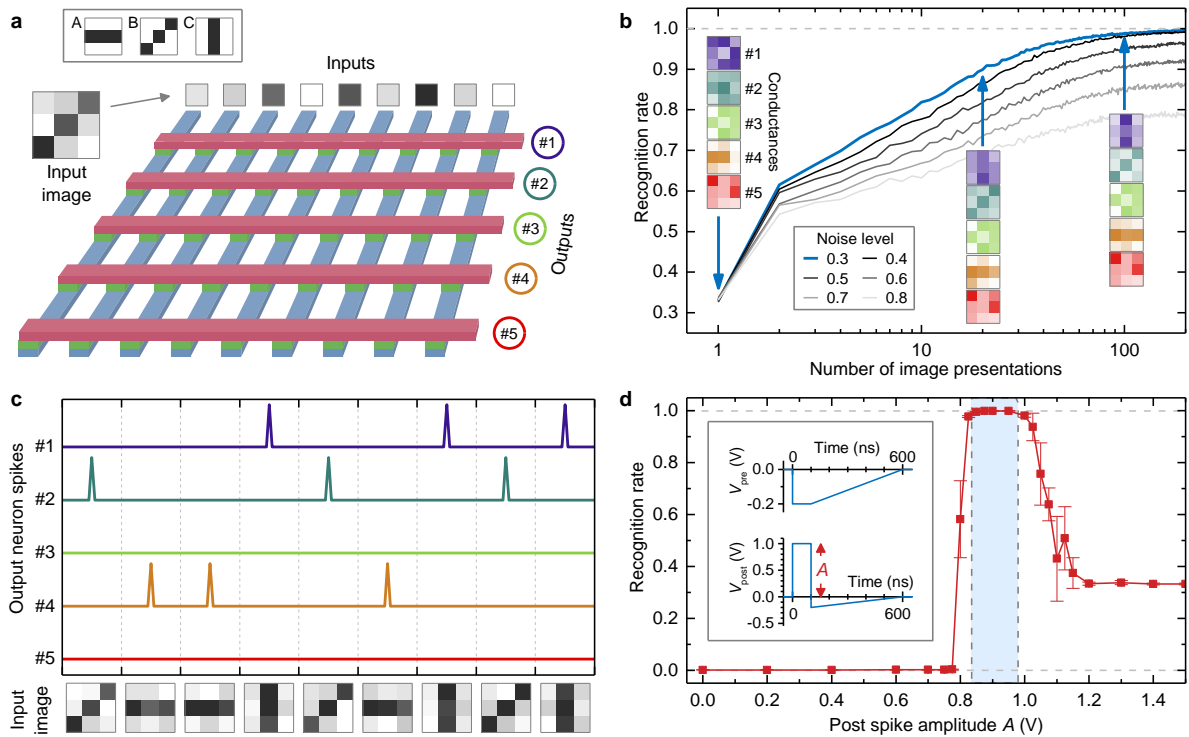

*Figure 32. Unsupervised learning with ferroelectric synapses. (a) Simulated spiking neural network comprising 9 input neurons connected to 5 output neurons by an array of ferroelectric memristors. The inputs are noisy images of the patterns to recognize: horizontal (A), diagonal (B) and vertical (C) bars in 3 × 3 pixel images. (b) Recognition rate as a function of the number of presented images for different noise levels. The colored images are conductance maps of the memristors in each line and show their evolution for a noise level of 0.3 (blue line). (c) Behavior of the network after successful learning. Neurons 1, 2 and 4 emit spikes when inputs C, B and A are presented, respectively. (d) Evolution of the final recognition rate as a function of the amplitude of the post-synaptic spike (106).*

In summary, we have established that spike-timing-dependent plasticity can be harnessed from intrinsically inhomogeneous polarization switching in ferroelectric memristors. Combining time-dependent transport measurements, ferroelectric domain imaging, and effective-Hamiltonian-based atomistic molecular dynamics simulations, we show that the ferroelectric switching underlying resistance changes in these devices can be described by a nucleation-limited model. Using this physical model, we can reliably predict the conductance evolution of ferroelectric synapses with varying neural inputs. These results pave the way toward low-power hardware implementations of billions of reliable and predictable artificial synapses in future brain-inspired computers.

**Related publication:**
*Learning through ferroelectric domain dynamics in solid state synapses*
S. Boyn, J. Grollier, G. Lecerf, B. Xu, N. Locatelli, S. Fusil, S. Girod, C. Carrétéro, K. Garcia, S. Xavier, J. Tomas, L. Bellaiche, M. Bibes, A. Barthélémy, S. Saïghi, V. Garcia
**NATURE COMMUNICATIONS 8, 14736 (2017)**

## 3.7 Electric control of functional properties with multiferroics

### 3.7.1 Interfacial magnetoelectric coupling with multiferroic tunnel junctions

Coupling ferroelectrics with magnetic materials may enable a control of magnetism by an electric field. Interfacial magnetoelectric effects have aroused large interest over the past few years (*131*). Significant changes of the magnetic properties at the ferroelectric/metal interface can appear due to screening effects (*132–135*), changes of the interface bonding (*136–139*) and changes of the interfacial magnetocrystalline anisotropy (*140*). The reversible electric field created by the ferroelectric is predicted to induce interesting features such as (i) a modification of the magnetization of an adjacent ferromagnet (*132*, *141*), (ii) the stabilization of ferromagnetism in a paramagnetic metal (*133*) or (iii) ferromagnetic/antiferromagnetic phase transitions (*135*).

In spintronics, a key building block is the magnetic tunnel junction where two ferromagnetic electrodes sandwich an ultrathin layer of dielectric (*142*). In these devices, the tunnel resistance has two distinct states corresponding to the antiparallel or parallel configurations of the magnetic electrodes (which are controllable by a magnetic field). This effect, called tunnel magnetoresistance (TMR), is based on spin-dependent electron tunneling that is extremely sensitive to the spin-dependent electronic properties of the ferromagnet/dielectric interface (*143*). Hence, measurements of TMR can be used to probe interfacial changes of the magnetic properties. The interplay between ferroelectricity and ferromagnetism can be investigated through transport measurements in artificial multiferroic tunnel junctions (Figure 33) where a ferroelectric tunnel barrier is sandwiched between two ferromagnetic electrodes (*144*). The two ferroic order parameters (ferromagnetism and ferroelectricity) give rise to four distinct resistance states due to the combined TMR and TER effects. In addition, interface magnetoelectric coupling between the ferroelectric and the ferromagnetic can be probed by measuring the modulation of the spin-dependent tunneling current (through the TMR). Gajek *et al.* were the first to demonstrate the existence of a four resistance state memory using a multiferroic (ferroelectric and ferromagnetic) tunnel barrier of $La_{0.1}Bi_{0.9}MnO_3$ sandwiched between $La_{0.67}Sr_{0.33}MnO_3$ and Au electrodes (*75*). Artificial multiferroic tunnel junctions are in principle easier to realize because multiferroics that are both ferromagnetic and ferroelectric are rare, and it is difficult to decouple their magnetization when used as a tunnel barrier from that of the ferromagnetic electrode.

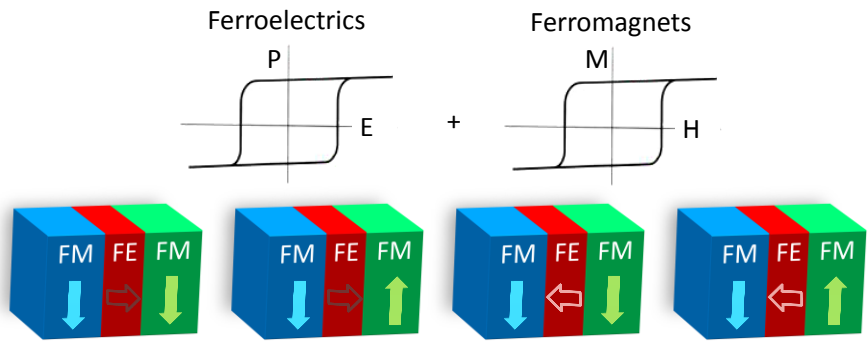

*Figure 33. Multiferroic tunnel junctions in which a ferroelectric (FE) tunnel barrier is combined with ferromagnetic (FM) electrodes. The two ferroic orders give rise to four logic states.*

Ab initio calculations performed on Fe/BaTiO$_3$ interfaces showed a large interface magnetoelectric coupling resulting from a modification of the ferromagnet/ferroelectric interface bonding upon switching of the ferroelectric polarization (*136*). Sizeable modifications of the Fe and Ti-induced magnetic moments at the Fe/BaTiO$_3$ interface have been calculated by several groups (*136*, *138*, *139*). In order to probe this interfacial magnetoelectric coupling, we fabricated Fe/BaTiO$_3$ (1.2 nm)/La$_{0.67}$Sr$_{0.33}$MnO$_3$ nanoscale tunnel junctions (Figure 34a). The tunnel junctions in which Fe is exchanged coupled to a CoO/Co bilayer to enlarge its magnetic coercive field exhibit negative TMR at 4 K with well-defined antiparallel plateaus (Figure 34b). The sign of the TMR indicates that the Fe/BaTiO$_3$ interface tunnel spin-polarization is negative, as predicted by ab initio calculations (*136*, *138*, *145*).

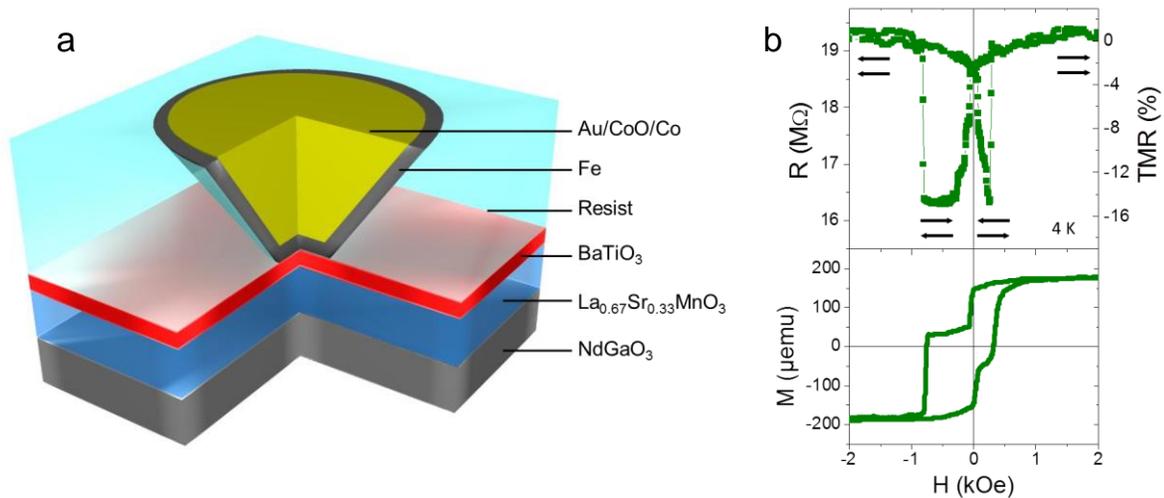

*Figure 34. (a) Sketch of the nanojunction defined by electrically controlled nanoindentation (146). (b) Top: R(H) recorded at -2 mV and 4.2 K showing negative TMR; bottom: magnetization vs. magnetic field recorded at 30 K with a SQUID magnetometer (57).*

The ferroelectric polarization of BaTiO$_3$ is switched with voltage pulses of +1/-1 V and 1 s duration, and the TMR is subsequently measured while sweeping the magnetic field and collecting the current at V$_{DC}$ = -50 mV in the different ferroelectric states (Figure 35). Large changes of the negative TMR are observed depending on the orientation of the ferroelectric polarization: the TMR is large (small) when the ferroelectric polarization points toward Fe (La$_{0.67}$Sr$_{0.33}$MnO$_3$) (*57*). The variation of the TMR with the ferroelectric polarization can reach 450%. Hence, the ferroelectric tunnel barrier provides a way to efficiently control the spin-polarization of the tunnel current, locally, in a non-volatile manner and potentially with a low energy (*57*).

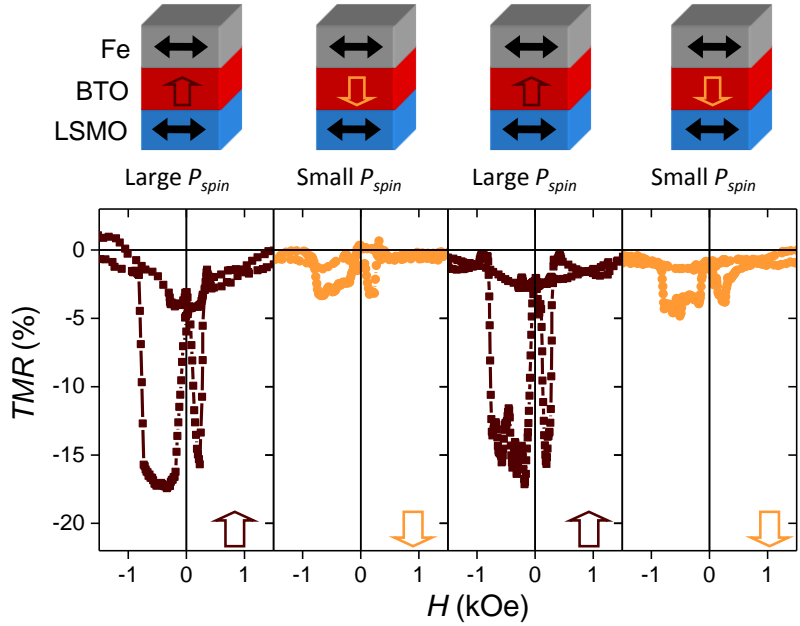

*Figure 35. TMR vs. magnetic field recorded at $V_{DC}$ = -50 mV and T = 4.2 K for an Fe/BaTiO$_3$/La$_{0.67}$Sr$_{0.33}$MnO$_3$ nanojunction after poling the ferroelectric up, down, up, down with voltage pulses of $\pm 1$ V and 1 s (57).*

In collaboration with the group of Sergio Valencia (Helmholtz Zentrum Berlin), we investigated the Fe/BaTiO$_3$ interface on similar heterostructures using the Synchrotron light source (BESSY). The X-ray absorption dichroic signal of ultrathin films of Fe (2 nm) is typical of metallic bcc iron and suggests a very limited oxidation at the Fe/BaTiO$_3$ interface (Figure 36a). Using a more surface-sensitive X-ray resonant magnetic scattering technique, we detected a small induced Ti magnetic moment in BaTiO$_3$ (Figure 36b). Hysteresis cycles with magnetic field of the dichroic signals of Fe, Mn, and Ti revealed that the dichroic signal of Ti is quasi-ferromagnetic and coupled to that of Fe (Figure 36c), in agreement with ab initio predictions (*136*, *138*, *145*). These experiments suggest that a multiferroic state can be induced at room temperature in BaTiO$_3$ through the interfacial coupling with a ferromagnetic metal such as Fe (*147*). In collaboration with the group of Alexandre Gloter (LPS, Orsay), high-resolution transmission electron microscopy investigations of the Fe/BaTiO$_3$ interface indicate a complex Fe/FeO/TiO$_2$/BaO interface (*145*). Ab initio calculations show that despite the presence of the FeO, a magnetic moment is induced in Ti. This interfacial magnetic moment is changing upon the orientation of the ferroelectric polarization suggesting a new type of interfacial magnetoelectric coupling, in agreement with the ferroelectric control of spin-polarization mentioned previously. Such interfacial magnetoelectric coupling probed by spin-dependent tunneling was also observed in Co/BaTiO$_3$/La$_{0.67}$Sr$_{0.33}$MnO$_3$ junctions (*147*). More recently, Radaelli *et al.* demonstrated that ferroelectric polarization reversal at the Fe/BaTiO$_3$ interface controls the magnetic interaction of the interfacial ultrathin FeO (*148*). This could explain our spin-dependent transport results in Fe/BaTiO$_3$/La$_{0.67}$Sr$_{0.33}$MnO$_3$ junctions: when the ferroelectric polarization points toward Fe, ferromagnetism in FeO promotes a significant spin-polarization while when it points away from Fe, antiferromagnetism in FeO results in a low effective spin-polarization.

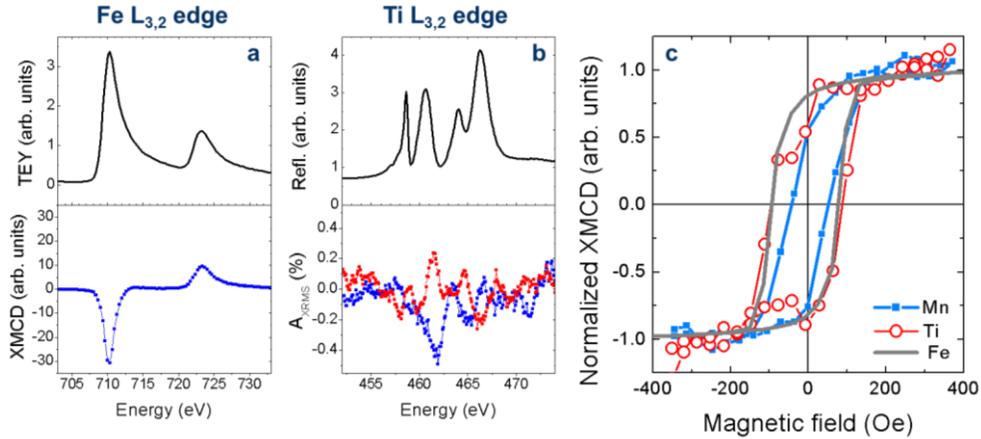

*Figure 36. Element specific magnetic signals at the Fe/BaTiO$_3$ interface.(a) Top: XAS spectrum for Fe; bottom: XMCD spectrum for Fe. (b) Top: XRMS spectrum for Ti; bottom: XRMS asymmetry spectra for Ti obtained for right (red) and left (blue) helicities of the incoming circularly polarized radiation. (c) XMCD and XRMS asymmetry vs. magnetic field for Mn, Fe, and Ti. To prevent oxidation, the Fe (2 nm) top electrode is capped with AlOx (1.5 nm)/Al(1.5 nm) (147).*

Pantel *et al.* showed that, with large area Co/PbZr$_{0.2}$Ti$_{0.8}$O$_3$ (3.2 nm)/La$_{0.67}$Sr$_{0.33}$MnO$_3$ tunnel junctions, it is even possible to reverse the sign of the TMR by switching the ferroelectric polarization (*149*). They performed magnetotransport measurements at V$_{DC}$ = 10 mV and 50 K and observed a change of the TMR from -3% to +4% when the ferroelectric polarization was pointing to the Co or the La$_{0.67}$Sr$_{0.33}$MnO$_3$, respectively. Although the TMR is not large, its relative variation with the ferroelectric polarization reaches -230%. Thus, the orientation of the ferroelectric polarization of the PbZr$_{0.2}$Ti$_{0.8}$O$_3$ tunnel barrier controls the sign of the current spin-polarization, providing potential for electrically-controlled spintronic devices (*149*). In the experiments by Yin *et al.*, the use of the interfacial La$_{0.5}$Ca$_{0.5}$MnO$_3$ layer was shown to induce large modifications of the TMR in La$_{0.67}$Sr$_{0.33}$MnO$_3$/La$_{0.5}$Ca$_{0.5}$MnO$_3$ (0.8 nm)/BaTiO$_3$/La$_{0.67}$Sr$_{0.33}$MnO$_3$ junctions while switching the ferroelectric polarization of the barrier (*80*). The polarization-induced metal/insulator phase transition in La$_{0.5}$Ca$_{0.5}$MnO$_3$ is accompanied by a ferromagnetic/antiferromagnetic transition. Hence, at low temperature (80 K), the TMR is about 100% (close to zero) when the ferroelectric polarization points toward (away from) La$_{0.5}$Ca$_{0.5}$MnO$_3$ corresponding to its ferromagnetic (antiferromagnetic) state. An interfacial magnetic phase transition driven by the ferroelectric polarization of the tunnel barrier is thus an efficient way to control the spin-polarization of the tunnel current: the variation of the TMR with the ferroelectric polarization orientation reaches 500% in this specific case.

All the above-mentioned experiments of ferroelectric control of spin-polarization use at least one ferromagnetic electrode of La$_{0.67}$Sr$_{0.33}$MnO$_3$, thereby limiting the experiments to operation below room temperature (*150*). Moreover, thermally-activated inelastic tunneling processes through defects in the ferroelectric tunnel barrier can limit both TER and TMR at higher temperature (*80*, *83*, *151*). Nevertheless, Yin et al. showed a four-state memory at room temperature with La$_{0.67}$Sr$_{0.33}$MnO$_3$/Ba$_{0.95}$Sr$_{0.05}$TiO$_3$ (3.5 nm)/La$_{0.67}$Sr$_{0.33}$MnO$_3$ tunnel junctions, albeit with extremely small contrasts of resistance (<1%) and no apparent variations of the spin-dependent signal with the ferroelectric polarization orientation (*152*). Systems with solely transition metals or their alloys as bottom and top electrodes should be investigated to look for efficient ferroelectric control of spin-polarization at room temperature.

> **Related publications:**
> *Ferroelectric control of spin-polarization*
> V. Garcia, M. Bibes, L. Bocher, S. Valencia, F. Kronast, A. Crassous, X. Moya, S. Enouz-Vedrenne, A. Gloter, D. Imhoff, C. Deranlot, N. D. Mathur, S. Fusil, K. Bouzehouane, A. Barthélémy
> **SCIENCE 327, 1106-1110 (2010)**
> *Interface-induced room-temperature multiferroicity in BaTiO$_3$*
> S. Valencia, A. Crassous, L. Bocher, V. Garcia, X. Moya, R. O. Cherfi, C. Deranlot, K. Bouzehouane, S. Fusil, A. Zobelli, A. Gloter, N. D. Mathur, A. Gaupp, R. Abrudan, F. Radu, A. Barthélémy, M. Bibes
> **NATURE MATERIALS 10, 753-758 (2011)**
> *Atomic and Electronic Structure of the BaTiO$_3$/Fe Interface in Multiferroic Tunnel Junctions*
> L. Bocher, A. Gloter, A. Crassous, V. Garcia, K. March, A. Zobelli, S. Valencia, S. Enouz-Vedrenne, X. Moya, N.D. Mathur, C. Deranlot, S. Fusil, K. Bouzehouane, M. Bibes, A. Barthélémy, C. Colliex, O. Stephan
> **NANO LETTERS 12, 376-382 (2012)**

### 3.7.2 Using ferroelectrics to control functional properties

During the post-doc of Daniel Sando, we investigated the optical properties of BiFeO$_3$ thin films. According to previous studies, epitaxial thin films of BiFeO$_3$ can be deposited on a variety of substrates imposing a wide range of biaxial strain varying from -3% to 1%. This strain mainly influences the critical temperature for the ferroelectric state and lowers it toward the Néel temperature that is virtually not sensitive to strain (*153*). In addition, strain induces a destabilization of the spin cycloid existing in bulk BiFeO$_3$ and leads to a pure antiferromagnetic state (*154*). Here we apply epitaxial strain engineering to tune the optical response of BiFeO$_3$ thin films. Ellipsometry measurements show a decrease of the refractive index with epitaxial strain in the wavelength window of 500 to 800 nm. These results indicate a large elasto-optic coefficient for BiFeO$_3$, greater than most of the materials used in commercial modulators. The increase of compressive strain leads to the stabilization of the super-tetragonal phase of BiFeO$_3$ (T-phase) over the bulk-like pseudo-rhombohedral phase (R-phase). These two phases show significant differences in optical absorption with a 0.3 eV larger optical bandgap for the T-phase than for the R-phase. Within the collaboration with the group of Laurent Bellaiche, we explained these differences by a modification of the electronic configurations between the two phases related to a change in coordination. For specific growth conditions on LaAlO$_3$ substrates, it is possible to stabilize an ordered structure in which the two phases coexist with a peculiar topography associated (Figure 37a). Moreover, the application of an electric field enables the reversible transformation of these phases from R+T to T (*155*). Combining scanning probe techniques with optical microscopy, we demonstrated that this phase transformation was associated to a reversible modulation of the optical absorption of the films (Figure 37), leading to electrochromic effects in BiFeO$_3$ (*156*).

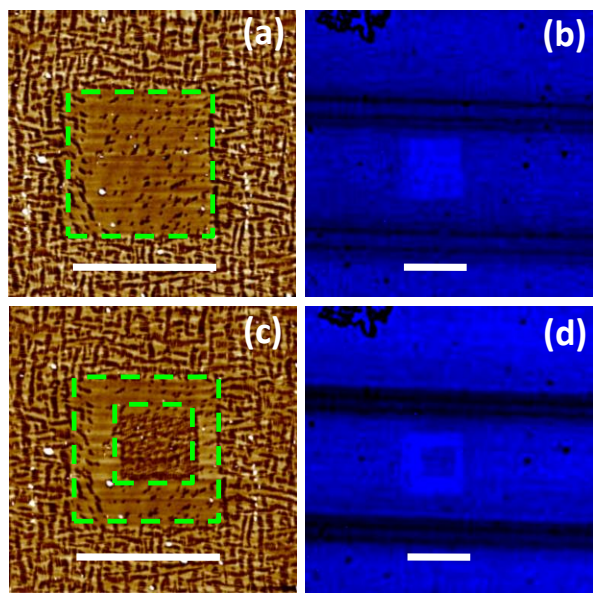

*Figure 37. Electrochromism in BiFeO$_3$ thin films. (a) AFM topography image after poling a 10 × 10 µm² square, locally transforming the R + T phase into T-like BiFeO$_3$. (b) Transmission optical image acquired in the same region with a dielectric filter centered at 420 nm (bandwidth 10 nm). (c) AFM topography image after poling the central 5 × 5 µm² region with an opposite voltage, restoring R + T structures. (d) Transmission optical image of the area using the 420 nm filter. The scale bars are 10 µm (156).*

In the context of functional oxides, a significant effort is devoted to the research on intrinsic or artificial multiferroics for a non-volatile and low energy control of magnetization with an electric field (*157*). During the PhD of Ryan Chérifi, we explored the combination of FeRh, which undergoes a first order antiferromagnetic-ferromagnetic transition near room temperature, with a ferroelectric material. The ferroelectricity of BaTiO$_3$ substrates was used to modify the metamagnetic transition of epitaxial thin films of FeRh, thus toggling electrically between the antiferromagnetic and magnetic states slightly above room temperature (Figure 38) (*158*). Combining structural, magnetic, and *ab initio* studies, we concluded that strain modifications related to the electric control of ferroelastic domains in the BaTiO$_3$ substrate are mainly governing this magnetic phase transition. These results correspond to a magnetoelectric coefficient one order of magnitude larger than any other reports and open the path to the use of ferroelectric materials for magnetic storage or spintronics.

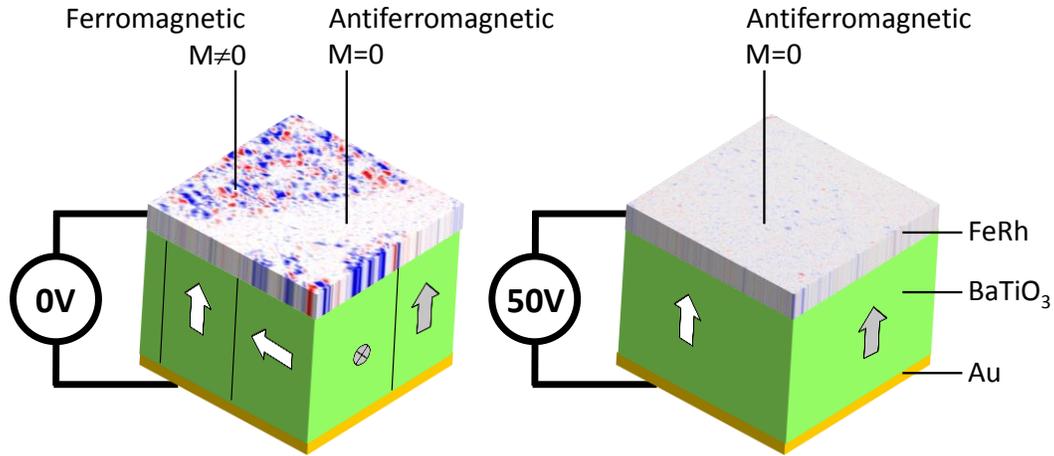

*Figure 38. Sketch of the electric control of magnetism in FeRh. Applying voltage to the BaTiO$_3$ substrate changes the ferroelastic domain configurations affecting the strain-sensitive magnetic order in FeRh films. The surfaces are the actual XMCD-PEEM magnetic images of FeRh collected at the Fe L$_3$ edge and 385 K (158).*

**Related publications:**
*Large elasto-optic effect and reversible electrochromism in multiferroic BiFeO$_3$*
D. Sando, Y. Yang, E. Bousquet, C. Carrétéro, V. Garcia, S. Fusil, D. Dolfi, A. Barthélémy, P. Ghosez, L. Bellaiche, M. Bibes
NATURE COMMUNICATIONS 7, 10718 (2016)
*Electric-field control of magnetic order above room temperature*
R.O. Cherifi, V. Ivanovskaya, L.C. Phillips, A. Zobelli, I.C. Infante, E. Jacquet, V. Garcia, S. Fusil, P.R. Briddon, N. Guiblin, A. Mougin, A.A. Ünal, F. Kronast, S. Valencia, B. Dkhil, A. Barthélémy, M. Bibes
NATURE MATERIALS 13, 345-351 (2014)

### 3.7.3 Electric control of non-collinear antiferromagnetic order

Nearly 90% of known magnetic substances have dominant antiferromagnetic interactions, resulting in no or very small magnetization, and most are insulators (*159*). This strongly complicates their investigation, especially when the magnetic order needs to be mapped at the nanoscale. While magnetic force microscopy (*160*) or X-ray photoemission electron microscopy (*161*, *162*) can reach a spatial resolution of a few tens of nm, their sensitivities are not compatible with the detection of weak magnetic signals commonly involved in antiferromagnets. Spin-polarized scanning tunneling microscopy can resolve the magnetic moments of single atoms (*163*, *164*) but is only applicable to conductive systems. Therefore, the nanoscale spin texture cannot be directly imaged in the vast majority of magnetic materials. This is increasingly problematic since magnetic materials with complex antiferromagnetic orders show very appealing functionalities (*165–171*), which are absent in ferromagnets, and could be exploited to develop a new generation of low-power spintronic devices (*172*).

For instance, BiFeO$_3$ (*173*), in which antiferromagnetism and ferroelectricity coexist with high ordering temperatures, appears as a unique platform for spintronic (*168*) and magnonic devices (*174*). However, while the nanoscale structure of its ferroelectric domains has been widely investigated with PFM, revealing unique domain structures and domain wall functionalities (*102*, *175*, *176*), the corresponding nanoscale magnetic textures present in BiFeO$_3$ and their potential for spin-based technology still remain concealed. In this work we demonstrate the first real-space imaging of the spin cycloid and electric field manipulation of complex antiferromagnetic order in a BiFeO$_3$ thin film using a non-invasive scanning magnetometer based on a single nitrogen-vacancy (NV) defect in diamond.

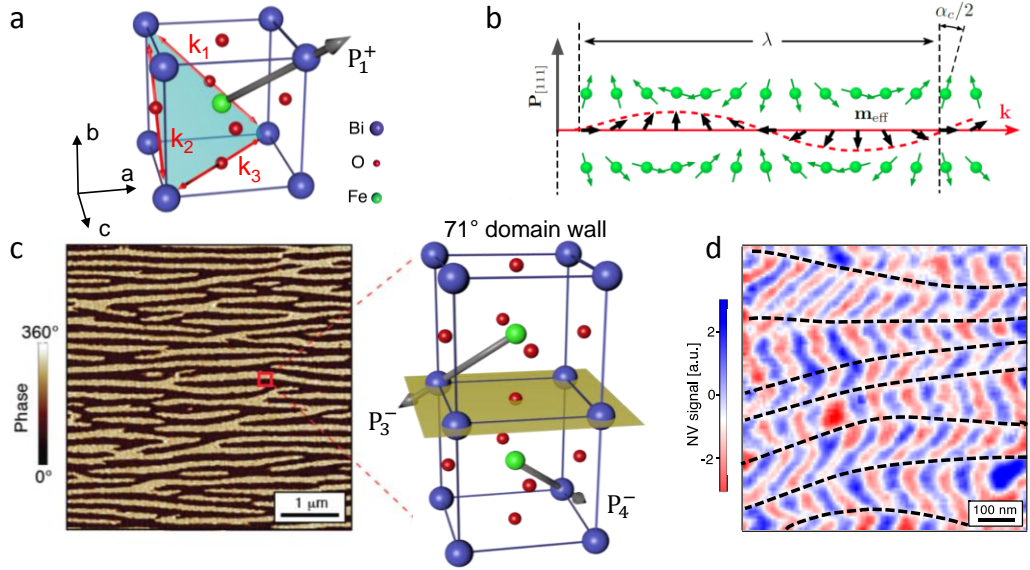

*Figure 39. Ferroelectric and magnetic textures of BiFeO$_3$ thin films. (a) Pseudocubic unit cell of BiFeO$_3$ showing one of the eight possible variants of ferroelectric polarization $P_1^+$ pointing along the [111] direction with the three corresponding propagation directions of the spin cycloid ($k_1$, $k_2$, $k_3$). (b) Schematic representation of the spin cycloid. Magnetoelectric coupling induces a cycloidal rotation of Fe$^{3+}$ spins (green arrows). The canted antiferromagnetic alignment between consecutive atomic layers, characterized by the angle $\alpha_C$, results in an effective magnetic moment $m_{eff}$ (black arrows) describing a cycloid with a wavelength $\lambda$. The propagation direction of the spin cycloid k is normal to the ferroelectric polarization vector P. (c) Ferroelectric striped domains in a 32-nm-thick film of BiFeO$_3$ mapped by in-plane PFM. The right panel sketches the two pristine variants of ferroelectric domains ($P_3^-$ and $P_4^-$) separated by 71° domain walls. (d) Magnetic field image recorded above the BiFeO$_3$ film while operating the NV magnetometer in dual-iso-B imaging mode. The dashed lines are guides to the eyes indicating the ferroelectric domain walls where sharp rotations of the magnetic propagation direction occur. The sketches in (a) and (c) are in top view with a slight tilt.*

Bulk BiFeO$_3$ crystallizes in a slightly-distorted rhombohedral structure but is commonly described by the pseudo cubic unit cell shown in Figure 39a. The displacement of Bi ions relative to FeO$_6$ octahedra gives rise to a strong polarization along one of the [111] directions (*173*). This system is complex as the eight possible polarization orientations $P_i^{\pm}$ give rise to three types of ferroelectric domain walls (71°, 109°, or 180°). From the magnetic point of view, BiFeO$_3$ was initially thought to be a conventional G-type antiferromagnet (*177*) but neutron diffraction later revealed a cycloidal modulation of the antiferromagnetic order (*178, 179*) with a characteristic period of $\lambda$ = 64 nm (Figure 39b). The rhombohedral symmetry of BiFeO$_3$ allows three equivalent propagation directions of the cycloid ($k_1$, $k_2$, and $k_3$) for a given variant of the ferroelectric domain (*180*) (Figure 39a). In thin films biaxial strain tends to destabilize the spin cycloid which leads to a modulation of the spin cycloid period (*154*). A 32-nm-thick film of BiFeO$_3$(001) was deposited by pulsed laser deposition on a DyScO$_3$(110) substrate, using an ultrathin buffer electrode of $SrRuO_3$. Epitaxial strain leads to a striped pattern of ferroelectric domains where only two variants of polarization coexist, separated by 71° domain walls (Figure 39c).

The corresponding magnetic structure of the $BiFeO_3$ thin film was investigated through stray field measurements using a scanning nano-magnetometer based on a single NV defect in diamond (*181–183*), in collaboration with the group of Vincent Jacques (Univ. Montpellier). This point-like impurity can be exploited for quantitative magnetic-field imaging at the nanoscale by recording Zeeman shifts of its electronic spin sublevels through optical detection of the electron spin resonance (ESR). In this study, a single NV defect located at the apex of a nanopillar of a diamond scanning probe is integrated into an AFM, which allows scanning the NV defect in close proximity to the sample (*184, 185*). At each point of the scan, optical illumination combined with microwave excitation enable measuring the ESR spectrum of the NV defect by recording its spin-dependent photoluminescence (PL) intensity. Any magnetic field emanating from the sample is then detected through a Zeeman shift of the ESR frequency. The resulting magnetic field sensitivity is a few $\mu T/\sqrt{Hz}$, while the spatial resolution is fixed by the distance $d$ between the sample and NV center (*183*). This distance is independently measured through a calibration process along the edges of a uniformly magnetized ferromagnetic wire (*186, 187*), leading to $d = 49.0 \pm 2.4$ nm. A typical magnetic image recorded above the surface of the film of $BiFeO_3$ is shown in Figure 39d. We observe a periodic variation of the magnetometer signal along the horizontal axis, which directly reveals the spatially oscillating magnetic field generated by the spin cycloid. Moreover, the propagation direction of the spin cycloid is periodically modified along the vertical axis. The resulting zig-zag shaped magnetic field distribution mimics the width (i.e., ∼ 100 nm) of ferroelectric domains (Figure 39c).

To gain further insights into the properties of the spin cycloid in this 32-nm-thick film of $BiFeO_3$, PFM was used to design a single ferroelectric domain (Figure 40a), taking advantage of the trailing electric field induced by the slow scan axis of the scanning probe (*188*). The magnetic field distributions recorded over such a ferroelectric monodomain exhibit a simple periodic structure (Figure 40c), indicating the presence of a single spin cycloid. Clearly, the (001) surface projection of the spin cycloid propagation direction is normal to that of the polarization. Among the three possible cycloid propagation directions, only $k_1$ is normal to the (001) projection of $P_1^+$, the other two lying at 45° (Figure 40b). We therefore conclude that the spin cycloid propagates along $k_1$, i.e. in the thin film plane. This result can be explained by considering that the epitaxial strain modifies the magnetic anisotropy along the growth direction (*154*). For $BiFeO_3$ thin films grown on $DyScO_3$, compressive strain stabilizes magnetic structures with their spins away from the [001] direction, which lifts the degeneracy between the three possible cycloidal propagation directions (*154*). Sinusoidal fits of magnetic profiles along the propagation direction of the spin cycloid result in a characteristic period of $\lambda = 70.6 \pm 1.4$ nm. The slightly enhanced period compared to the bulk value is interpreted by the small biaxial compressive strain induced by the substrate (*154*). This result illustrates that the local magnetoelectric interaction between neighboring atoms at the origin of the spin cycloid does not require thick films of $BiFeO_3$, i.e. with thicknesses well above its characteristic wavelength (*189*).

After demonstrating that the polarization and the cycloid propagation direction are intimately linked, we electrically manipulated this cycloid using the magnetoelectric coupling by defining another ferroelectric domain with an in-plane component of the polarization rotated by 90° ($P_4^-$). The magnetic image shows that the propagation direction of the spin cycloid is also rotated by 90° with a very similar wavelength ($\lambda = 71.4 \pm 1.4$ nm), corresponding to the propagation direction $k_1'$ lying again in the (001) plane. These experiments illustrate how magnetoelectric coupling can be used to efficiently control and manipulate the antiferromagnetic order in a BiFeO$_3$ thin film. They also confirm that the abrupt rotations of the antiferromagnetic order observed in Figure 39d are occurring at ferroelectric domain walls.

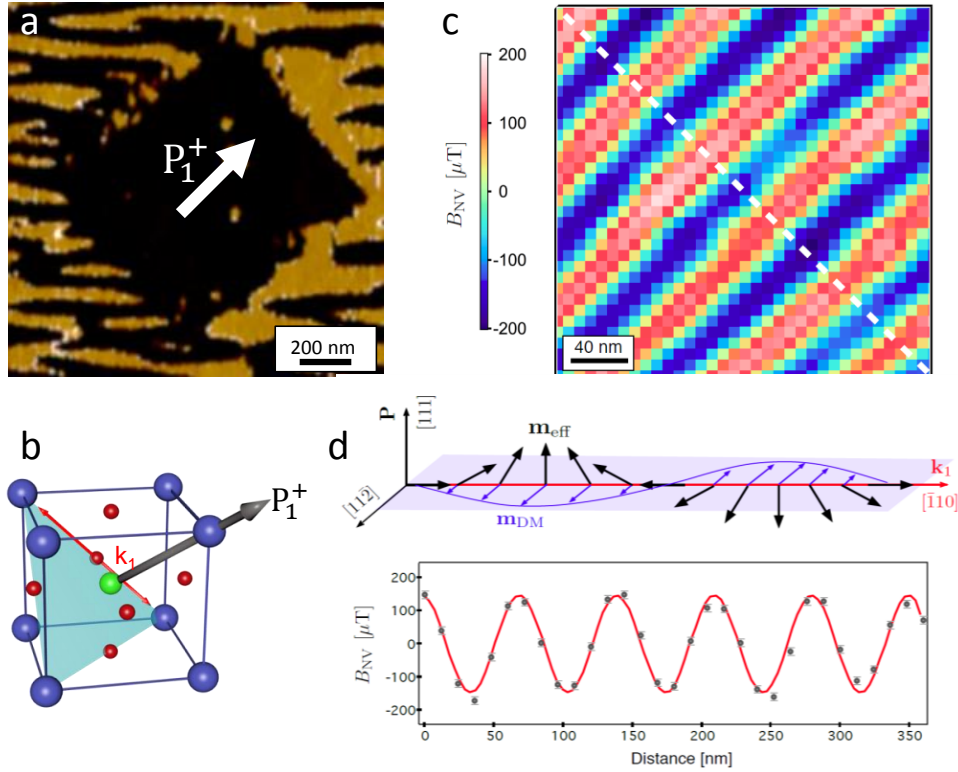

Figure 40. Electric control of the spin cycloid in BiFeO$_3$ and quantitative analysis. (a) In-plane PFM image of a ferroelectric monodomain with $P_1^+$ polarization. (b) Sketch of the ferroelectric polarization vector together with the propagation vector of the spin cycloid $k_1$. (c) Fully-quantitative magnetic field distribution $B_{NV}$ recorded above the ferroelectric domain displayed in a, showing a single cycloid with propagation vector $k_1$. (d) (top) Schematic representation of the spin density wave (SDW) corresponding to an uncompensated magnetic moment $m_{DM}$ (blue arrows) oscillating along the $[11\bar{2}]$ direction, i.e. perpendicular to both the polarization vector and $k_1$. The uncompensated moment due to the pure cycloid $m_{eff}$ is shown with black arrows. (bottom) Profile of the magnetic field distribution along the cycloid propagation direction (dashed line in (c)). The symbols are the experimental data and the red solid line is the result of a fit using the analytical formula of the stray field produced by the sample for $d$ = 49 nm, $m_{eff}$ = 0:07 $\mu_B$, $\lambda$ = 70 nm, $t$ = 32 nm. The only free parameter is $m_{DM}$.

Finally, a quantitative analysis of the magnetic field image (Figure 40c) recorded above the ferroelectric domain can be done. In order to understand the magnetic field modulation in the range of ∼ 140 µT, we computed the stray field produced by the film of BiFeO$_3$. To this end, the antiferromagnetic order is modeled by a rotating uncompensated magnetization vector $m_{\text{eff}}$ describing a cycloid in the plane defined by the ferroelectric polarization vector and $k_1$ (Figure 39b). The uncompensated magnetic moment per Fe is given by $m_{eff} = m_{Fe}\sin(\alpha_C/2)$, where $m_{Fe} = 4.1$ µB is the measured magnetic moment of Fe atoms in BiFeO$_3$ at room temperature (*190*) and $\alpha_C$ is the canting angle between antiferromagnetically coupled Fe atoms. This angle is directly deduced from the measured cycloid wavelength leading to $\alpha_C = 2°$ and $m_{eff} = 0.07$ $\mu_B$. The Dzyaloshinskii-Moriya (DM) interaction resulting from the alternate rotation of the FeO$_6$ octahedra along the [111] direction is another source of deviation from a pure antiferromagnetic order in BiFeO$_3$ (*191*, *192*). In the homogeneous G-type state obtained at high magnetic fields (> 20 T), this effect is known to generate a weak and uniform magnetization in BiFeO$_3$. In the cycloidal state, this magnetization is converted into a spin density wave (SDW) oscillating in the $[11\bar{2}]$ direction, which leads to a periodic wiggling of the cycloidal plane (*193*). This SDW can be simply modeled by an additional uncompensated magnetic moment $m_{DM}$ whose magnitude oscillates along the cycloid propagation direction (Figure 40d) (*193*). The value of the SDW amplitude still remains debated: it is often considered small (0.03 $\mu_B$) (*194*) or even negligible (*195*), but polarized neutron scattering studies revealed a maximum amplitude of 0.09 $\mu_B$ in bulk BiFeO$_3$ (*193*).

We performed an analytical calculation of the stray field produced above the BiFeO$_3$ sample in which $m_{DM}$ is the only free parameter. The fit of the experimental data (Figure 40d) leads to $m_{DM} = 0.16 \pm 0.06$ $\mu_B$ where the overall uncertainty mainly reflects the imperfect knowledge of the probe-to-sample distance. Consequently this study suggests a stronger DM interaction than all reported values from the literature. This enhanced DM interaction could result from the abrupt broken inversion symmetry at the sample surface that propagates in the film by exchange interaction. This observation opens many perspectives for the investigation of emergent interface-induced magnetic interactions resulting from a local breaking of inversion symmetry.

In summary, we report the first real-space imaging and electric-field control of the cycloidal antiferromagnetic order in a BiFeO$_3$ thin film using a scanning NV magnetometer operating under ambient conditions. These results open new perspectives for unraveling intriguing phenomena occurring in multiferroic materials like BiFeO$_3$, from magnetoelectric coupling (*168*), peculiar properties induced by surface symmetry breaking, to conduction and magnetotransport properties at ferroelectric domain walls (*102*, *196*). On a broader perspective, NV magnetometry appears as a unique tool to study the antiferromagnetic order at the nanoscale. In this way, similar investigations could be extended to a myriad of non-collinear antiferromagnetic materials, or to the domain walls of regular antiferromagnets, open an exciting avenue toward the development of low-power antiferromagnetic spintronics (*172*).

**Related publications:**
*Real-space imaging of non-collinear antiferromagnetic order with a single spin magnetometer*
I. Gross, W. Akhtar, V. Garcia, L.J. Martinez, S. Chouaieb, K. Garcia, C. Carrétéro, A. Barthélémy, P. Appel, P. Maletinsky, J.-V. Kim, J.-Y. Chauleau, N. Jaouen, M. Viret, M. Bibes, S. Fusil, V. Jacques
**SUBMITTED (2017)**

# 4. PERSPECTIVES

## 4.1 Electric control of a Mott insulator

Controlling strongly-correlated electronic states and inducing metal/insulator transitions by electric-field effect is the key objective of Mott-tronics. Conceptualized by IBM in 1998 (*197*), Mott transistors could surpass conventional metal-oxide field effect transistors (MOSFETs) in terms of OFF/ON ratios and power consumption. With ferroelectric gates, they could also lead to fast and high-endurance non-volatile memories. Two key challenges must be addressed to meet the long-standing goal in Mott-tronics of a non-volatile, reversible, electronically-driven transition between a metallic and an insulating state: (i) identify a channel material in which a metal-insulator transition occurs at very low doping level; (ii) combine it with a switchable gate material capable of accumulating and depleting large carrier densities.

In the framework of my recent ANR FERROMON project, we will address both challenges and investigate a model system consisting of epitaxial perovskite heterostructures combining a Mott insulator, (Ca,Ce)MnO$_3$, with a (magnetic) ferroelectric, BiFeO$_3$. BiFeO$_3$/(Ca,Ce)MnO$_3$ heterostructures are of high crystalline quality and we demonstrated a large electrical response at room temperature induced by ferroelectric switching in both planar (*198*) and vertical FTJ (*9*, *62*, *91*) devices. The rich phase diagram of (Ca,Ce)MnO$_3$ offers great potential for new exciting effects where ferroelectric field effect could not only drive metal/insulator but also magnetic phase transitions. The three main objectives of the project will be to (i) drive electronic and/or magnetic phase transitions in strongly correlated oxides by ferroelectric field effect, (ii) understand at the atomic level the interplay between ferroelectricity and electronic properties at oxide interfaces, (iii) exploit ferroelectric domain dynamics to control electronic and/or magnetic properties at the nanosecond scale.

### 4.1.1 Electron-doped manganites

Perovskite manganites display sharp phase transitions when they are subjected to external stimuli such as magnetic fields, electric fields or light (*199*). The origin of these phase changes is the competition between different ground states which are intimately coupled to the spin, charge and orbital ordering of the Mn-3*d* states (*200*, *201*).

Bulk CaMnO$_3$ crystalizes in the paraelectric P*bnm* orthorhombic ground-state structure where the tilting of oxygen octahedra produces a small distortion from the cubic structure of the perovskite. It is a G-type antiferromagnetic insulator (G-AFI) (*202*) mainly governed by superexchange interactions, although Dzyaloshinskii-Moriya interactions at distorted Mn-O-Mn bonds cause spin canting, leading to a non-collinear G$_x$A$_y$F$_z$ order with weak ferromagnetic moment of 0.04 µB/Mn (*203*). The chemical substitution of Ca$^{2+}$ by Ce$^{4+}$ in Ca$_{1-x}$Ce$_x$MnO$_3$ manganites gives two electrons in the Mn-3*d* (e$_g$) band which induces a transition to a metallic state around x = 5% (*202*). Thus, the metal-insulator transition occurs at much lower carrier concentrations than in widely investigated hole-doped manganites (*204*). This makes electron-doped manganites very appealing in view of integrating them as channel materials in oxide-based field effect transistors (FETs) where metal-insulator transitions could be triggered by an electric field. The delocalization of carriers resulting from Ce doping generates ferromagnetic double-exchange (*205*) and the transition to a metallic state is accompanied by the onset of a canted magnetic state (cG-AFM) (*202*). Higher Ce doping induces an orbital/charge ordering transition to a C-type antiferromagnet (C-AFI) (*206*). In thin films, the epitaxial strain imposed by the substrate has a profound effect on the phase diagram of (Ca,Ce)MnO$_3$ (*207*). While compressive strain (-0.5% on YAlO$_3$ substrates) stabilizes the cG-AFM metallic state at much lower doping than in the bulk (Figure 41a), tensile strain (+0.8% on LaSrAlO$_4$ substrates) suppresses the metal-insulator transition (*208*). This suggests that both electron doping and lattice distortions are critical parameters governing the ground states of (Ca,Ce)MnO$_3$.

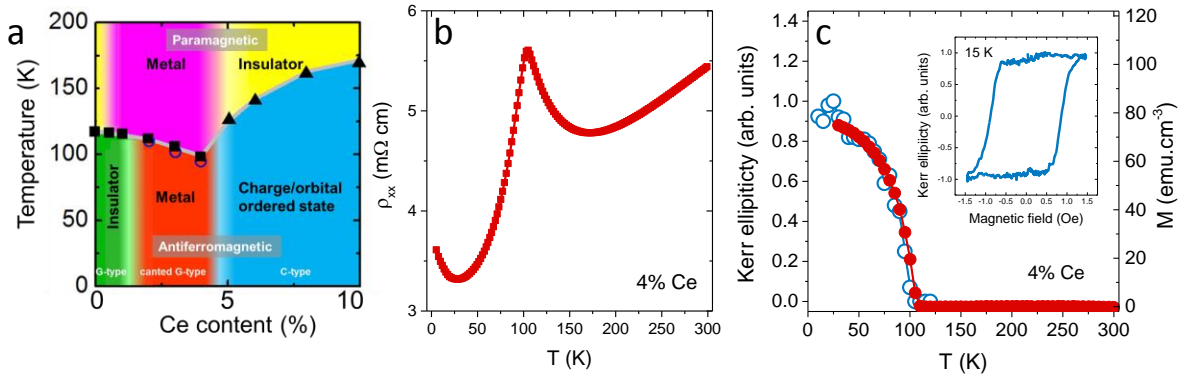

*Figure 41. (a) Phase diagram of (Ca,Ce)MnO$_3$ thin films grown on YAlO$_3$(001) substrates (207). Transport and magnetic properties of a (20 nm) Ca$_{0.96}$Ce$_{0.04}$MnO$_3$ thin film grown on YAlO$_3$(001). (b) Longitudinal resistivity vs. temperature. (c) Out-of-plane magnetization as a function of temperature detected by Kerr ellipticity (open blue symbols) and SQUID magnetometry (full red symbols). The inset shows a hysteresis cycle at 15 K of the Kerr ellipticity with the out-of-plane magnetic field.*

We recently started to investigate the transport and magnetic properties of epitaxial films of Ca$_{0.96}$Ce$_{0.04}$MnO$_3$ thin films on YAlO$_3$(001). The resistivity of these metallic films shows a cusp at 110 K that signals the transition to a weak-ferromagnetic state (Figure 41b). The weak-ferromagnetic nature of the films with perpendicular anisotropy (*207*) is revealed by the temperature dependence of the magnetization (0.4 $\mu_B$/Mn) and of the Kerr ellipticity (collaboration with the group of Gervasi Herranz (ICMAB)) up to 110 K (Figure 41c). The Hall resistivity of the films (Figure 42a) shows unusual signatures. While a linear behavior signals the normal Hall effect for electron carriers at high temperature, a hysteresis component develops below the Néel temperature following the magnetization measured with Kerr ellipticity: it corresponds to the anomalous Hall effect (AHE) of a perpendicularly magnetized sample. Below 80 K, a third component appears in the form of a peak centered around 1T. Its peculiar shape is reminiscent of the topological Hall effect (THE) observed in skyrmion systems (*209–213*). To get more insight into the nature of the spin configurations responsible for the THE in the films, low-temperature magnetic force microscopy (MFM) measurements were performed in collaboration with the group of Weida Wu (Rutgers) (Figure 42b). During magnetization reversal, a large density of small bubble domains (~200 nm) peaks in the same field range as the THE, indicating that the THE is created by the specific spin texture associated with the bubbles. The large amplitude of the THE ($\mu\Omega$.cm) is interpreted by a Berry phase mechanism (*209*) in which the bubbles carry a topological charge, hence called skyrmion-bubbles (*214*). In addition, the strong sensitivity of the THE with the carrier density seems to indicate additional correlation effects near the metal-insulator transitions, and offers exciting perspectives for its non-volatile control in heterostructures combining (Ca,Ce)MnO$_3$ with ferroelectrics.

Further investigations are required to shed light on this system, exploring the high sensitivity of scanning NV magnetometry for example, as the bubble density seems to be higher than detected by MFM. These results would represent the first observation of a skyrmionic state in single films of complex oxides, opening a broad range of exciting possibilities for the control and coupling of skyrmions.

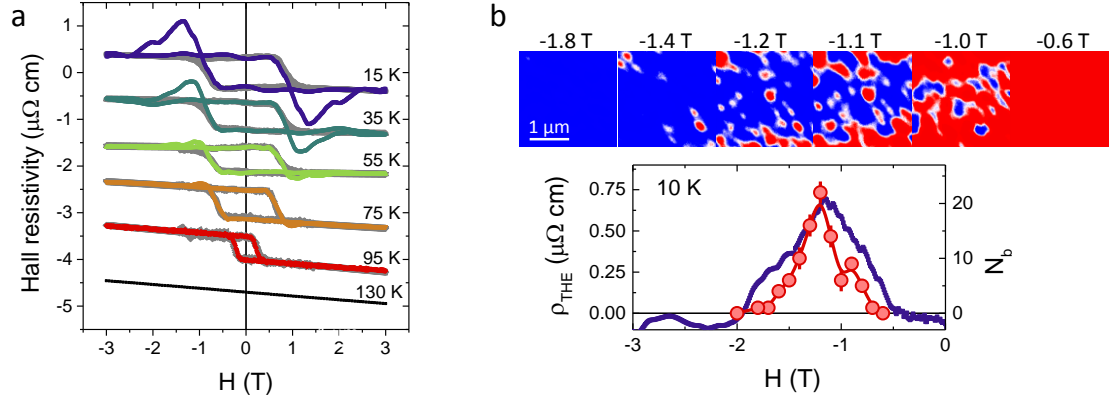

*Figure 42. Hall effect and micromagnetic structure of a Hall-bar-patterned $Ca_{0.96}Ce_{0.04}MnO_3$ (20 nm) thin film grown on $YAlO_3$(001). (a) Hall effect at different temperatures together with Kerr ellipticity data (grey lines). In addition to the AHE, THE is observed around 1 T. (b) MFM images at 10 K for different magnetic fields, after applying a positive perpendicular field of +3 T. Field dependence of the THE (blue symbols) and the number of observed bubbles in the MFM images (red symbols) at 10 K.*

In certain conditions, the strain-induced polar instability of $CaMnO_3$ thin films can result in a ferroelectric character which, combined with the antiferromagnetic state, makes the material multiferroic (*215*, *216*). In oxide perovskites, ferroelectric instabilities are competing against the tilt of oxygen octahedra (AFD instabilities). First-principles calculations show that strong AFD instabilities are responsible for the orthorhombic ground state of $CaMnO_3$, while a weak ferroelectric instability already exists at its equilibrium volume (*215*). Increasing the volume of the unit cell of $CaMnO_3$ favors the ferroelectric instabilities over the AFD ones which are volume independent. Thus, it is predicted that growing thin films under biaxial tensile strain should produce a ferroelectric transition: a ferroelectric state with an in-plane polarization is calculated for 2% of tensile strain. Interestingly, the ferroelectric instability is driven by the motion of Mn cations, illustrating that ferroelectricity and magnetism are not exclusive. An efficient alternative to increase the volume and stabilize a ferroelectric state in $CaMnO_3$ is chemical substitution on the A-site. Replacing Ca by a larger cation (Sr or Ba) strongly favors the ferroelectric distortion of the Mn and suppresses the tilts of oxygen octahedra. Experiments on single crystals of $Sr_{1-x}Ba_xMnO_3$ confirm these predictions (*215*) with the appearance of a ferroelectric state at room temperature with a sizeable polarization (> 4.5 μC/cm²) (*217*). In addition, the strong depression of the ferroelectric distortion below the magnetic ordering temperature suggests a very large magnetoelectric coupling (*217*). We aim at exploring these strategies to induce multiferroicity in thin films.

In summary, $CaMnO_3$ appears as a very flexible template to generate a broad range of physical properties through three main handles: (i) epitaxial strain, (ii) Ce doping, (iii) isovalent substitution on the A-site.

### 4.1.2 Ferroelectric control of electronic/magnetic phase transitions

Ferroelectric-field effect is equivalent to a switchable chemical doping. The large ferroelectric polarization in the supertetragonal phase of BiFeO$_3$ makes it, in principle, possible to accumulate or deplete carrier densities on the order of $3 \times 10^{14}$ cm$^{-2}$ in a non-volatile way. This is two orders of magnitude higher than what can be achieved with standard dielectric gate oxides, and comparable to values found with ionic liquids (*218*). In compressively strained (Ca,Ce)MnO$_3$ thin films grown on YAlO$_3$, very low carrier densities ($3 \times 10^{13}$ cm$^{-2}$ or 0.04 e$^-$/u.c.) are required to induce an insulator to metal transition (Figure 41a). Thus, in BiFeO$_3$/(Ca,Ce)MnO$_3$ bilayers, a purely electrostatic and reversible metal-insulator transition by ferroelectric polarization reversal is accessible.

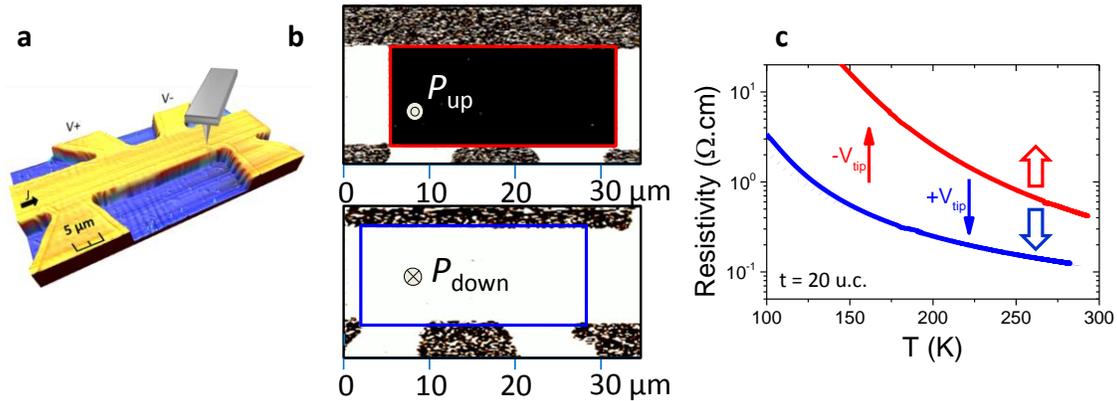

*Figure 43. (a) Sketch of the BiFeO$_3$/CaMnO$_3$ ferroelectric field-effect transistor. (b) Configuration of the ferroelectric gate imaged by PFM after local switching with the AFM tip. (c) Temperature dependence of the resistivity of the CaMnO$_3$ channel in the two states of ferroelectric polarization of the gate (198).*

In the literature, electrostatic modulations of the conduction of a CaMnO$_3$ channel was demonstrated using an ionic liquid gate, showing a large ON/OFF conductance ratio of 1200 at 50 K and 4 at 300 K (*208*). In BiFeO$_3$/CaMnO$_3$ bilayers, our first experiments evidenced large modulations of the resistivity of up to 10 at 200 K and 4 at 300 K (Figure 43) (*198*). The combination of STEM-EELS experiments with ab initio calculations revealed the importance of interfacial electronic reconstructions at the BiFeO$_3$/CaMnO$_3$ interface that minimize the field-induced carrier modulations despite the large polarization of the ferroelectric (*54*). While these investigations reveal that the field effect is only efficient over 3 unit-cells (Figure 10), Fe-FET transport experiments suggest the presence of a dead layer between CaMnO$_3$ and the YAlO$_3$ substrate (*198*), which prevents the realization of ultrathin electronically-active layers of CaMnO$_3$. These results show the great potential of Mott insulators such as CaMnO$_3$ as this is the largest non-volatile room-temperature modulation of resistivity achieved in a Fe-FET. However, progress must be made in the control and understanding of the interface properties in order to enhance the gating efficiency. Then, reversible and non-volatile metal-insulator transitions will be achievable at room temperature, with large associated OFF/ON ratios. Moreover, we should be able to control the magnetic ground state of the channel, corresponding to a purely electronic magnetoelectric effect that has never been reported before. In particular, we wish to control the skyrmion-bubbles formation and their associated Hall signal with the ferroelectric gate. Finally, solid-state ferroelectric gating will enable the operation of devices at the nanosecond scale (as was demonstrated with FTJs (*9*, *62*)) and the control of electronic and magnetic response at the nanometer scale using scanning probe microscopy.

## 4.2 Ferroelectric/antiferroelectric devices for neuromorphic computing

### 4.2.1 Antiferroelectric thin films

Ferroelectric materials possess a spontaneous electric polarization that can be reversed by the application of an external electric field. This polarization is the sum of individual electric dipoles in each unit cell of the material that align parallel to each other. In contrast, antiferroelectric materials, in which adjacent electric dipoles align antiparallel to each other (Figure 44a), show no spontaneous polarization, but rather double hysteresis loops of polarization vs. electric field (Figure 44b).

More precisely, an antiferroelectric material is similar to a ferroelectric one in that its structure is obtained through distortion of a nonpolar high-symmetry reference phase. In addition, there must be an alternative low energy ferroelectric phase by a polar distortion of the same high-symmetry reference phase. An applied electric field must induce a first-order phase transition from the antiferroelectric to the ferroelectric phase, producing a characteristic double hysteresis loop of polarization vs. electric field (*219*). Complementarily, antiferroelectrics are often described as materials that exhibit a structural phase transition at zero field between two nonpolar phases with a strong dielectric anomaly around this transition (*220*, *221*). In systems in which the two phases have different unit-cell volumes, this transition is accompanied by large nonlinear strain responses. In addition, antiferroelectric materials display large electrostriction coefficients and giant electrocaloric effects (*222*, *223*), leading to potential applications as high-energy storage capacitors, electrocaloric refrigerators and high-strain actuators.

Currently the latest developments of artificial neural networks in hardware rely on electronic synapses made of memristors and neurons fabricated with CMOS. The use of two different technologies for synapses and neurons complicates their interconnection and possible integration within a single wafer. Here we propose a concept of electronic neurons with similar materials as for ferroelectric synapses. Indeed, by inserting an antiferroelectric tunnel barrier between two electrodes (*224*), we aim at realizing a relaxation oscillator that emulates the behavior of a spiking neuron (*225*). Considering a circuit containing a serial resistance, a capacitance, and the unipolar resistive memory based on the antiferroelectric tunnel junction, the application of a dc voltage will lead to an oscillating output voltage mimicking the behavior of the neuron. In order to reach this goal, ultrathin films of antiferroelectric need to be closely investigated. However, the field of antiferroelectric thin films is still in its infancy compared to ferroelectric ones and requires a better understanding of the underlying mechanisms to set the basis for the design of future nanoelectronic devices.

Besides the use of lead-based materials such as archetypical $PbZrO_3$ (Figure 44) (*226*), we will explore a possible way to make antiferroelectric films via chemical substitution on the A-site of ferroelectric $BiFeO_3$. The chemical substitution of $Bi^{3+}$ ions with smaller rare earth isovalent cations influences both the tilt pattern of oxygen octahedra and the long range ferroelectric interaction. For example, (Bi,Sm)FeO₃ thin films exhibit a morphotropic phase boundary (MPB) in which a structural transition is accompanied by a ferroelectric to antiferroelectric transition (*227*). On a more exploratory perspective, domain walls in antiferroelectric materials are expected to show peculiar properties as exemplified by the appearance of a polar state at twin boundaries of CaTiO₃ crystals (*228*). In addition, ferroelectricity was recently suggested at antiphase boundaries of antiferroelectric PbZrO₃ single crystals (*229*). These ferroelectric domain walls can be viewed as nanoscale (1-10 nm) objects carrying information.

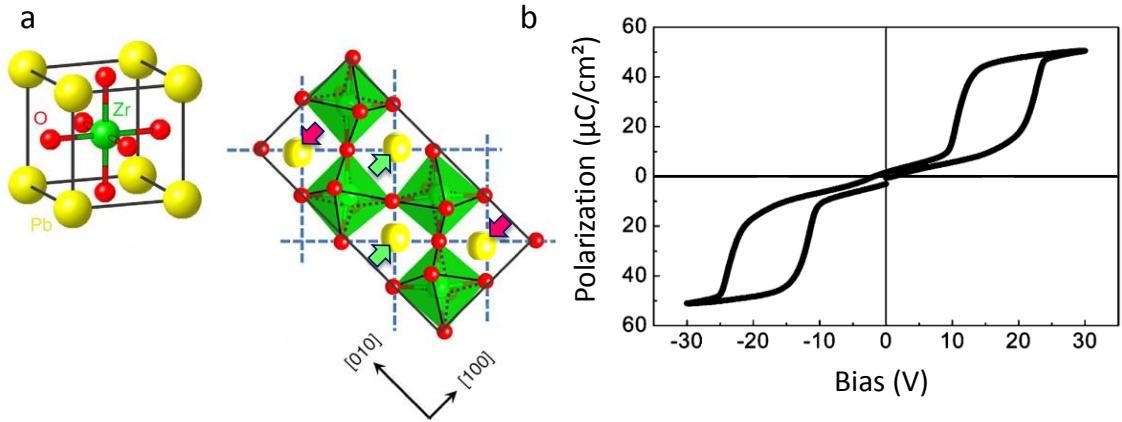

*Figure 44. (a) Structure of the archetypical antiferroelectric: PbZrO₃. (left) Cubic unit cell of the perovskite.(right) Lead displacements and antiphase rotations of oxygen-octahedra in the orthorhombic cell (229). (b) Double hysteresis polarization vs. voltage loop in a 385 nm thick antiferroelectric PbZrO₃ film (230).*

To summarize, this activity on antiferroelectric thin films will be both fundamental (interplay between strain and tilt patterns, phase transitions, functional domain walls) and applied to the realization of antiferroelectric tunnel junctions to emulate the behavior of spiking neurons.

### 4.2.2 Neural networks with ferroelectric synapses

Conventional image sensors suffer from significant limitations imposed by their principle and operation. Acquiring visual information frame-by-frame limits the temporal resolution of the data, leading to distorted information from fast moving objects. This results in vast amounts of redundancy unnecessarily inflating data rate and volume, and yields relatively poor intra scene dynamic range. The neuromorphic event-based approach to vision and image sensing is recently gaining substantial attention as it proposes solutions to all the previously mentioned issues encountered with conventional technology. Neuromorphic vision systems are driven by events happening within the scene in view, as in biology, and unlike conventional image sensors, which are triggered by artificially-created timing and control signals. As a result, each pixel in a sensor array samples its visual input individually and adapts its sampling rate to the dynamics of the input signal itself. The output of such a sensor is a time-continuous stream of pixel data, delivered at unprecedented temporal resolution, containing zero redundancy, and encoding orders of magnitude higher dynamic range than conventional imagers (*231–233*).

Bio-inspired event-driven vision and image sensors have reached a maturity level where they start being applied to an increasing variety of machine vision tasks. However, the event-based asynchronous output of these sensors has been processed using conventional computing devices such as central processing units (CPUs). This way of processing is obviously non ideal and does not allow to fully benefit from the unique characteristics of the sensor. The logical way would be to process the asynchronous event output with an event-based computational network based on neurons and synapses. In collaboration with the group of Sylvain Saïghi (Univ. Bordeaux), the Institut de la Vision and ChronoCam, the ANR MIRA project proposes to explore this concept by building a demonstrator in which a bio-inspired event-driven pixel array will be connected to an array of silicon-based neurons interconnected by ferroelectric synapses to solve an exemplary high-performance vision task.

Following the demonstration of spike-timing-dependent plasticity with ferroelectric tunnel junctions, governed by nucleation-limited switching dynamics, we aim at integrating such artificial synapses in spiking neural networks. Indeed, spike-timing-dependent plasticity is at the heart of unsupervised learning in spiking neural networks (*234*). This important result demonstrates that the ferroelectric memristor is an excellent building block for developing neuromorphic electronics based on memristive synapses (*94*). STDP has been demonstrated for different types of memristors (*101*, *120–126*) but the ferroelectric memristor is advantageous because of its speed, large OFF/ON ratio, retention, and endurance. To go beyond the single synapse to the circuit level, it is necessary to densely integrate memristors in crossbar arrays that can be interconnected with CMOS neurons (*116*). Fabricating functional crossbar arrays of memristors is at the state of the art. We have recently fabricated 81 × 10 crossbar arrays of ferroelectric memristors with encouraging preliminary results. There is no published demonstration yet of unsupervised learning with crossbar arrays of memristors. This is indeed very challenging, in particular because it requires devices with high reproducibility and endurance. In the MIRA project we want to demonstrate that ferroelectric memristors are ideal building blocks to demonstrate unsupervised learning in large scale neuromorphic circuits. For this purpose, we intend to demonstrate that these devices can be processed in large scale and dense crossbar arrays, and interfaced with CMOS neurons and sensors for bio-inspired visual processing.

In the European H2020 ULPEC project (starting early 2017), the long term goal is to develop advanced vision applications with ultra-low power requirements and ultra-low latency. The output of the project is a demonstrator connecting a neuromorphic event-based camera to a high speed ultra-low power asynchronous visual data processing system (spiking neural network with memristive synapses). The project consortium includes the partners mentioned from the MIRA project as well as an industrial end-user (Bosch) which will investigate autonomous and computer assisted driving. For this kind of applications, vision and recognition of traffic event must be computed with very low latency and low power. Our main role will be to use our expertise with ferroelectric memristors to develop crossbar arrays of these ferroelectric memristors on silicon within this project. The oxide heterostructures on silicon will be elaborated by our partners in IBM Zurich, Univ. of Twente, and Twente Solid State Technology and we will probe their ferroelectric and transport properties using scanning probe microscopy and pulse-assisted electrical transport. The best identified systems on silicon will then be integrated as crossbar arrays with CMOS using wafer bonding.

*Appl. Phys. Lett.* **87**, 222114 (2005).

145. L. Bocher *et al.*, Atomic and electronic structure of the BaTiO$_3$/Fe interface in multiferroic tunnel junctions. *Nano Lett.* **12**, 376–382 (2012).

146. K. Bouzehouane *et al.*, Nanolithography Based on Real-Time Electrically Controlled Indentation with an Atomic Force Microscope for Nanocontact Elaboration. *Nano Lett.* **3**, 1599–1602 (2003).

147. S. Valencia *et al.*, Interface-induced room-temperature multiferroicity in BaTiO$_3$. *Nature Mater.* **10**, 753–758 (2011).

148. G. Radaelli *et al.*, Electric control of magnetism at the Fe/BaTiO$_3$ interface. *Nature Commun.* **5**, 3404 (2014).

149. D. Pantel, S. Goetze, D. Hesse, M. Alexe, Reversible electrical switching of spin polarization in multiferroic tunnel junctions. *Nature Mater.* **11**, 289–93 (2012).

150. V. Garcia *et al.*, Temperature dependence of the interfacial spin polarization of La$_{2/3}$Sr$_{1/3}$MnO$_3$. *Phys. Rev. B*. **69**, 52403 (2004).

151. Z. Wen, L. You, J. Wang, A. Li, D. Wu, Temperature-dependent tunneling electroresistance in Pt/BaTiO$_3$/SrRuO$_3$ ferroelectric tunnel junctions. *Appl. Phys. Lett.* **103**, 132913 (2013).

152. Y. W. Yin *et al.*, Coexistence of tunneling magnetoresistance and electroresistance at room temperature in La$_{0.7}$Sr$_{0.3}$MnO$_3$/(Ba, Sr)TiO$_3$/La$_{0.7}$Sr$_{0.3}$MnO$_3$ multiferroic tunnel junctions. *J. Appl. Phys.* **109**, 07D915 (2011).

153. I. C. Infante *et al.*, Bridging multiferroic phase transitions by epitaxial strain in BiFeO$_3$. *Phys. Rev. Lett.* **105**, 57601 (2010).

154. D. Sando *et al.*, Crafting the magnonic and spintronic response of BiFeO$_3$ films by epitaxial strain. *Nature Mater.* **12**, 641–646 (2013).

155. R. J. Zeches *et al.*, A Strain-Driven Morphotropic Phase Boundary in BiFeO$_3$. *Science*. **326**, 977–980 (2009).

156. D. Sando *et al.*, Large elasto-optic effect and reversible electrochromism in multiferroic BiFeO$_3$. *Nature Commun.* **7**, 10718 (2016).

157. S. Fusil, V. Garcia, A. Barthélémy, M. Bibes, Magnetoelectric Devices for Spintronics. *Annu. Rev. Mater. Res.* **44**, 91–116 (2014).

158. R. O. Cherifi *et al.*, Electric-field control of magnetic order above room temperature. *Nature Mater.* **13**, 345–51 (2014).

159. J. M. D. Coey, Noncollinear spin structures. *Can. J. Phys.* **65**, 1210–1232 (1987).

160. U. Hartmann, Magnetic force microscopy. *Annu. Rev. Mater. Res.* **29**, 53–87 (1999).

161. A. Locatelli, E. Bauer, Recent advances in chemical and magnetic imaging of surfaces and interfaces by XPEEM. *J. Phys. Condens. Matter*. **20**, 93002 (2008).

162. N. Wu *et al.*, Imaging and Control of Surface Magnetization Domains in a Magnetoelectric

# 6. Acknowledgments